\documentclass[12pt,a4paper]{article}

\usepackage{verbatim}
\usepackage{enumitem}
\usepackage[left=2.3cm,top=2.4cm,right=2.3cm,bottom=2.4cm]{geometry}
\usepackage[english]{babel} 
\usepackage{pdflscape} 
\usepackage{indentfirst}

\usepackage{setspace} 
\usepackage{ae,aecompl}
\usepackage[T1]{fontenc} 
\usepackage{epstopdf}
\usepackage{everypage}  
\usepackage{lipsum} 
\usepackage{authblk} 
\usepackage[super]{nth} 
\usepackage{xcolor}

\usepackage{authblk}

\usepackage{graphicx} 
\usepackage{float}
\usepackage{color,soul}
\usepackage{adjustbox} 
\usepackage{rotating} 
\usepackage{subfigure} 

\usepackage{amsmath,amsfonts,bm,amssymb} 
\usepackage{latexsym}
\usepackage{calc}

\usepackage[flushleft]{threeparttable} 
\usepackage{booktabs} 
\usepackage{array} 
\usepackage{longtable} 
\usepackage{multirow} 
\usepackage{placeins} 

\usepackage{siunitx} 
\usepackage{makecell} 

\usepackage[toc,page]{appendix}
\usepackage[authoryear]{natbib}
\bibpunct{(}{)}{;}{}{,}{,} 
\usepackage{hyperref}
\hypersetup{
     colorlinks=true,
     linkcolor=blue,
     filecolor=blue,      
     urlcolor=blue,
     citecolor=blue,
     pdftitle={Overleaf Example},
     pdfpagemode=FullScreen,
     }

\usepackage[nobiblatex]{xurl} 
\usepackage[flushleft]{threeparttable}

\title{Leveraging Knowledge Networks: Rethinking Technological Value Distribution in mRNA Vaccine Innovations}

\author[1]{Rossana Mastrandrea\thanks{Email: rossana.mastrandrea@unito.it}}
\author[2,6]{Fabio Montobbio}
\author[2]{Gabriele Pellegrino}
\author[3,4]{Massimo Riccaboni}
\author[5]{Valerio Sterzi}

\affil[1]{\small Dept. of Management, University of Turin, Italy}
\affil[2]{\small Dept. of Political Economy, Universit\`{a} Cattolica, Milano,Italy}
\affil[3]{\small IMT School for Advanced Studies, Lucca, Italy}
\affil[4]{\small IUSS, Pavia, Italy}
\affil[5]{\small Bordeaux School of Economics (BSE), Univ. Bordeaux, CNRS, UMR 6060}

\date{}

\begin{document}

\maketitle

\begin{abstract}


This study examines the roles of public and private sector actors in the development of mRNA vaccines, a breakthrough innovation in modern medicine. Using a dataset of 151 core patent families and 2,416 antecedent (cited) patents, we analyze the structure and dynamics of the mRNA vaccine knowledge network through network theory. Our findings highlight the central role of biotechnology firms, such as Moderna and BioNTech, alongside the crucial contributions of universities and public research organizations (PROs) in providing foundational knowledge. We develop a novel credit allocation framework, showing that universities, PROs, government and research centers account for at least 27\% of the external technological knowledge base behind mRNA vaccine breakthroughs—representing a minimum threshold of their overall contribution. Our study offers new insights into pharmaceutical and biotechnology innovation dynamics, emphasizing how Moderna and BioNTech’s mRNA technologies have benefited from academic institutions, with notable differences in their institutional knowledge sources.

\end{abstract}



\section{Introduction}

The timely development of vaccines against COVID-19 is an extraordinary scientific achievement and a fundamental step in the fight against the pandemic. While some vaccines (such as Vaxzevria from AstraZeneca and the University of Oxford and JNJ-78436735 from Johnson \& Johnson) were based on relatively established technologies, the most effective vaccines, still in use, come from a new mRNA vaccine platform developed in record time. The mRNA technology is transforming vaccinology and opening up new opportunities for the development of new therapies in many therapeutic areas such as oncology, cardiovascular diseases, cystic fibrosis \citep{androsavich2024,pardi2018,sahin2014}. This rapid development of mRNA vaccines would not have been possible without prior investment in basic research on mRNA, in methods to improve the efficiency of mRNA delivery (e.g. lipid nanoparticles) and in pharmacological modifications to reduce its instability and innate immunogenicity \citep{pardi2018,martin2020mrna,slaoui2020}. As governments and international institutions have faced the challenge of ensuring access to vaccines for the entire global population, understanding the processes and actors that have driven this important biomedical innovation offers important insights for policy beyond the COVID-19 pandemic \citep{agarwal2022,florio2022extent,florio2023mapping,d2024can}.

This paper analyses the different roles of the public and private sectors as a source of breakthrough innovation in the life sciences, with a focus on the development of mRNA vaccines for COVID-19. Such breakthrough innovations are fundamental to our society as they drive new developments in science and technology and provide important opportunities to increase social welfare and economic growth \citep{hall2005,phene2006, kaplan_double-edged_2015, kuhn_structure_1962}.
Typically, breakthrough innovations challenge the existing technological landscape, open up new technological trajectories enabling business growth and new business development. 

In order to gain a comprehensive understanding of the role played by the different actors in the development of mRNA vaccines, three different research objectives are pursued in this article.
The first objective is to assess the role of different types of organizations within the mRNA patent landscape. We investigate how universities, research centers, biotech companies and pharmaceutical companies have contributed to the development of mRNA vaccines. We distinguish between organizations that have contributed directly to the mRNA vaccine platform (``Core") and those that have developed enabling technologies that support the development of mRNA vaccines (``Antecedent”).

The second objective is to identify the key players in the development of the mRNA vaccine platform by relying on network analysis. We construct a comprehensive network of relationships between different types of organizations based on backward patent citations, the so-called \textit{mRNA vaccine knowledge network}. We use centrality measures - such as \textit{in-} and \textit{out-strength}, \textit{betweenness centrality} and \textit{hub \& authority scores} - to assess the importance of different actors in the mRNA vaccine knowledge network. This analysis allows us to identify the most influential actors and understand their importance based on local and global network characteristics.

Finally, the third objective is to explore how credit for mRNA vaccine discovery could be redistributed by leveraging the structural properties of the mRNA vaccine knowledge network. We introduce a novel network measure that allows us to assess the contribution of different actors to the development of specific technologies. With this measure, we can fairly distribute the merits in mRNA vaccine discovery by considering the contributions of cited patents as the main sources of knowledge. In this way, we aim to create a more nuanced understanding of how knowledge networks can be used to identify the most important contributions to the development of breakthrough innovations. 

We leverage a new and carefully selected large-scale dataset with a Core set of 151 patent families covering the mRNA vaccine platform (published from 2006 to 2020) and their citations to earlier patents, totaling 3\,625 patent documents (2\,416 patent families). For each patent document, we performed a manual check of the name of the assignee to ensure harmonization where applicable and to identify the type of assignee (e.g. pharmaceutical company, biotech firm, public research organization). Patent applications in the field of mRNA vaccine technology have increased significantly in the last seven years, while the basic knowledge cited in the prior art dates back to the mid-1980s. We find that biotech company patents dominate both patents directly related to mRNA vaccine technology (Core) and patents that enabled the development of the mRNA vaccine platform (Antecedent). However, universities also hold a significant share of patents in both groups, about 24\% in the Core group and 19\% in the Antecedent group. Remarkably, pharmaceutical companies are practically not represented in the Core group (4.6\%) and hold 17.5\% of patents in the Antecedent group.

Based on the network analysis, we find that universities and research centers such as MIT, California University, British Columbia University and Max Planck play an important role as `authorities' and `bridges' (authority and betweenness centrality) and biotech companies such as Curevac and Moderna have a particularly high hub centrality. Finally, our analysis shows that the main contribution to the discovery of Moderna and BioNTech vaccines comes from the biotech sector: 42.7\% and 45.2\% respectively. This is followed by universities, which are credited with 19.3\% and 21.7\% of the contribution of the knowledge network to the discovery of the Moderna and BioNTech vaccines respectively. This distribution underlines the importance that both the private and public sectors play in the mRNA knowledge landscape.

This paper adds to the existing literature by providing a comprehensive analysis of the roles played by various entities—universities, research centers, Biotechnology firms, and pharmaceutical companies—in the development of mRNA vaccine technology. Building on prior studies that have examined the dynamics of breakthrough innovations, we extend this research by integrating patent analysis with network theory to map the complex relationships that  underpinned the mRNA vaccine platform. While earlier research has emphasized the role of academic research and public funding in driving technological breakthroughs
 \citep{mansfield_academic_1991, sampat2009academic}, our study provides new insights into how these contributions are distributed across various actors.

In doing so, we introduce a novel approach to redistributing credit for technological innovations, addressing a significant gap in the existing literature on intellectual property rights and credit allocation. Traditional metrics, such as patent counts or simple citation numbers, often fail to capture the nuanced contributions of different entities in complex innovation ecosystems. By leveraging the structural properties of the mRNA vaccine knowledge network, we propose a more equitable method for allocating credit. This approach accounts for the significant, yet often underappreciated, role of public institutions in generating the foundational knowledge on which private firms rely. This aligns with the broader discourse on how innovation policy should more fairly distribute the economic and reputation rewards of scientific discovery, particularly in fields where public and private sectors both play essential roles \citep{mowery2004ivory, david2001public}.

The rest of the paper is organized as follows. Section 2 outlines the background of this study by introducing the specific characteristics of the mRNA ecosystem and discussing the contribution of academic research to the breakthrough innovation. Section 3 presents the data and key concepts of complex network theory and describes the methodology of the study. Section 4 presents the results. In the final section, we draw a conclusion and discuss some policy implications.

\section{Literature and Background}

Our contribution in this paper is based on three main strands of research. The first strand deals with the network origins of innovation in the biotechnology sector \citep{orsenigo2001}. The second research strand deals with the emergence of breakthrough innovations as a recombination process \citep{trajtenberg1997,fleming2001}. The last line of research provides a historical reconstruction of the most important inventions that paved the way for the discovery of mRNA vaccines. This literature is mainly descriptive and collects evidence for the key discoveries that enabled the mRNA revolution.

\subsection{The network of R\&D collaborations in the bio-pharmaceutical sector}

In biotechnology, the development of breakthrough innovations depends on close collaboration between companies and the scientific community, as well as complex interactions between public and private institutions \citep{owen2002comparison}. These interactions have been analyzed in an extensive literature, which shows that the link between innovation breakthroughs and scientific activities is facilitated, for example, by the engagement of star scientists and geographical proximity. Several studies also show that co-location of scientists working in both research institutions and companies significantly improves collaboration and research productivity \citep{phene2006,cockburn1998,cockburn2000,gambardella1995,zucker_intellectual_1998, zucker2002,powell_interorganizational_1996,orsenigo2001,mckelvey2003,pammolli2021}. This evidence is provided in the context of the introduction of biotechnology as a fundamental platform for drug discovery that emerged from the groundbreaking advances in genetics and molecular biology in the 1970s and 1980s (referred to as the new ``genetics paradigm'') \citep{gittelman2016, ng_drugs_2004}. The development of biotechnology paralleled the shifts in industrial organization and the establishment of new institutions. Key aspects of this model include the strengthening of intellectual property rights that facilitate technology transfer and the creation of new companies (e.g. Bayh-Dole Act), the rise of start-ups that use venture capital to drive early-stage research and are often driven by prominent academic entrepreneurs \citep{zucker2002, franzoni2022}, and the role of large pharmaceutical companies that license new targets and compounds from smaller companies and focus their resources on applied research, regulatory approval, manufacturing and commercialization.


There are significant direct effects when public research institutions are directly involved in the invention and hold patents, and significant indirect effects when private companies rely on innovative research in conjunction with publicly funded research programs. In the new genetic paradigm, public investment has been critical from the beginning. The development of rDNA technology and the mapping of the human genome are two compelling examples of how substantial public investment over many years has generated a wealth of commercial opportunities, technology transfer, licensing, and significant entry of new companies into the industry \citep{hughes2001, gittelman2016}.

The available empirical evidence, which includes both direct and indirect influences, supports the notion that public sector research has a significant impact on drug development \citep{mansfield_academic_1991, mansfield_academic_1995, mansfield1998, toole_does_2007,toole_impact_2012, narin1992status, narin_increasing_1997,cockburn1998, franzoni2022}. Basic research in academic institutions or public research centers provides the scientific basis for drug discovery by uncovering disease mechanisms, therapeutic strategies, drug targets and prototype compounds. In addition, the public sector makes an important contribution at the interface between the academic and commercial sectors through direct collaborations and partnerships with industry and through efforts to promote technology transfer and translational science in the public sector \citep{owen2002comparison}.

There is compelling evidence that basic research funded by the National Institutes of Health (NIH) has an economically and statistically significant impact on the development of new drugs in the US \citep{stevens2011role, dimasi2003price, kaitin1993role}. \citet{sampat2009academic} shows that 7.7\% of all U.S. Food and Drug Administration (FDA) approvals and 10.6\% of NMEs are based on academic patents. In addition, \citet{sampat2011respective} analyze 379 drugs approved between 1988 and 2007 and find that 48\% were associated with a patent based on prior art generated in the public sector, indicating a significant indirect influence of government funding. They emphasize that while nine percent of the drugs had a direct public sector patent, a much larger proportion had patents that cited either a public sector patent or a government publication. They also show that the influence of the public sector was more pronounced for the most innovative drugs \citep{patridge2015analysis, galkina_cleary_contribution_2018}.

Assessing the role of the public sector in drug development is particularly important in the context of mRNA vaccines, as they have a significant impact on health. Pfizer's Comirnaty, for example, generated nearly \$40 billion in sales in 2022, representing 38\% of the company's total sales. There are also relevant oppositions and infringement actions \citep{Montobbio_et_al_2024}\footnote{e.g. \url{https://www.iam-media.com/article/gsk-becomes-latest-mrna-patentee-enter-covid-ip-wars-increasing-pressure-pfizerBioNTech}} and important patent disputes between NIH and Moderna\footnote{
\url{https://www.nytimes.com/2021/11/09/us/moderna-vaccine-patent.html}} and between NIH and Pfizer\footnote{\url{https://www.iam-media.com/article/biontech-facing-fresh-mrna-patent-spat-in-vaccine-royalties-dispute}}. The question is whether governments should try to recoup the profits from the commercialization of drugs they have helped to develop. This policy question, which dates back to the debate surrounding the Bayh-Dole Act in the 1980s, concerns whether and to what extent the government should claim a portion of the profits from drugs for which government patents exist or that were developed through government-funded research and development. If the sources of new medicines are significantly publicly funded, it can be argued that private companies should not receive the lion’s share of the profits. There is also a risk that taxpayers pay twice for the new drugs: once through taxes with publicly funded R\&D and another time through high market prices driven by market power.

\subsection{At the origin of breakthrough innovation}

A breakthrough innovation is traditionally defined as a rare and disruptive technological change that can lead to a radical shift in the prevailing technological paradigm \citep{kuhn_structure_1962,dosi1982technological}. To operationalize the concept of breakthrough innovation, researchers have developed a set of metrics based on patent citations \citep{phene2006}. A first approach is to consider the most frequently cited patents as breakthrough innovations \citep{phene2006,ahuja2001}. More recently, backward citations have been used to identify patents that recombine basic and distant knowledge as potential breakthrough innovations \citep{dahlin2005,verhoeven2016,silvestri2018}. Although they are a reliable measure of technological radicalness, patent-based measures have some well-known pitfalls when it comes to identifying breakthrough innovations \citep{capponi2022}. Some much-cited patents do not lead to effective innovation from an economic and social point of view. Citations are also an imperfect measure of technological relevance, as citations can be distorted by competitive and social forces. \citet{kaplan_double-edged_2015} depart from a citation based metric of the innovation value and use topic modeling as a method to identify breakthrough ideas. In this article, we take a different approach by looking at mRNA vaccines as a prototypical example of a breakthrough innovation with enormous societal impact. Starting from an undeniable breakthrough innovation, we use patent citations to trace the foundations of mRNA vaccines as a recombinant innovation and the contribution of different actors to the discovery of mRNA technology.

\subsection{Patent Citations Networks and Credit Allocation}


This paper exploits the assumption that patent citations reveal how new inventions build upon the previous ones. By examining which patents cite which others, you can trace the evolution of mRNA vaccine technology and identify the key contributions. The network of patent citations can highlight the most influential innovators in the field. The existing economics literature provides well-documented evidence that the number of citations often correlates with the value of patents \citep{jaffe_patent_2017, hall_market_2005}. In addition, judges and legal experts often use patent citations in patent litigation to assess the value and impact of patents.\footnote{For example, in 2010, Oracle sued Google for alleged infringement of patents and copyrights related to application programming interfaces. The defendant’s damages expert used Patent Citation Analysis to evaluate and rank the value of three of the patents at issue in comparison to 22 other patents \citep{malaspina2019patent}; in "Finjan v. Blue Coat Systems, 2013", the court expressly recognized that “[...] qualitative analysis of asserted patents based upon forward citations may be probative of a reasonable royalty in some instances” \citep{malaspina2019patent}.} 

In addition network theory provides a natural approach to describe complex systems with many interacting actors such as the ecosystem of mRNA vaccine innovation \citep{barabasi2016network,newman2018networks}. There is a long tradition of studies using network analysis to study technological change \citep[see e.g.][]{orsenigo2001}. Network analysis has also been successfully used to study citation networks in various contexts \citep{hummon1989connectivity,kajikawa2007creating, ding2009pagerank,radicchi2011citation, wallace2012small, golosovsky2017growing, guan2017impact} including the patent landscape \citep{li2007patent, cho2011patent, erdi2013prediction, van2017patent, mariani2019early, chakraborty2020patent}. The use of patent citation networks has been validated in different papers to address different issues. For example, \citet{barbera2011mapping} use patent citations to analyze the technological trajectories in surgical prostheses. In the wind power technology, a patent citation network is used to evaluate how a product's design affects the inventive activity \citep{huenteler2016product}. \citet{malhotra2021new} study the lithium-ion battery technology and show how the emergence of new use environments shape knowledge generation and product innovation. Finally \citet{iori_direction_2022} use patent citation networks to analyze how the role of government grants and patents by Federal Agencies and State Departments has influenced the development of artificial intelligence. 

Building on this literature, we follow a different approach to analyze mRNA vaccines, focusing on the contribution of different (types of) organizations. In particular we leverage the structural property of the mRNA knowledge network and present a new method for evaluating contributions to a breakthrough technological innovation. In so doing we provide an attempt to improve on credit attribution research. Conventional measures, like patent counts or raw citation frequencies oversimplify the complex dynamics within innovation ecosystems, failing to accurately reflect the diverse contributions of various actors. This aligns with broader discussions regarding the fair distribution of economic and reputational benefits arising from scientific discovery, especially in sectors like vaccine development where both public and private investment are critical \citep{mowery2004ivory, david2001public}.

\subsection{The mRNA Innovation Ecosystem}

The mRNA revolution in vaccinology has attracted the attention of the scientific community and is considered a prime example of a breakthrough innovation. There are a number of extremely interesting research articles and inspiring reports on how different scientific trajectories have progressed independently and contributed to the development of the new mRNA vaccines \citep{sahin2014, pardi2018, dolgin2021covid, dolgin2021tangled, zuckerman_shot_2021, fauci_story_2021,veugelers2021mrna,barbier2022, franzoni2022}.\footnote{\url{https://www.nytimes.com/2022/01/15/health/mrna-vaccine.html; https://www.ft.com/content/b2978026-4bc2-439c-a561-a1972eeba940; https://www.forbes.com/sites/nathanvardi/2021/08/17/ \\covids-forgotten-hero-the-untold-story-of-the-scientist-whose-breakthrough-made-the-vaccines-possible/}}. This detailed account of technological and scientific developments confirms that the main features of the mRNA ecosystem are consistent in that the vaccine is the result of the convergence of three major scientific pathways:

\begin{enumerate}
    
    \item \textbf{The discovery of mRNA in 1961}: This discovery provided the impetus for a branch of research that remained a scientific backwater for decades. The main problems were the instability of mRNA and innate immunogenicity. A team of scientists at the Pennsylvania University led by Katalin Karikò and Drew Weissman found a way to take up the mRNA into the cells without triggering an immune response.\footnote{Their work \citep{kariko2005suppression} was initially rejected by two prominent journals - Nature and Science - and finally published in a relatively lower impact journal, Immunity. Two fundamental milestones in the development of the mRNA vaccine are: (a) a patent on the “prefusion of Coronavirus spike protein” (by the Department of Health and Human Services, The Scripps Research Institute, and Dartmouth College) and (2) a patent by Katalin Kariko and Drew Weissman at the University of Pennsylvania (the research was funded by the National Institute of Health) on ``RNA containing modified nucleosides and methods of use thereof''. The 2023 Nobel Prize in Physiology or Medicine was awarded jointly to Katalin Karikó and Drew Weissman for their discoveries that enabled the development of these effective vaccines.}
    \item \textbf{Improving mRNA delivery efficiency}: Small biotech companies (such as Inex, Protiva, Tekmira, Arbutus and many others, – often based in Vancouver) developed methods to treat the fatty coats to protect the delivery of genetic material. A major advance was made in 2004 when, after eight years of research, Ian MacLachlan found a suitable mixture of lipids to form a nanoparticle that would protect the genetic material and prevent it from escaping. This innovation was suitable for manufacturing and enabled drug manufacturers to scale up production.
    \item \textbf{The U.S. government's effort to find a vaccine to prevent AIDS}: At the Vaccine Research Center (VRC) of the U.S. National Institute of Allergy and Infectious Diseases, a group led by Peter Kwong has been trying to target the ``spikes'' on H.I.V. viruses that allow them to invade cells. Although H.I.V. vaccines were not successful, this work paved the way for research into the spike protein of the coronaviruses that cause Middle East Respiratory Syndrome (MERS) and Severe Acute Respiratory Syndrome (SARS). Another team led by VRC scientist Barney Graham focused on the development of a vaccine against the respiratory syncytial virus.\footnote{\citet{kirchdoerfer2016pre} and \citet{pallesen_immunogenicity_2017} are major breakthroughs in this direction.}
   
\end{enumerate}

Using Anthony Fauci’s words: ``So, when the genetic sequence of the SARS-CoV-2 became available, Graham's team lost no time in joining their long-time collaborators at Moderna to develop an RNA vaccine using a stabilized, prefusion spike protein as the immunogen. Pfizer and BioNTech, where Karikó was working, also used the RNA platform that she and Weissman had perfected and the immunogen designed by Graham to develop an RNA vaccine. Additional companies also used Graham's immunogen in other vaccine platforms that had been evolving for years, to make SARS-CoV-2 vaccines.'' \citep{fauci_story_2021}, p.109.

\section{Data and Methodology}

\subsection{mRNA Vaccine Patents}\label{Data}

\subsubsection*{Data collection \& organization}

Our analysis begins by defining a \textit{core}  set of patent families related to the mRNA vaccine platform. First, we consider 113 patent applications identified by \cite{martin2020mrna} and published between January 1, 2010 and April 1, 2020. We also consider 86 patents identified by \citet{gaviria2021network}, which examined the patent portfolios of eight organizations involved in mRNA vaccine technology research: BioNTech, Moderna, Curevac, Arbutus, Acuitas, University of British Columbia (UBC), University of Pennsylvania and Arcturus. After excluding ten patents without additional information, our final dataset (referred to as the Core dataset) comprises 189 patents corresponding to 151 different DOCDB families. Using PATSTAT, we create an additional dataset (the so-called Antecedent dataset), which contains information on all backward citations of the Core dataset and comprises a total of 3\,625 patent documents (2\,416 DOCDB families). In addition, we supplement both datasets with further bibliographic information from PATSTAT, including the filing dates and the names of the applicants. The latter information is particularly important to identify all public and private contributors to the advancement of mRNA technology.
A notable limitation of many patent records arises from inconsistencies in the naming of companies and organizations in patent documents. Due to the discretion given by patent offices to applicants in specifying the name of the patent assignees, there are significant differences in the presentation of company names. Although PATSTAT provides a standardized version of patent assignee names, there are still significant discrepancies in the names provided. Combined with the lack of a unique identifier for the assignee in the patent data, this makes it difficult to categorize patents belonging to the same organization.

To solve this problem, we apply different strategies. First, we apply proven methods from previous studies (see e.g. \citeauthor{bessen09}, \citeyear{bessen09}) to standardize frequently occurring terms in company names. Common examples are typical acronyms attached to company names, such as INC, AG, Ltd, GmbH, Co Kg and others.In addition, we carry out manual checks of the assignee's name to ensure harmonization where necessary. This harmonization and disambiguation process results in a list of 1\,173 distinct assignees, which are then categorized according to the type of institution or organization they represent. Identified categories include ``University'' (including hospitals), ``Research Center'', ``Government'', ``Public Research Organizations - PRO'', ``Firm: Biotechnology'', ``Firm: Pharma'', ``Firm: Other Med'', ``Firm: Other non-Med'', and ``Individual''.
We divide the time span into three periods of five years according to the registration date of the citing patent to characterize the technological progress at regular intervals: $2006-2010$, $2011-2015$, $2016-2020$.

\subsubsection*{Data description}

\begin{figure}[!ht]
\centering
\subfigure[Core ]
{\includegraphics[width=0.49\textwidth]{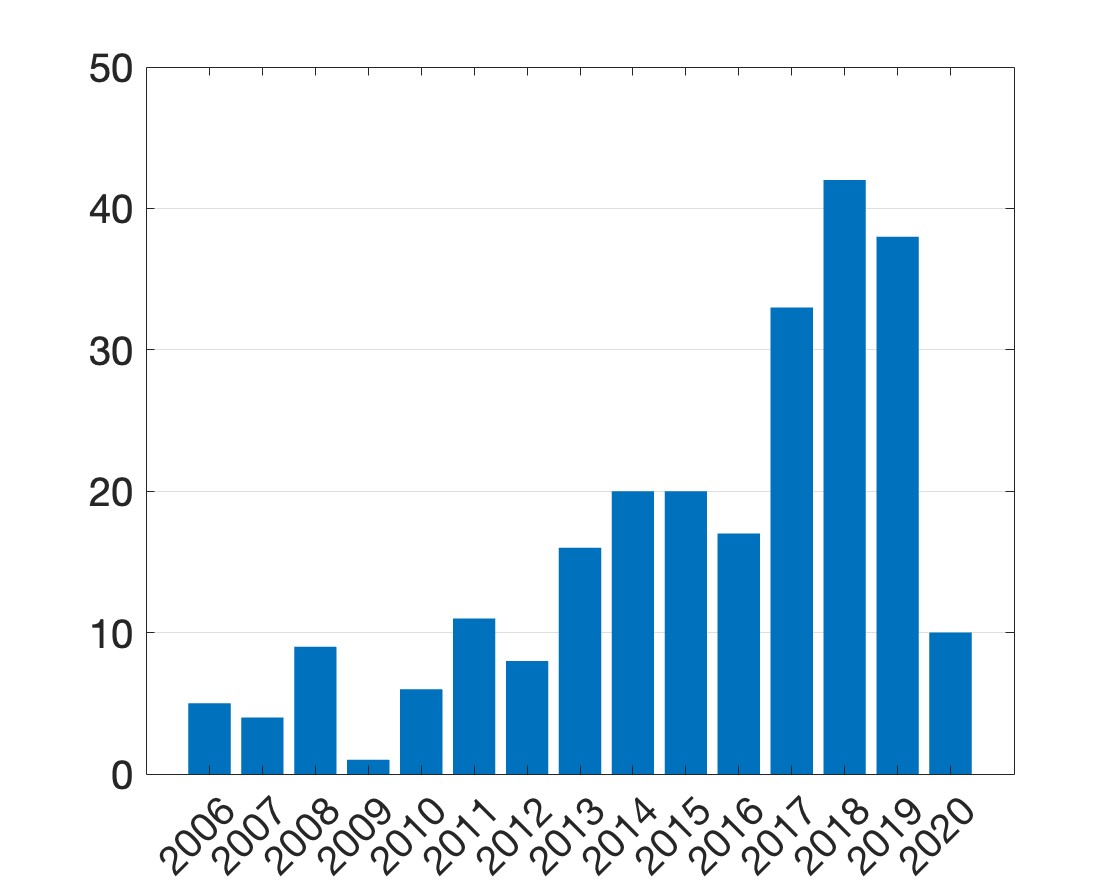}}
\hspace{-5mm}
\subfigure[Antecedents]
{\includegraphics[width=0.51\textwidth]{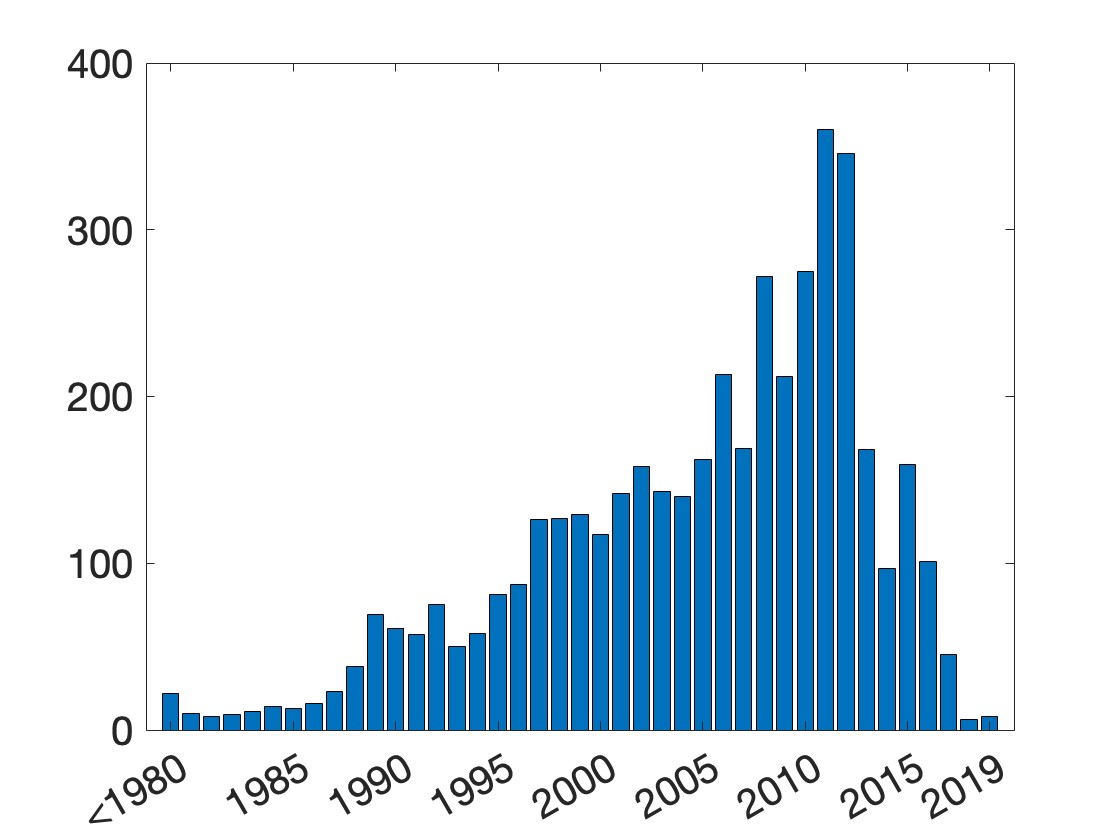}}
\caption{\textbf{Number of patents by  priority year} (a) Patents belonging to the Core mRNA vaccine collection; (b) patents classified as Antecedents.}
\label{CoreAnt}
\end{figure}

Figures \ref{CoreAnt} (a)-(b) show the number of patents by priority year in the Core and Antecedent datasets. The number of patent applications in the field of mRNA vaccine technology has increased significantly in the last seven years, while the knowledge cited in the prior art goes back further, to the mid-1980s. Figure \ref{CoreAnt} (a) shows that the Core dataset collects information on patents up to 2020.\footnote{In Section 2.3 we show that when the genetic sequence of SARS-CoV-2 became available, the mRNA technology to produce the vaccine was ready. Therefore, we are interested in patents up to 2020. Nevertheless, the patent race for mRNA vaccines has intensified in recent years. This is confirmed by a keyword search in the Patent Lens database: 228 patents filed between 2021 and 2022, for example, contain the keywords `mRNA AND vaccine' in their abstracts.}

\begin{figure}[!ht]
\centering
\subfigure[Core ]
{\includegraphics[width=0.5\textwidth]{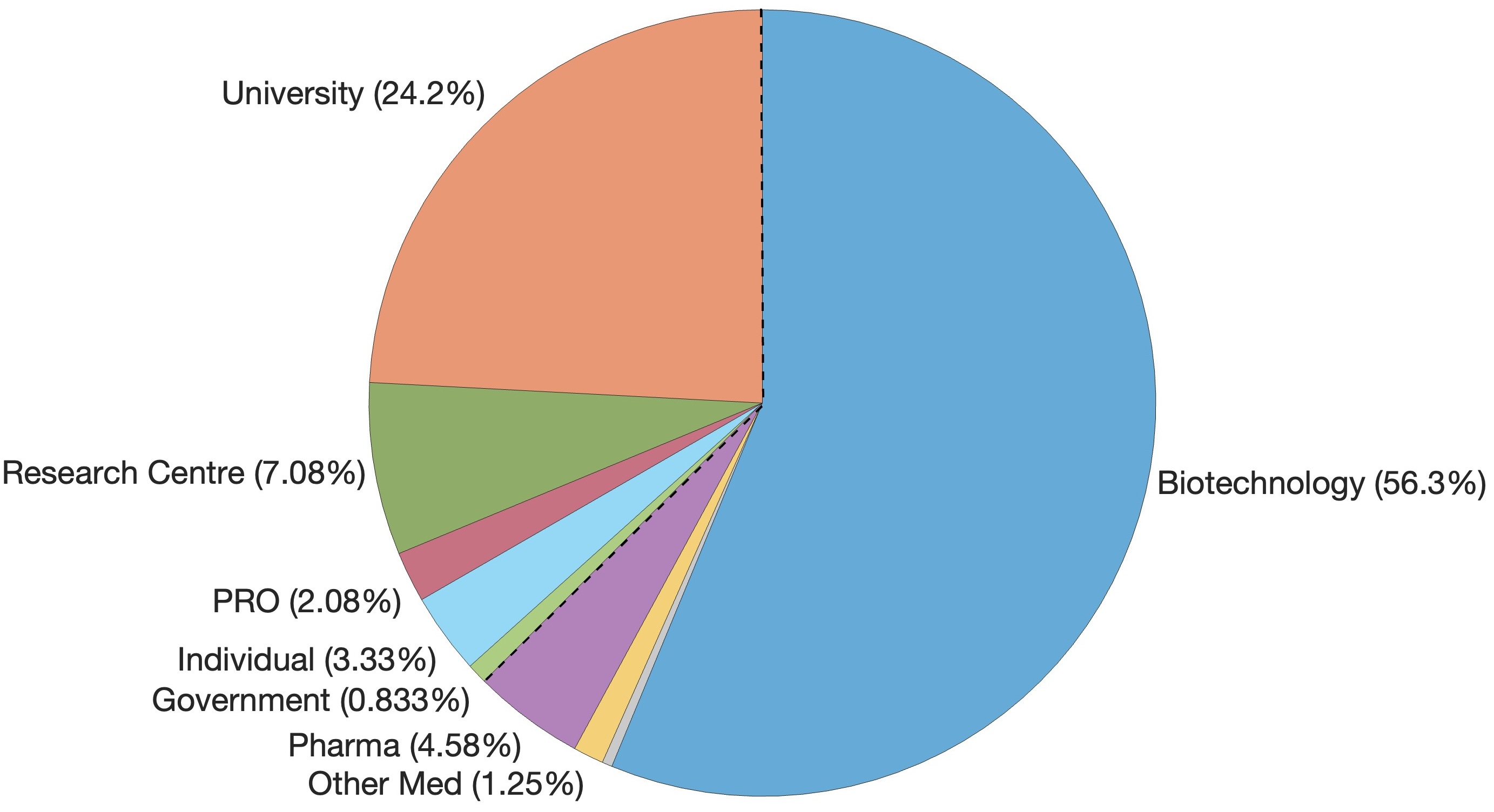}}
\hspace{-5mm}
\subfigure[Antecedents]
{\includegraphics[width=0.5\textwidth]{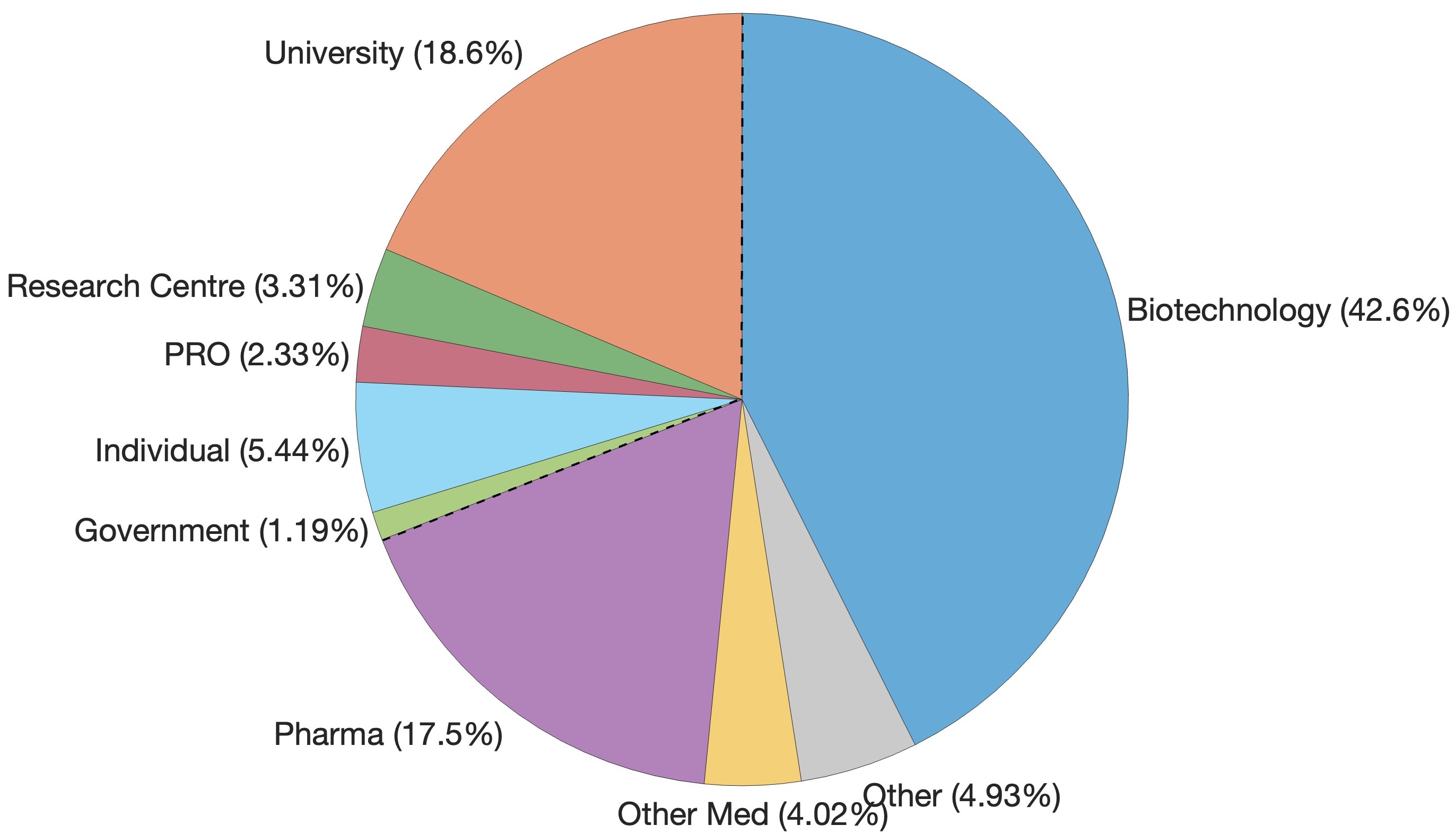}}
\caption{\textbf{Patents sector pie-charts.} The pie charts show the percentage of Core (a) and Antecedents (b) patents by type of organization.
}
\label{pie}
\end{figure}

The compiled dataset allows us to depict the types of assignees involved in mRNA vaccine technology. The pie chart in Figure \ref{pie} shows how the Core and Antecedents patents are distributed across the nine different categories, with the industrial patents (on the right) separated from the non-industrial patents by a dotted line. A look at the Core dataset shows that the largest share of patents ($56\%$) was filed by biotech companies. This is not surprising, as the biotech sector plays a central role in the realization of mRNA vaccine technology. On the other hand, only a minority of patents ($4.5\%$) were filed by pharmaceutical companies. Looking at the research institutions, the important role of universities is striking, as around $24\%$ of all patents were filed by an academic institution, 7\% by research centers and 2\% by public research institutions. Finally, around $3\%$ of the patents included in the Core dataset appear to have at least one assignee as solo inventor.\footnote{The most frequent assignees include Sahin Ugur (founder of BioNTtech), Hoerr Ingmar (founder of Curevac) and Pieter Cullis, Michael Hope and Thomas Madden, who have been working together for many years on the development of LNP systems that enable RNA- and DNA-based drugs.}

When looking at the Antecedents dataset, some differences are noticeable compared to the Core dataset. In particular, the contribution of pharmaceutical companies seems to be more important, as they produced about $17.5\%$ of the total patents in this sample, while biotech companies also have a relevant presence in this group ($43\%$). We can also note that the number of patents originating from industry is only slightly higher in the Antecedent patents ($69\%$) than in Core patents ($62.5\%$). 

The Tables \ref{tab:tab_core} and \ref{tab:tab_ant} show the number and corresponding percentage of companies and institutions belonging to the nine categories of organizations in the three periods analyzed (for the group of Antecedents, the first period includes patents filed before 2010). As expected, the Core size increases while the number of Antecedents decreases over time.
The presence of patents originating from biotech companies dominates in both groups, especially in the last period for both the Core and Antecedent group (2006-2010). It is worth noting that in the first period in the Core database, universities are the category with the highest number of patents. Also in the Antecedent database the weight of universities is higher in the first period. It is also interesting to note the decreasing presence of patents for pharmaceutical Antecedents from the first to the last period.

\begin{table}[!ht]
  \centering
  \small
    \caption{Sector: Core dataset}
\scalebox{1}{ 
\begin{threeparttable}  
    \begin{tabular}{lS[table-format=3.0]S[table-format=3.0]|S[table-format=3.0]S[table-format=3.0]|S[table-format=3.0]S[table-format=3.0]}
\toprule
   \textbf{\textit{Sector}}&\textbf{\textit{2006-2010}}&\textbf{\textit{Percent}} &\textbf{\textit{2011-2015}}&\textbf{\textit{Percent}} &\textbf{\textit{2016-2020}}&\textbf{\textit{Percent}}\\  
   \hline
   \hline
Biotechnology&9 & 36\% & 37 &49\% &89&64\%\\
Government &0 & 0\%&0 &0\% & 2&1\%\\
Individual &0 &0\%&2&3\%& 6&4\% \\
Other &0& 0\%&0 & 0\% &1&1\%\\
Other Med&2 & 8\% &0 & 0\% &1&1\%\\
PRO &1 & 4\%&1& 1\% &3&2\%\\
Pharma &2 & 8\%&2& 3\%& 7&5\%\\
Research Centre&1 & 4\% &8 & 11\%&8&6\%\\
University&10 & 40\%&25& 33\% &23&16\%\\
\hline \hline
Total & 25 & 100\% & 75&100\%&140&100\%\\
   	\bottomrule
		  \end{tabular}

  \end{threeparttable}}
   \label{tab:tab_core}
\end{table} 

\begin{table}[!ht]
  \centering
  \small
    \caption{Sector: Antecedents dataset}
\scalebox{1}{ 
\begin{threeparttable}  
    \begin{tabular}{lS[table-format=3.0]S[table-format=3.0]|S[table-format=3.0]S[table-format=3.0]|S[table-format=3.0]S[table-format=3.0]}
\toprule
   \textbf{\textit{Sector}}&\textbf{\textit{Before 2010}}&\textbf{\textit{Percent}} &\textbf{\textit{2011-2015}}&\textbf{\textit{Percent}} &\textbf{\textit{2016-2020}}&\textbf{\textit{Percent}}\\  
   \hline
   \hline
Biotechnology&1177 & 38\% & 567&50\% &121&76\%\\
Government &35 & 1\%&12&1 &5&3\%\\
Individual &194 &6\%&43&4\%& 1&1\% \\
Other &180& 6\%&35&3\%&1&1\%\\
Other Med&153 & 5\% &20&2\% &3&2\%\\
PRO &74 & 3\%&25&2\%&2&1\%\\
Pharma &554 & 18\%&201&18\%&12&8\%\\
Research Centre&100 & 3\% &41&4\%&4&2\%\\
University&620 & 20\%&185&16\%&11&6\%\\
\hline \hline
Total & 3087 & 100\% & 1129 & 100\% & 160 & 100\%\\

   	\bottomrule
		  \end{tabular}

  \end{threeparttable}}
  \label{tab:tab_ant}
\end{table}

\subsection{Methodology}\label{Method}

\subsubsection*{The mRNA vaccine knowledge network}

A network is completely described by a set of \textit{nodes} connected by \textit{edges} that indicate the existence of interactions between them. In addition, a \textit{weight} can be assigned to each link to quantify the intensity of the connection. In the case of the mRNA knowledge network, the nodes can represent patents, entities owning them\footnote{It can be a private or public entity as we explain in the section \lq\lq Data collection and organization\rq\rq{}.}, and categories to which these entities belong (such as different types of organizations in the private and public sectors). In the first case, the links simply show the direction of citations between patents, whereas when looking at entities and sectors, the edges are weighted by the total number of backward or forward citations. The mRNA vaccine knowledge network consists of $1\,184$ organizations, $5\,168$ links  and a total volume of $81\,358$ citations.

We investigate basic network properties (see \ref{app:AppA}) such as the \textit{density} (i.e., the ratio between the total number of node connections and all possible connections), the \textit{degree} (i.e., the total number of edges pointing to/from a node), the \textit{strength} (i.e., the total number of backward/forward citations of a node) to elucidate the structural organization of the mRNA vaccine network. Moreover, we focus on some centrality measures - such as \textit{betwenness centrality}, \textit{hub \& authority score} - to capture information about the role of nodes and identify key players based on local and global network properties. Lastly, we conduct a community detection analysis to uncover groups of nodes that are more densely connected to each other than to the rest of the network according to the amount of exchanged citations. This approach is particularly well-suited for characterizing the various technological trajectories that encompass the mRNA technology landscape  (details can be found in \ref{app:AppB}).

\subsubsection*{Credit allocation}\label{Allocation}

Using network properties, we propose a method to identify key players that contributed to breakthrough innovations among the different types of actors involved in mRNA knowledge networks. We investigate how the potential value of the innovation breakthrough, such as the profits of companies and institutions involved in the commercialization of COVID-19 mRNA vaccines, would be redistributed using the structural properties of the network  and the relative technological credit assigned to a specific player. This proposed distribution takes into account the contributions of the cited patents that serve as sources of knowledge for the innovation and development of these vaccines. The motivation behind this approach is to measure the efforts of different (types of) actors in innovating and creating useful step-stones in this field over the years.

It is important to clarify that public research organizations and universities contribute to the innovation process in various direct and indirect ways. For instance, they promote the creation and dissemination of basic knowledge, develop human capital and provide important research infrastructures. These contributions significantly drive innovation breakthroughs. However, our credit allocation method, which is based on the network of patent citations, only captures the direct technological contributions to the field. Consequently, the broader role of universities and public institutions in this context is significantly underestimated and should be considered as a floor or minimum threshold of their overall contribution.

Our methodology  incorporates information on the relevance of different patent assignees with respect to (i) the whole knowledge network and (ii) the company that developed and marketed mRNA vaccines by taking into account the volume of citations. Practically, we focus on centrality measures of nodes both at global and local level involving edge-weights, geodesic distances and iterative attribution of importance. First, we introduce a damping factor $\beta^d$ with $\beta \in (0,1)$ and $d$ representing the distance from node $i$ (i.e., out-going direction) to discount the effect of contributions in terms of forward citations. Second, we compute node importance using three different measures (details can be found in \ref{app:AppC}) : \begin{enumerate}
\item \textit{Markov}, the probability of observing a path of length $d$ that starts from node $i$ and reaches node $j$, calculated using the concept of Markov chains and the powers of the transition probability matrix applied to the knowledge network ;
\item \textit{Markov + Katz}, the probability introduced at point $1.$ multiplied by the normalized in-Katz centrality of node $j$ at the global level;
\item \textit{Markov + PageRank}, the probability introduced at point $1.$ multiplied by the normalized PageRank centrality of node $j$ at the global level.
\end{enumerate}





The three approaches to allocating credit\footnote{From now on, we’ll use the term "credit" to highlight the technological and scientific contributions of a company, paving the way for real recognition—whether it’s in the form of acknowledgment or even monetary rewards—for their role in driving innovation forward.} to a node differ in the network properties that are considered to identify key contributors to breakthrough innovations. The Markov rule can be seen as a local measure of the importance of a node, as it depends only on the distance between node $i$ and node $j$, weighted by the proportion of backward citations, i.e. the ratio between the edge weight between the two nodes and the total number of backward citations of the initial node. In other words, the distance between the two nodes takes into account not only the number of paths, but also the proportion of forward citations along the path. On the other hand, the rules \textit{Markov+Katz} and \textit{Markov+PageRank} also contain a measure for the global importance of the node. While the in-Katz centrality assigns importance to a node if it is linked from other important nodes or if it is strongly linked (i.e. it receives several citations from a node), the PageRank dilutes the importance that emanates from other important nodes according to their out-degree. In particular, important nodes with high out-degree transfer less importance to their links than nodes with low out-degree. However, both approaches combine local importance (through the weighted distance of nodes $i$ via the random walker) and global importance considering all links in the network (via Katz or PageRank centrality). It's worth noting that this method can be applied to any network $G$ and any node $i$ for which credit redistribution is required.\footnote{The method is additive, it is possible to compute the credit allocation starting from each node of interest separately and then sum the different allocations to obtain the overall credit assigned to a specific node.}

At the end of the procedure, the percentages assigned to each node (summing to $100$) can be interpreted in terms of its relative technological and scientific contribution to node $i$'s patents. To figure out how this method can be used to assign the technological credit, assume that we consider the final products of Moderna and Biontech. Suppose also that the value of these vaccines can be measured by these companies' profits. It is then possible to calculate the ideal share of profit of each contributor, based on the share of technological credit. This could be done by multiplying the profit for this ideal share.\footnote{In \ref{app:AppC} we show an example to illustrate how the approach could be used in this persepective.}



\section{The mRNA knowledge network: descriptive evidence}

In this section, we present the results of the analysis of the patent citation network of the mRNA vaccine. In the first part (4.1) we present the aggregate results showing how the different types of actors contribute to the mRNA knowledge network. In the second part (4.2) of the analysis we move to the the micro level property of the network and show how specific firms and institutions shape the development of the network. 

\subsection{Sectors}

\begin{figure}[!ht]
\centering
\subfigure[2006-2010]
{\includegraphics[width=0.3\textwidth]{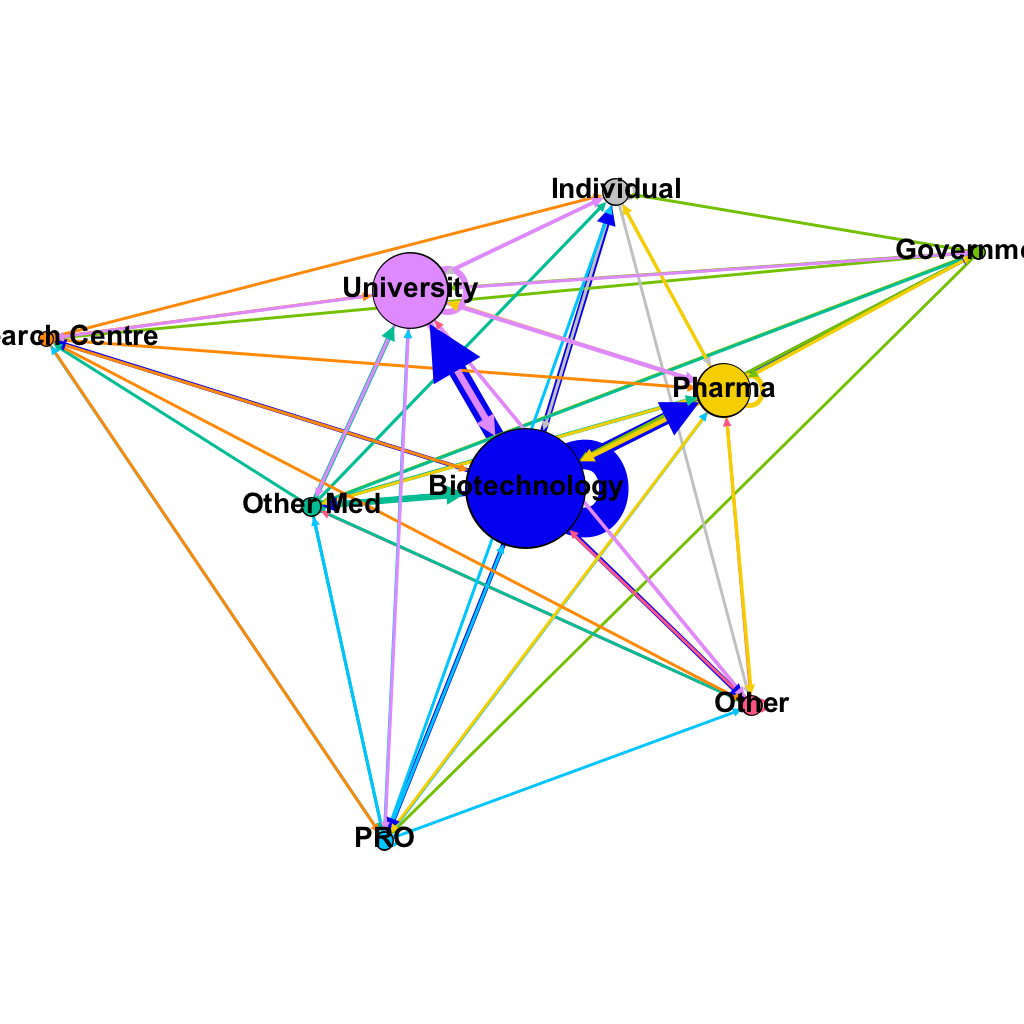}}
\subfigure[2011-2015]
{\includegraphics[width=0.3\textwidth]{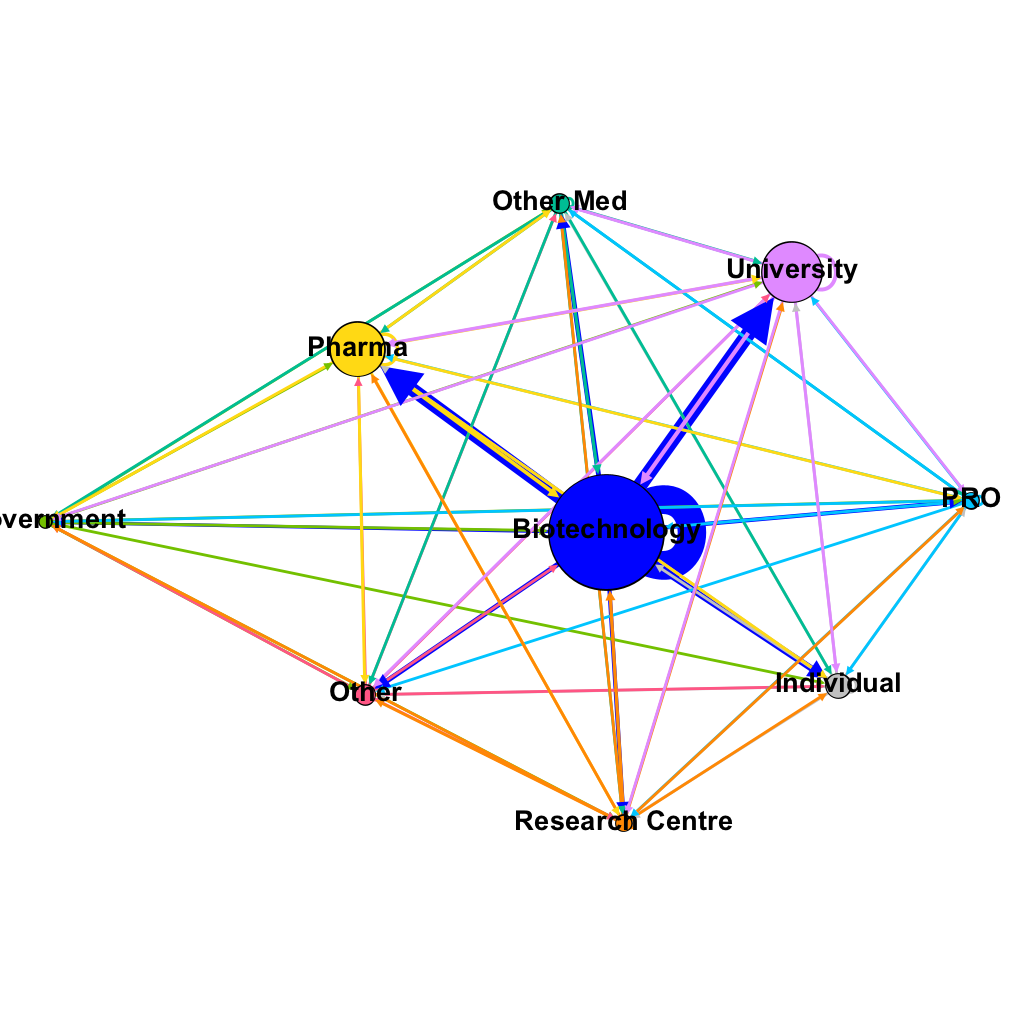}}
\subfigure[2016-2020]
{\includegraphics[width=0.3\textwidth]{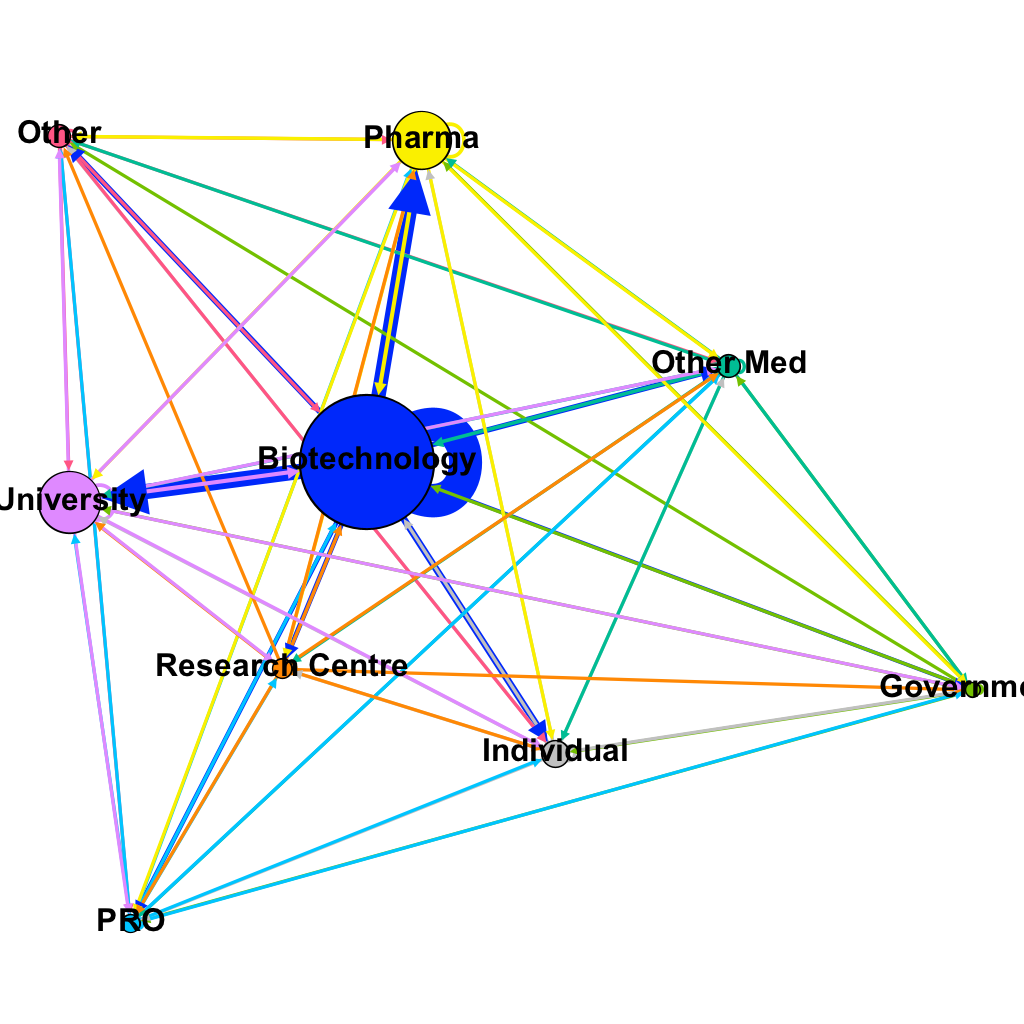}}
\hspace{55mm}
\subfigure[2006-2010]
{\includegraphics[width=0.3\textwidth]{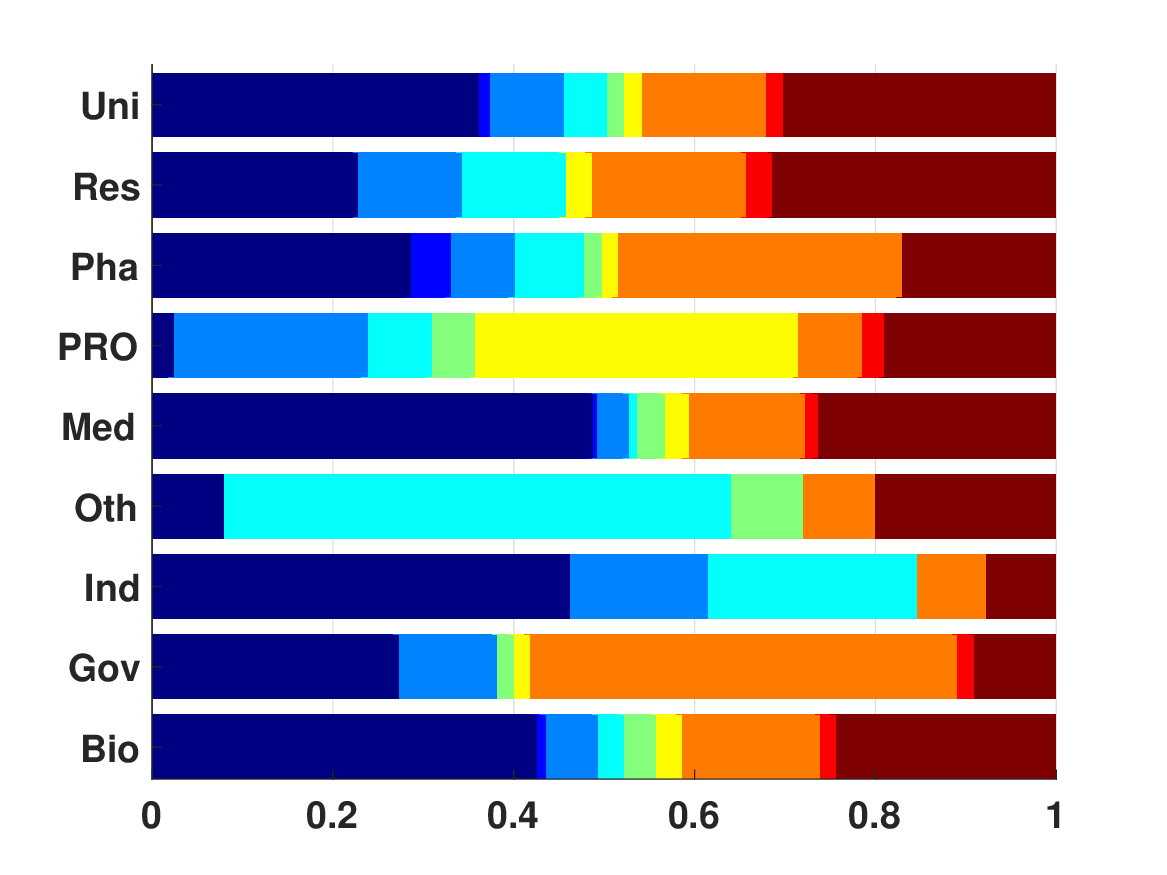}}
\subfigure[2011-2015]
{\includegraphics[width=0.3\textwidth]{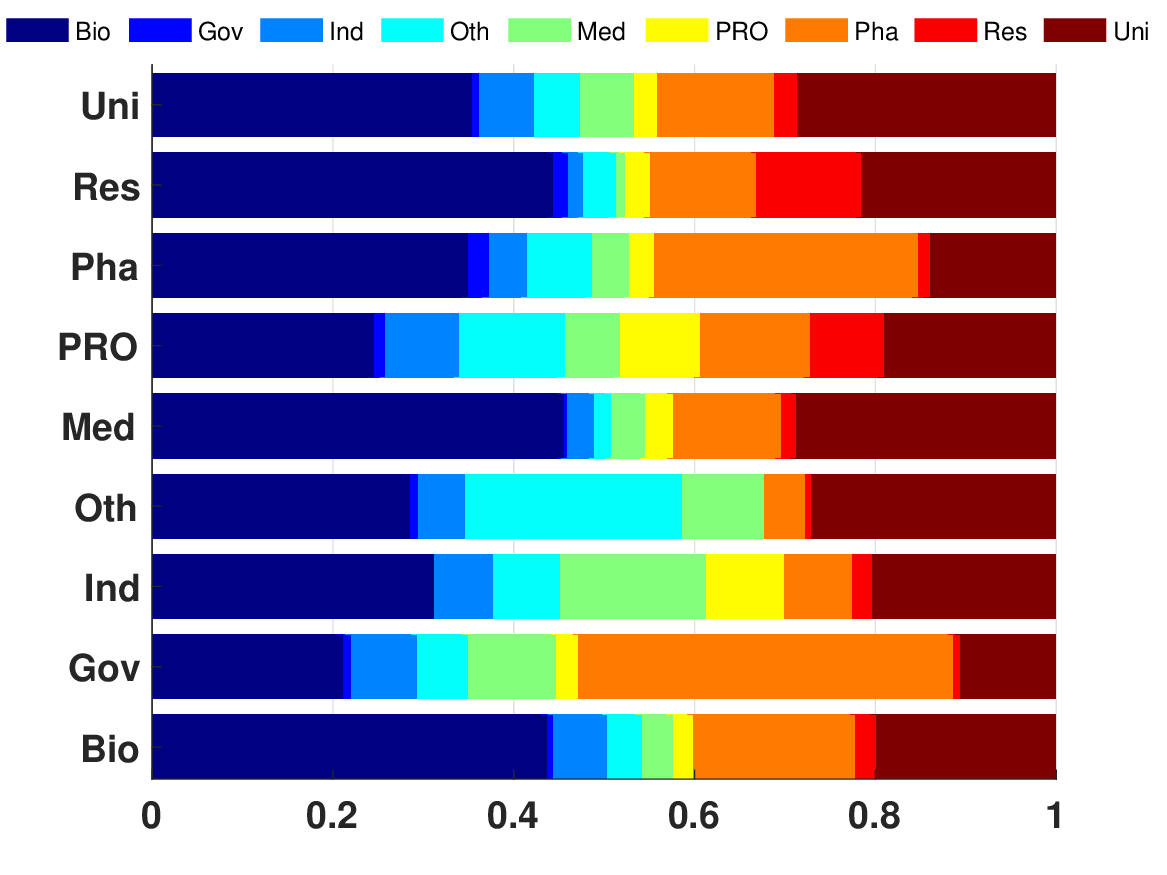}}
\subfigure[2016-2020]
{\includegraphics[width=0.3\textwidth]{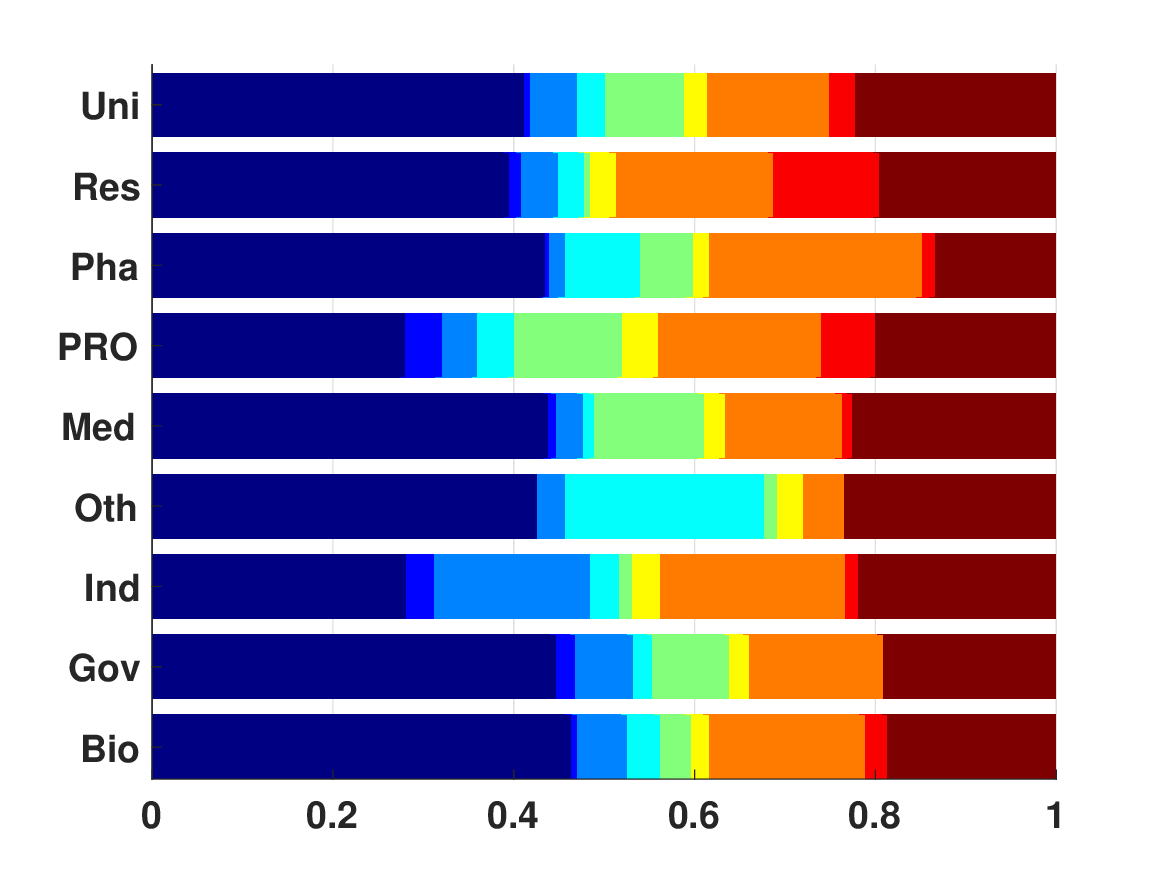}}
\subfigure[2006-2010]
{\includegraphics[width=0.3\textwidth]{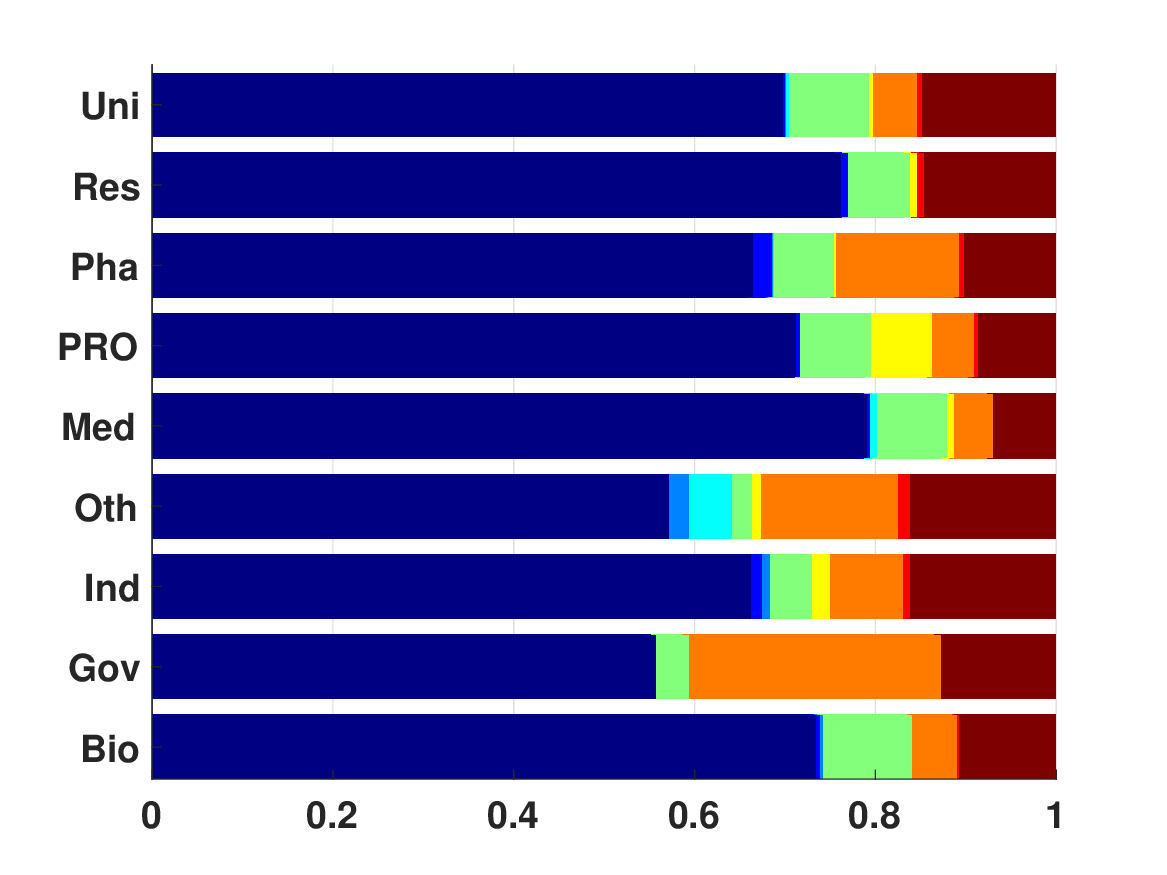}}
\subfigure[2011-2015]
{\includegraphics[width=0.3\textwidth]{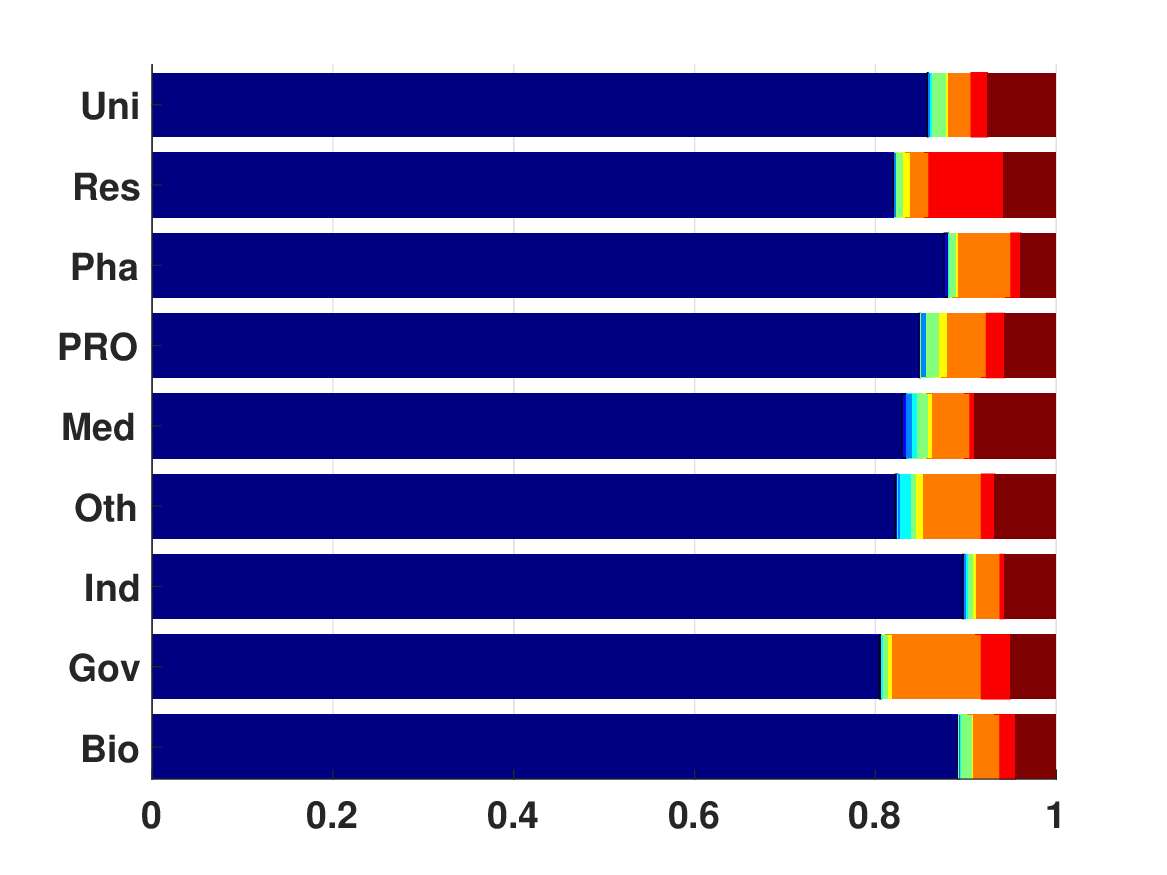}}
\subfigure[2016-2020]
{\includegraphics[width=0.3\textwidth]{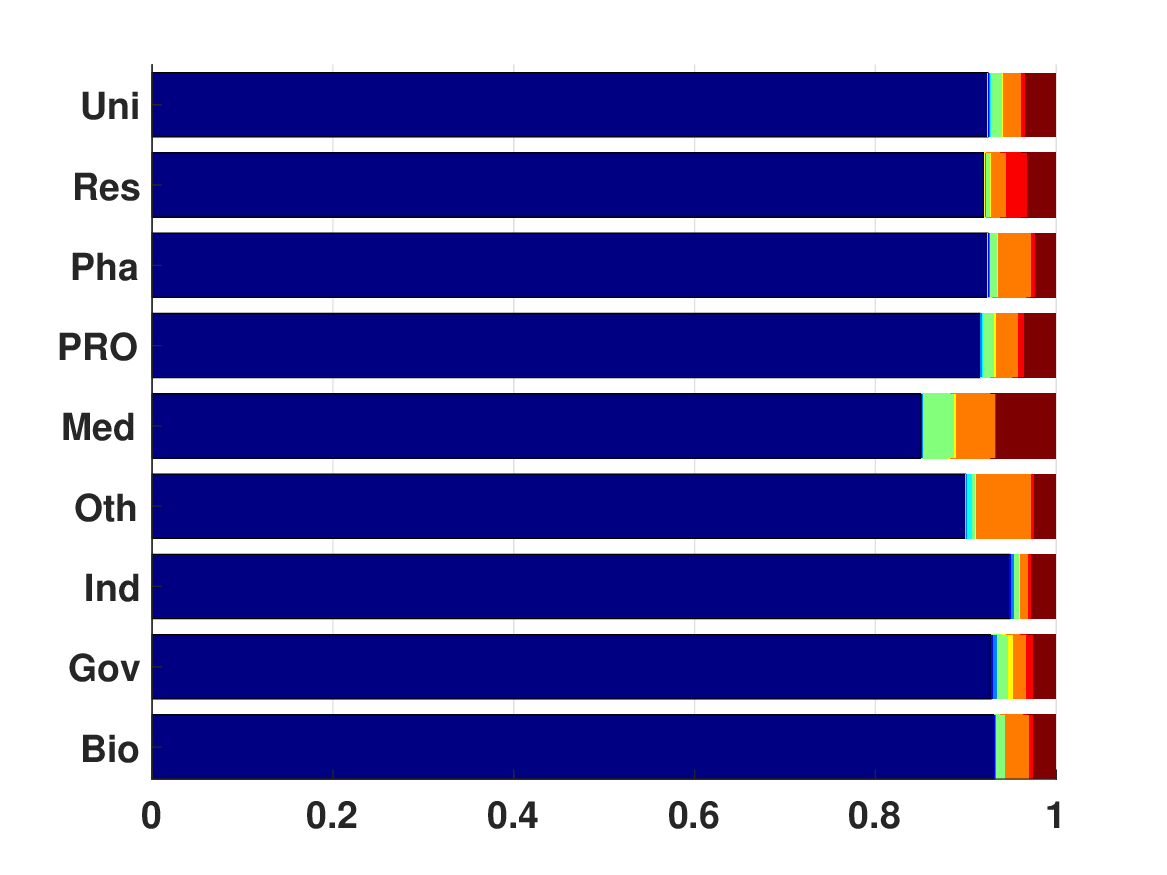}}
\caption{\textbf{Macroscale network.} (a)-(c) show the network of patent citations among the nine sectors indicated by the different colours: Biotechnology, Government, Individuals, Other, Other Medical, PRO, Pharma, Research Centre and University. Node size is proportional to the total number of backward citations; link thickness is proportional to the edge-weight (number of backward/forward citations) while its color is the same of the source node.  (d)-(f) report the total number of backward citations starting by the sector indicated by the row and reaching the other sectors (indicated by the different colors) normalized by the total number of backward citations;  (g)-(i) report the total number of forward citations pointing to  the sector indicated by the row and starting from the other sectors (indicated by the different colors) normalized by the total number of forward citations.}
\label{MacroNet}
\end{figure}

First, we consider the network of patent citations at sector level: nodes represent the 9 groups introduced before (Biotechnology, Government, Individuals, Other, Other Medical, PRO, Pharma, Research Centre, University) and edges are weighted by the total number of backward/forward  citations among them. In Figures \ref{MacroNet} (a)-(c) we show the snapshots of the networks in the three periods. Node size is proportional to out-strength (i.e. number of backward citations), while colors indicate the different sectors; link arrow shows the direction of citations (from source to target), while edge thickness is proportional to the total number of citations and its color is the same of the citing node (i.e., source). 

Network density goes from $0.83$ in the first period to $0.96$ in the last period, showing its maximum value in the second period ($0.97$). The greatest node is represented by Biotechnology followed by University and Pharma (reaching the same size in the third period). The thickest links starts from Biotechnology and are directed to the node itself (\textit{self-citations}), University and Pharma companies. The network is thus heterogeneous, primarily  driven  by the patenting activity of the Biotechnology companies, which often involves self-citations, as well as from  Universities and Pharmaceuticals companies. It is worth noticing the absence of connections from Research Centres, PROs and Individuals to the Government sector and from Pharma and Individual to Research Centres; while in the last period there are no citations from Government to Research Centres.

To investigate  the heterogeneous distribution of citations more thoroughly, in Figures \ref{MacroNet} (d)-(f) we show for each sector (indicated in the rows) the fraction of backward citations to the other sectors. We can see that the number of backward citations from the Biotechnology sector ($\sim40\%$) to itself almost doubles the quantity of citations to University and Pharma ($\sim 20\%$), even increasing in the third period. In general, we  observe a certain heterogeneity, with a clear abundance of citations directed to the Biotechnology sector, except for PRO and Other in the first period. Interestingly, in the initial  period, Biotechnology companies seem  to be less significant as a sources of knowledge compared to subsequent periods.  However,  Universities and Pharmaceutical companies also serve as key sources of citations  from  other sectors (especially for the Government sector in the first two periods).

In Figures \ref{MacroNet} (g)-(i) we show for each sector (indicated in the rows) the fraction of forward citations coming from the other sectors. We observe a significant presence of citations  from Biotechnology companies across all categories (with notable contributions from Pharma, University and Medical sector), especially in the last two periods. This trend reflects what shown in Table \ref{tab:tab_core} and Table \ref{tab:tab_ant}, where it emerges a  substantial impact of  Biotechnology patents, especially in the last period.

\subsection{Companies and institutions}

In Table \ref{Tab1} we report the main properties of the mRNA vaccine patent network in the three periods under study (see \ref{app:AppA} for the detailed definition of the indexes). The number of entities (nodes) in the network tripled from the first to the second period, revealing the presence of novel players in the field. The increase in the number of actors leads to a noticeable  rise in network sparseness, with the density decreasing from $0.01$ to $0.004$ in the second period and further to $0.003$ in the  third period. This outcome is confirmed by the relatively stable average degree, with increased  volatility in both  in-degree and out-degree, as evinced by the standard deviation nearly doubling from the first to the second period. 

\begin{table}[!ht]
\centering
\small
\begin{tabular}{l | c c c } 
& \textbf{2006-2010} & \textbf{2011-2015} & \textbf{2016-2020}\\ [1ex] 
\hline  
 \textbf{Number of nodes} & 318 & 1005 & 1064 \\ [1.5ex] 
 \textbf{Number of links} & 1152 & 3937& 3520\\[1.5ex] 
 \textbf{Volume of citations} & 7935 & 63411 & 63059\\[1.5ex] 
 \textbf{Density} & 0.01 & 0.004 & 0.003 \\[1.5ex]
\textbf{Average in-degree (std)}& 3.6 (3.9) & 3.9 (6.2) & 3.3 (5.1)\\[1.5ex]
\textbf{Average out-degree (std)}  & 3.6 (13.7) & 3.9 (27.9) & 3.3 (26.7)\\[1.5ex]
\textbf{Max in-degree} & 22 & 50&41\\[1.5ex]
\textbf{Max out-degree}& 212 & 782 &783\\[1.5ex]
\textbf{Average in-strength (std)}& 25 (78) & 63 (221) & 59 (254)\\[1.5ex]
\textbf{Average out-strength (std)}& 25 (85) & 63 (1334) & 59 (1445)\\[1.5ex]
\textbf{Max in-strength}& 719 & 3500 & 4982\\[1.5ex]
 \textbf{Max out-strength}& 2954 & 42138 & 46951\\[1.5ex]
 \hline
\end{tabular}
  \caption{Network binary properties}
  \label{Tab1}
\end{table}

The total number of backward citations per node appears generally more variable than the number of its forward citations in all periods. Indeed, the maximum observed out-degree is five times bigger than the maximum in-degree in the first period, fifteen times in the second and almost twenty times in the third period. The number of out-degree is highly heterogeneous with a big amount of zeros (i.e., companies just receiving but not sending citations) increasing over time ($68\%,67\%, 77\%$, respectively), and a few nodes citing a high number of different companies. In the first period, only 3\% of organizations cited at least 10\% of the total organizations (Alnylam Pharmaceuticals, British Columbia University, Cellscript Inc, Curevac, Epicentre, Inex-Tekmira-Arbutus, Protiva Biotherapeutics, Proviva, and the University of Pennsylvania). In the second and third periods, this percentage drops to 1\%. The organizations citing at least 10\% of the total organizations in the second period are Alnylam Pharmaceuticals, BioNTech, British Columbia University, Curevac, Inex-Tekmira-Arbutus, Moderna, and Tron. In the third period, they are British Columbia University, Curevac, Inex-Tekmira-Arbutus, and Moderna.


The global volume of citations significantly increases from the first to the second period, while slightly reduces in the third one. This reflects the large expansion of the network after 2010 and the fact that more recent patents (third period) receive fewer citations because of truncation. Both the distributions of the forward and backward number of citations appear heterogeneous - especially for the former, showing very high standard deviations -  with a few nodes sending/receiving most of the citations.

\begin{figure}[!ht]
\centering
{\includegraphics[width=0.9\textwidth]{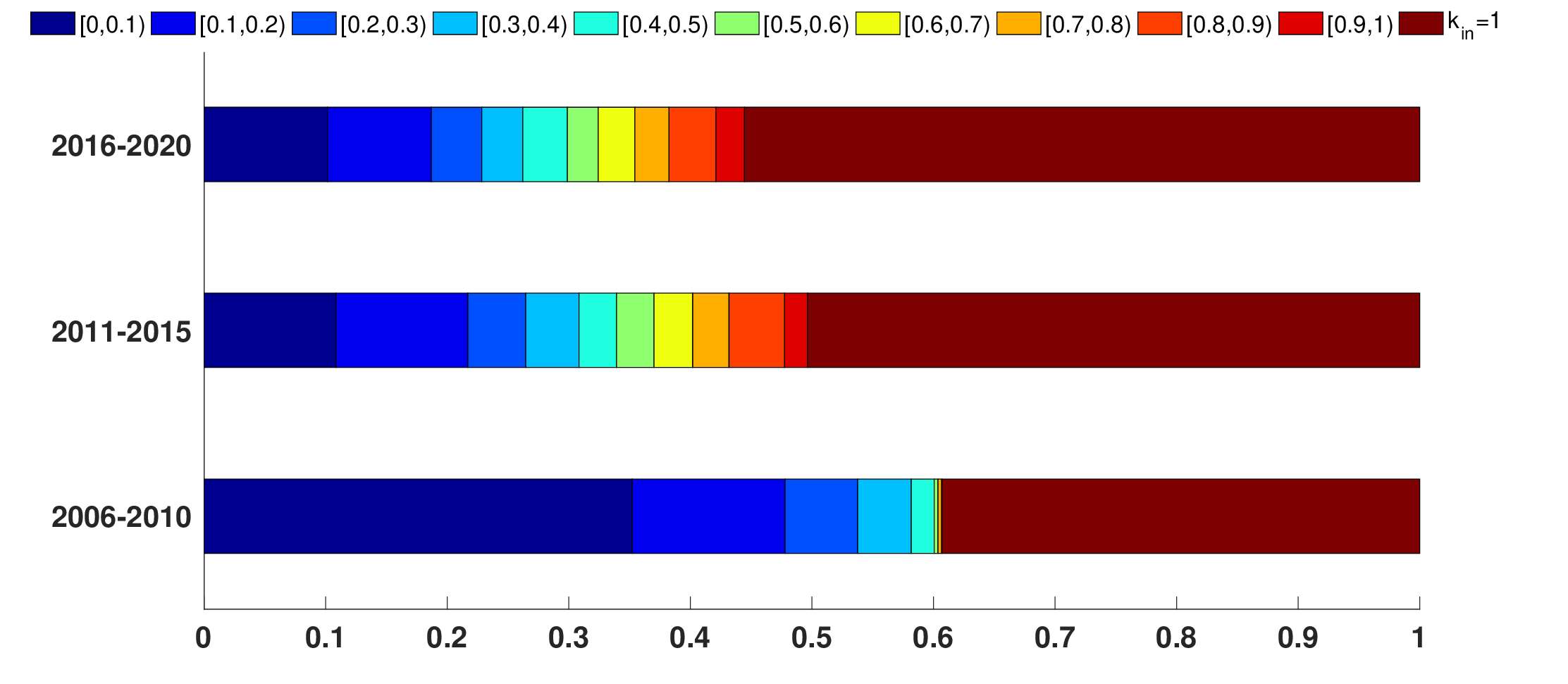}}
\caption{\textbf{Herfindahl Index of node in-strength}. Frequency of HHI values to quantify how the total number of forward citations is distributed among all incoming links of a node. }
\label{HHI}
\end{figure}

To better quantify the heterogeneity in the distribution of forward citations per node, we introduce the Herfindahl-Hirschman Index (HHI) as $h_i=\sum_{j=1}^{k_i} (w_{ji}/s^{in}_i)^2$. In doing so, we first compute the share of forward citations for each node as the ratio between edge-weights of its incoming links and their sum (i.e., the node in-strength), then we sum up their squares and normalize the result in order to have $h_i \in [0,1], \forall i=1, \dots, N$. Hence, $h_i=0$ implies that weights are equally distributed among links, while $h_i \rightarrow 1$ signals a high concentration of weight associated to one link.  In fig.\ref{HHI} we report the frequency of HHI values divided into ten ranges of length 0.1, along  with the frequency of in-degree equal to one (i.e., $h_i=1$ by definition). It highlights noticeable  differences, particularly between the first and the subsequent two periods with: (i) an increasing  number of nodes with an in-degree of one (i.e, nodes being cited only by one node) over time;  and (ii) a rise in higher  HHI values, indicating  a significant concentration of citations among a  few other entities (i.e., higher concentration of forward citations).

In summary, the number of actors in the network grows substantially from the first to the second period, revealing the emergence  of new actors in the field. Concurrently, the flow of knowledge grows after 2010, accompanied by  a diverse and expanding set of actors. At the same time, the knowledge network shows that a restricted number of entities are responsible for a substantial amount of backward citations, suggesting they are central in aggregating knowledge from different institutions.  In fact, there is a noticeable concentration in forward citations, indicating  that certain actors play a crucial role in the innovation process. Consequently, the next section will explore more in depth the centrality of the different entities within  the network. 

\paragraph*{Centrality measures}

We examine the roles of patent assignees by analyzing centrality measures to identify the key players during each of the three periods under study. We choose four different quantities:  (i) in-strength; (ii) out-strength; (iii) betweenness centrality; (iv) hub \& authority scores (see \ref{app:AppA} for more details). They highlight different properties of nodes in assigning the level of importance. The \textit{in-} and \textit{out-} strength are local properties focusing on the total number of forward and backward citations; they measure the total influence a node receives from other nodes in the network (in-strength) and the total influence a node provides to other nodes in the network (out-strength). The betweenness centrality looks at the global network organization taking into account weights, link directions and shortest paths lengths; it measures the importance of a node in terms of its ability to act as a bridge or intermediary in the network. Finally the hub \& authority scores compute the centrality of a node in a recursive way taking into account the global network organization and the link directionality. Hub scores measure the value of a node as a source of information. In other words, a strong hub is a node that points to many other nodes that are considered authoritative. Authority scores measure the value of a node as a reliable source of information. In other words, a strong authority is a node to which many hubs point.

\begin{figure}[!ht]
\centering
\subfigure[In-strength ranking]
{\includegraphics[width=0.48\textwidth]{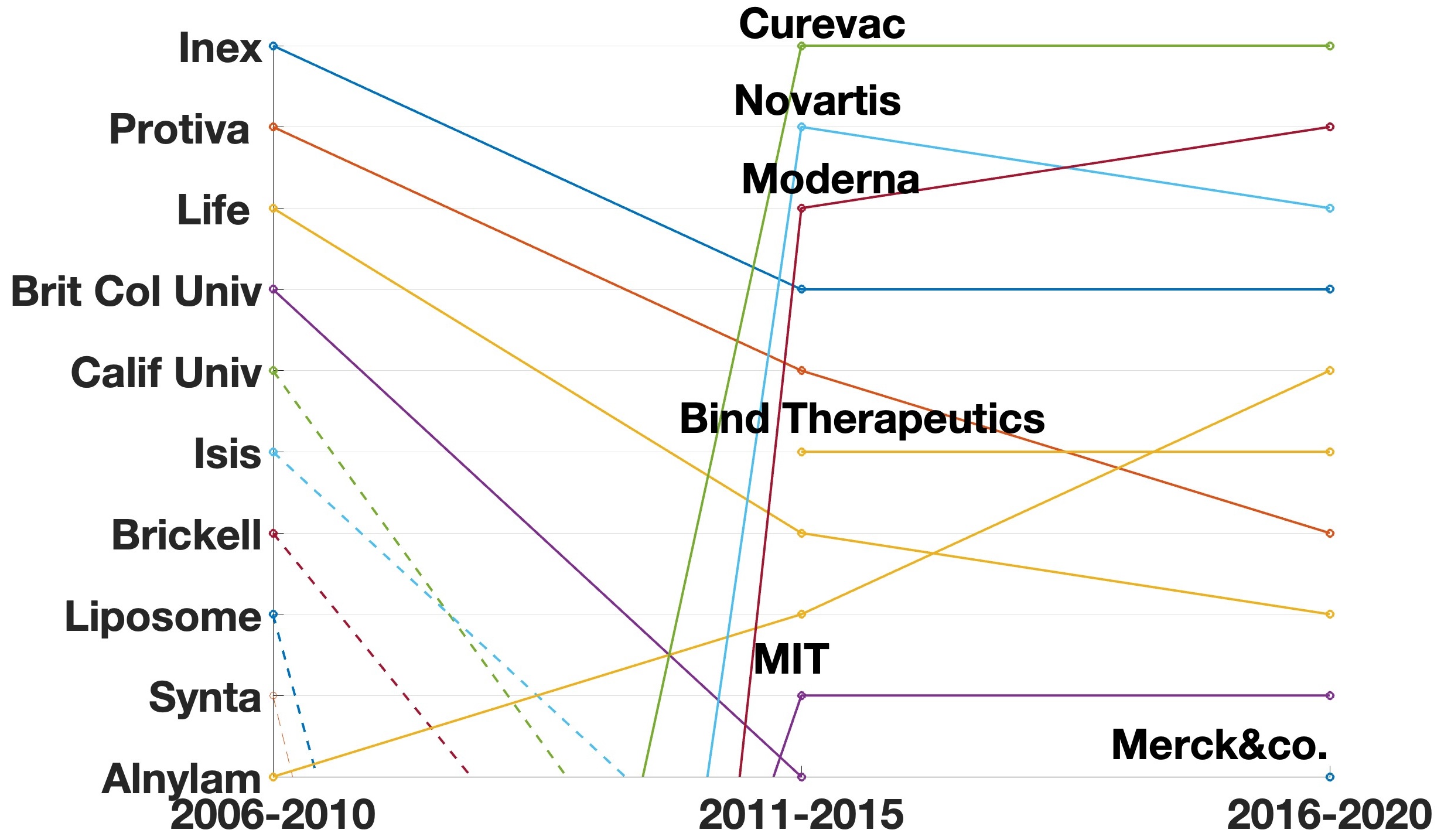}}
\hspace{-3mm}
\subfigure[Out-strength ranking]
{\includegraphics[width=0.43\textwidth]{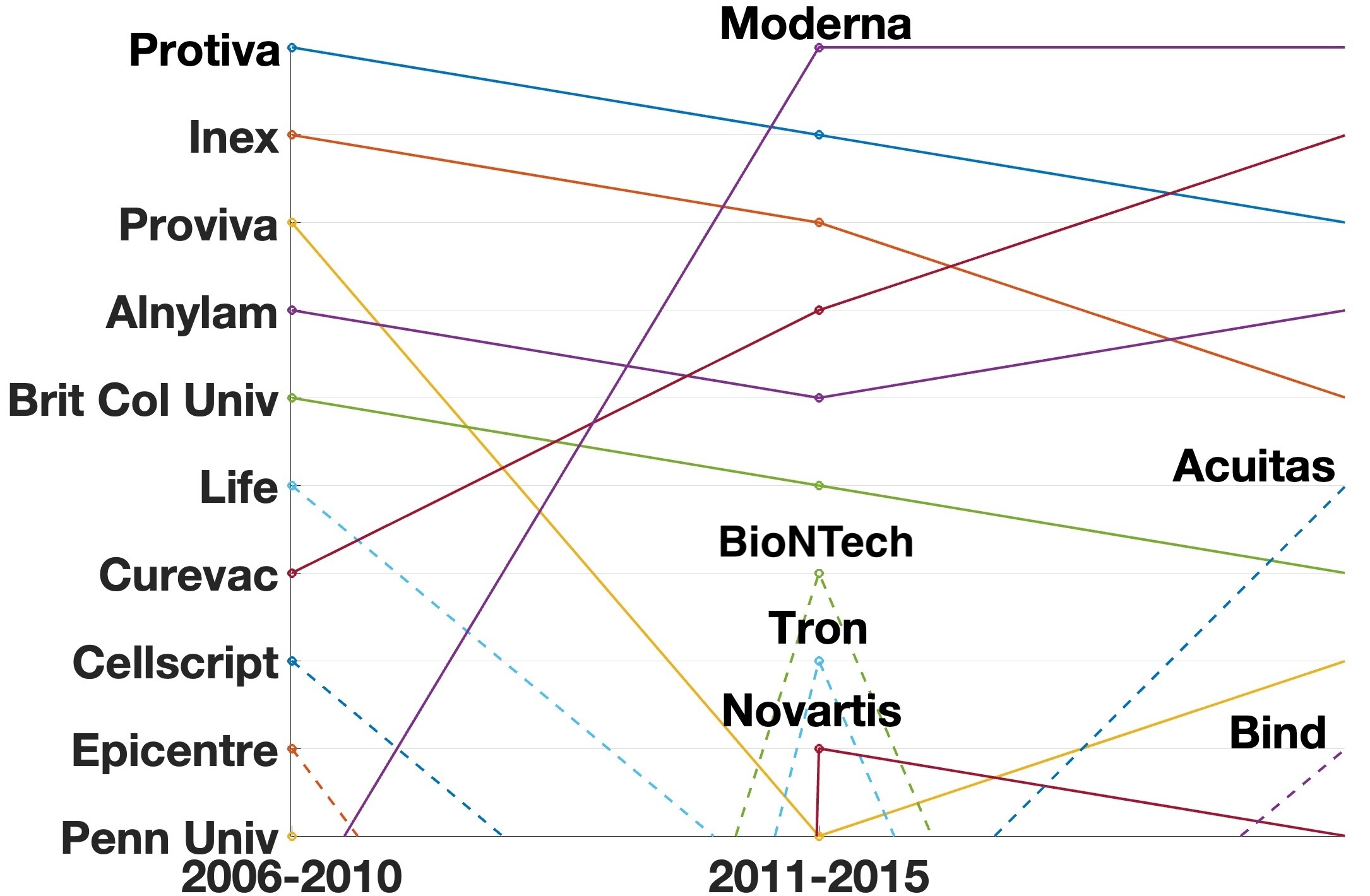}}
\subfigure[Betweenness centrality]
{\includegraphics[width=0.5\textwidth]{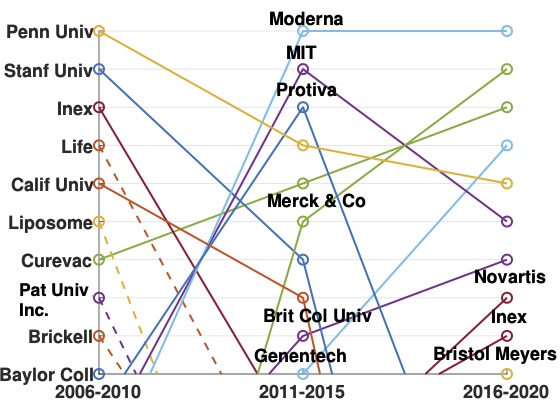}}
\caption{\textbf{Top ten rankings for centrality measures} (a) In-strength; b) out-strength; (c) betweenness centrality. The dotted lines indicates entities that appear only in one ranking.}
\label{Rankstr}
\end{figure}

In Figure \ref{Rankstr} we show the top ten nodes for the first three centrality measures. 
The rankings for the first period differ significantly from those of the subsequent two periods across the three centrality measures. For the in-strength centrality, we observe that some actors disappear from the ranking while new ones emerge. Notably, some universities  played a prominent role as a source of knowledge in the first period, whereas  Moderna, Curevac and Novartis gained importance after 2010.
In terms of out-strength centrality,  there is  stability in terms of  actors involved, though their  position change over the three periods. The betwenness centrality ranking show a certain degree of variation in both the  actors listed and  their positions. Universities and research organizations play a relevant role: Pennsylvania University, Stanford University and California University are prominent in the first period, while MIT and British Columbia University, become significant from the second period onward. Additionally, some Pharma companies, such as  Merck, Novartis and to a lesser extent Bristol Meyers increase their betweenness centrality in the second and third periods.
It is worth noting that  after 2010, Moderna and, to a lesser extent, Curevac exhibit very high centrality across various  measures in the mRNA knowledge network.

\begin{figure}[!ht]
\centering
\subfigure[Hub score]
{\includegraphics[width=0.5\textwidth]{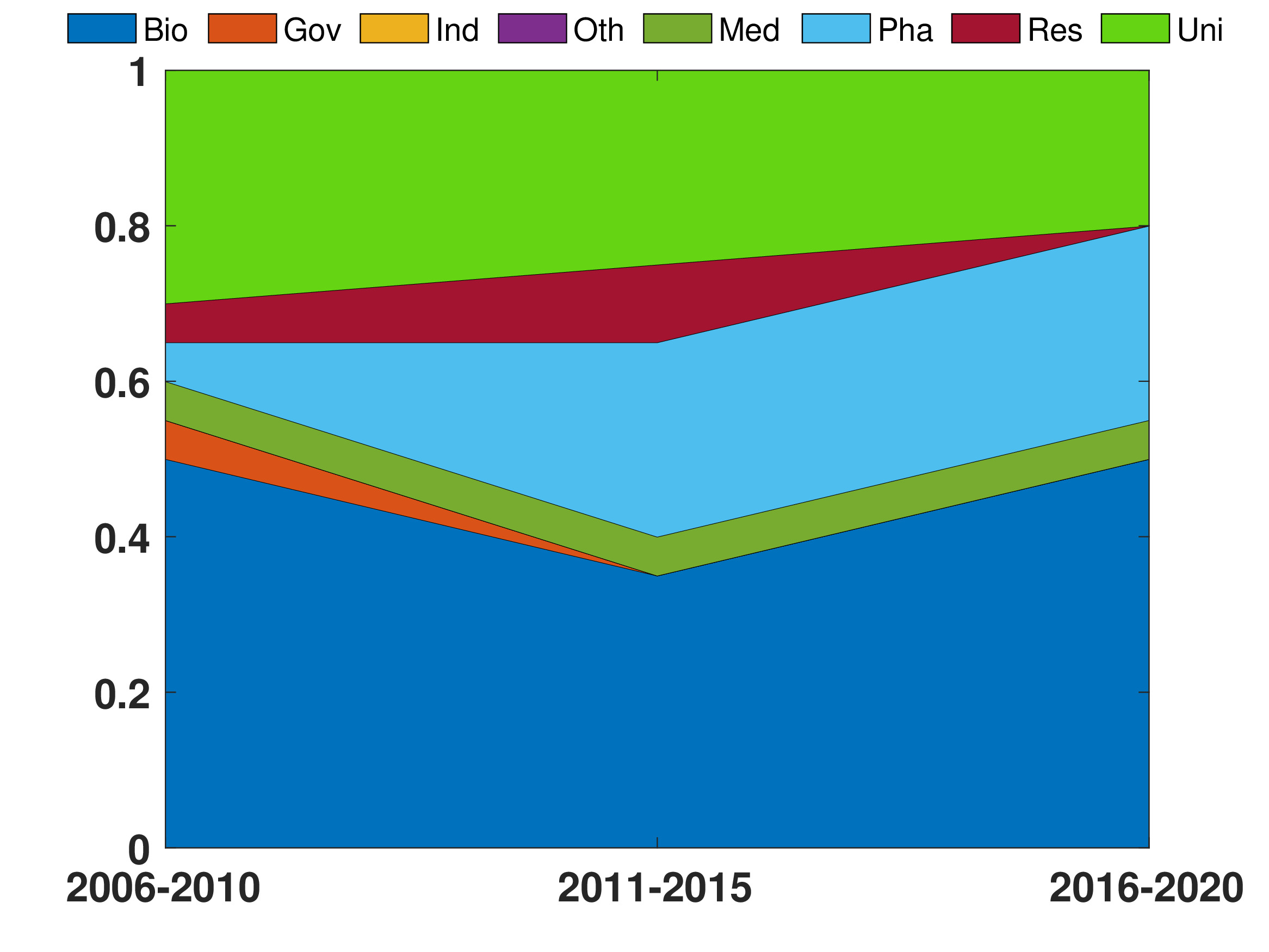}}
\hspace{-5mm}
\subfigure[Authority score]
{\includegraphics[width=0.5\textwidth]{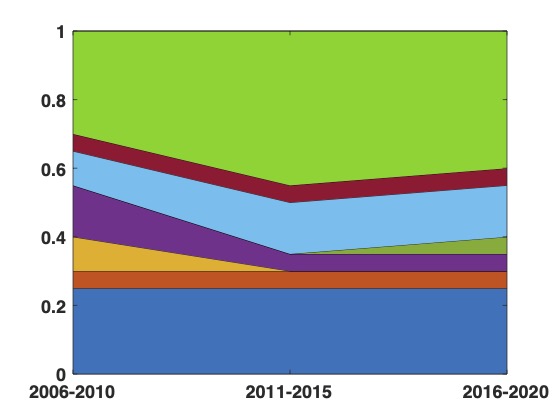}}
\caption{\textbf{Hub and Authority scores.} Frequency of company type in the top 20 ranking for the Hubs and Authority score (PRO and Universities have been merged in University).}
\label{RankHA}
\end{figure}

In Figure \ref{RankHA}, we select the top 20 entities according to their Hub and Authority scores and we categorize them by sector. The figure illustrates the proportion of each sector by showing the share of entities from each sector across the three periods. Table \ref{HA} lists the names of the entities that rank in the top ten for these measures. Remarkably, Universities and Public Research Organizations (combined into a single sector labeled as University) have a significant presence in the top 20 ranking for Authority scores, with their representation increasing over time. Biotechnology companies also  maintain  a steady share in this ranking. Conversely, the Hub ranking is predominantly led by the Biotechnology sector, followed by the Pharma sector, which exhibits an upward trend, and Universities, which show a decreasing trend.

From Table \ref{HA} several notable patterns and trends emerge, highlighting the evolving roles of different types of entities, including Biotechnology companies, pharmaceutical firms, universities, and PROs. Universities and PROs consistently feature prominently in the Authority rankings, emphasizing their role as key sources of foundational knowledge in mRNA research. Specifically, institutions like MIT, CNRS, Max Planck, California University, and Massachusetts University frequently appear in the Authority rankings. This trend underscores the increasing importance of academic research in driving innovation in the mRNA field. In the last period (2016-2020), MIT, California University, British Columbia University and Max Planck have notably maintained their presence in the Authority rankings, indicating their sustained impact on the field. 

Biotechnology firms dominate the Hub rankings across all three periods, reflecting their pivotal role in actively developing and citing crucial patents. Companies such as Inex-Tekmira, Alnylam, Protiva, and Curevac appear consistently in the Hub rankings. Moderna is a key Hub after 2010. The presence of Biotechnology firms in the Authority rankings, particularly in the later periods, suggests their dual role in both generating and disseminating critical knowledge. Notably, Inex-Tekmira and Alnylam are listed in both the Hub and Authority rankings in the 2016-2020 period. British Columbia University is the most important Hub among universities and PROs.

Figure \ref{RankHA} and Table \ref{HA} show that pharmaceutical companies are less dominant than Biotechnology firms. The presence of Novartis, Life and Takeda,  particularly in the latest period, in the Hub and Authority rankings highlights their role in leveraging and building upon existing research to develop commercial applications.

Finally, it is important to highlight that some entities exhibit high centrality in either  Hubs or Authorities exclusively. For example, Curevac, Moderna, Inex-Tekmira and Proviva are prominent Hubs, reflecting their role in citing and linking various patents. Conversely, CNRS, Max Planck, California University, and Massachusetts University are primarily Authorities, indicating their role in generating highly influential patents.

\begin{table}[!ht]
\centering
\footnotesize
\begin{tabular}{c  c |c c|c c} 
 \hline
    \multicolumn{2}{c|}{\textbf{2006-2010}} & \multicolumn{2}{c|}{\textbf{2011-2015}} &  \multicolumn{2}{c}{\textbf{2016-2020}}\\
    \hline
    \textbf{Hub}&    \textbf{Auth} &    \textbf{Hub}&    \textbf{Auth} &     \textbf{Hub}&    \textbf{Auth} \\
\hline
  
Brit Col Univ &	Isis &	Moderna &	MIT &	Moderna	& Inex-Tekmira\\

Inex-Tekmira&	Life&	Brit Colu Univ &	Brickell &	Curevac	&MIT\\

Alnylam &	MIT & Inex-Tekmira & Wisc Res Found &	Brit Col  Univ & Brit Col Univ\\

 Protiva & 	Cnrs & 	Curevac &	Penn Univ &	Inex-Tekmira & Protiva\\
 
 Proviva &	Max Planck & Alnylam  & Max Planck	& Alnylam & Brickell\\
 
Curevac &	Calif Univ &	BioNTech	 & Inex-Tekmira &	Protiva &	Alnylam \\

Life  & Pat Univ Inc.&	Tron	 & Mass Univ & Novartis	&Novartis\\

Epicentre	 &Liposome& Protiva & Us Depa Health&	Proviva	& Calif Univ \\

Cellscript & Mass Univ& 	Proviva	 & Protiva &	MIT & Life \\

MIT & 	White Bio Res & Mainz Univ & Life &	Takeda &	Max Planck\\
    \hline
 
\end{tabular}
 \caption{Hub and authorities top ten rank for the three period under study}
 \label{HA}
\end{table}

\paragraph*{Analysis of communities.} 

Finally we have performed a community detection analysis. The details of the analysis can be found in \ref{app:AppB}. The analysis of citation networks,  over three distinct periods, reveals the following patterns. In the first period (2006-2010), the landscape is characterized by smaller, more focused communities. Key players include Cellscript, Epicentre, University of Pennsylvania (first community), and Alnylam (second community). These entities are instrumental in laying the groundwork for mRNA technology, often collaborating on fundamental research and early technological developments.

As the field matures (2011-2015), larger communities emerge. Moderna and Curevac take center stage, driving innovation in mRNA vaccines and therapeutics. Moderna, in particular, becomes a key player, with a distinctive periphery of cited entities. Meanwhile, a community focused on LNP technology, involving British Columbia University, Inex, Alnylam, Protiva, and Proviva, highlight the importance of efficient delivery mechanisms for mRNA therapeutics.

In the growth phase (2016-2020), the organization of the communities remains similar and the network expands. The first community includes beside Moderna also Novartis and the Univesity of Pennsylvania. The second community includes the LNP network, with British Columbia University, Inex, Alnylam, Protiva, and Proviva. 

Finally, the German hub of innovation, centered around BioNTech, Tron, and Mainz University, continues to grow and new players like Novartis and Merck entered the scene.

The evolution of mRNA patent communities from 2006 to 2020 reflects a trajectory of foundational innovation, consolidation of leadership, and significant expansion. Over nearly two decades, the technological focus has shifted from basic research and initial applications to vaccine developments, bringing together different technological trajectories from mRNA syntesis and stabilization methods to LNPs and efficient delivery. This evolution has been driven by an heterogeneous set of companies and universities. After 2010, the mRNA patent landscape saw a consolidation, with companies like Moderna and Curevac strengthening their leadership positions and  new global Pharma players emerging.  Geographically, the network has expanded from being predominantly North American to a more globally distributed landscape, with significant hubs in both North America (US and Canada) and Europe (mainly Germany).

\section{The mRNA knowledge network: Credit allocation}\label{CreditResult}

In the previous sections, we have shown that the mRNA knowledge network is characterized by a dynamic interplay between universities, PROs, Biotechnology firms, and Pharmaceutical companies. Biotechnology firms play a crucial role in the creation and diffusion of important innovations. Universities and PROs also make significant contributions, especially in the early phase. This interaction between different institutions has significantly advanced mRNA innovation and led to major breakthroughs such as the mRNA COVID-19 vaccines.

Our citation network analysis reveals that Biotechnology companies, which have significantly benefited from the commercialization of mRNA vaccines against COVID-19 (in particular BioNTech and Moderna), have thrived within a dynamic and innovative ecosystem. This ecosystem includes a variety of organizations that have played crucial roles. In this section, we propose a method to determine the relative contributions of different types of organizations to the final discovery. Our aim is to suggest an ideal redistribution of the hypothetical value of the innovation and acknowledge the technological sources (based on cited patents) that have been instrumental in the development of these vaccines over time. Our method consists of  a two-step process that combines information on the number of forward citations and the network distances of Moderna and BioNTech, the first two companies responsible for delivering the  mRNA vaccines. 

In doing so, we consider the three key features highlighted by the three approaches presented in Section \ref{Allocation} and detailed in \ref{app:AppC}. In assigning credit from node $i$ to node $j$ Rule 1 (\textit{Markov}) considers the \textit{local} importance of node $j$ in relation to node $i$, measured by citations and shortest paths; Rule 2 (\textit{Markov+Katz}) considers the \textit{local} importance outlined by Rule \#1, along with the \textit{global} importance of node $j$, determined by citations received from multiple or significant nodes; Rule 3 (\textit{Markov+PageRank}) considers the \textit{local} importance described in Rule \#1, combined with the \textit{global} importance of node $j$, assessed by citations received from multiple or important nodes discounting the effect of numerous connections. The primary difference of Rule 1 when compared to the other two rules is that it does not consider the general significance of the node in the overall progress of mRNA technology but only the importance in the progress of mRNA technologies used by Moderna (or Biontech). For this reason, in what follows we focus mainly on the first approach.

\begin{figure}[!ht]
\centering
\subfigure[Total allocation for sectors]
{\includegraphics[width=0.4\textwidth]{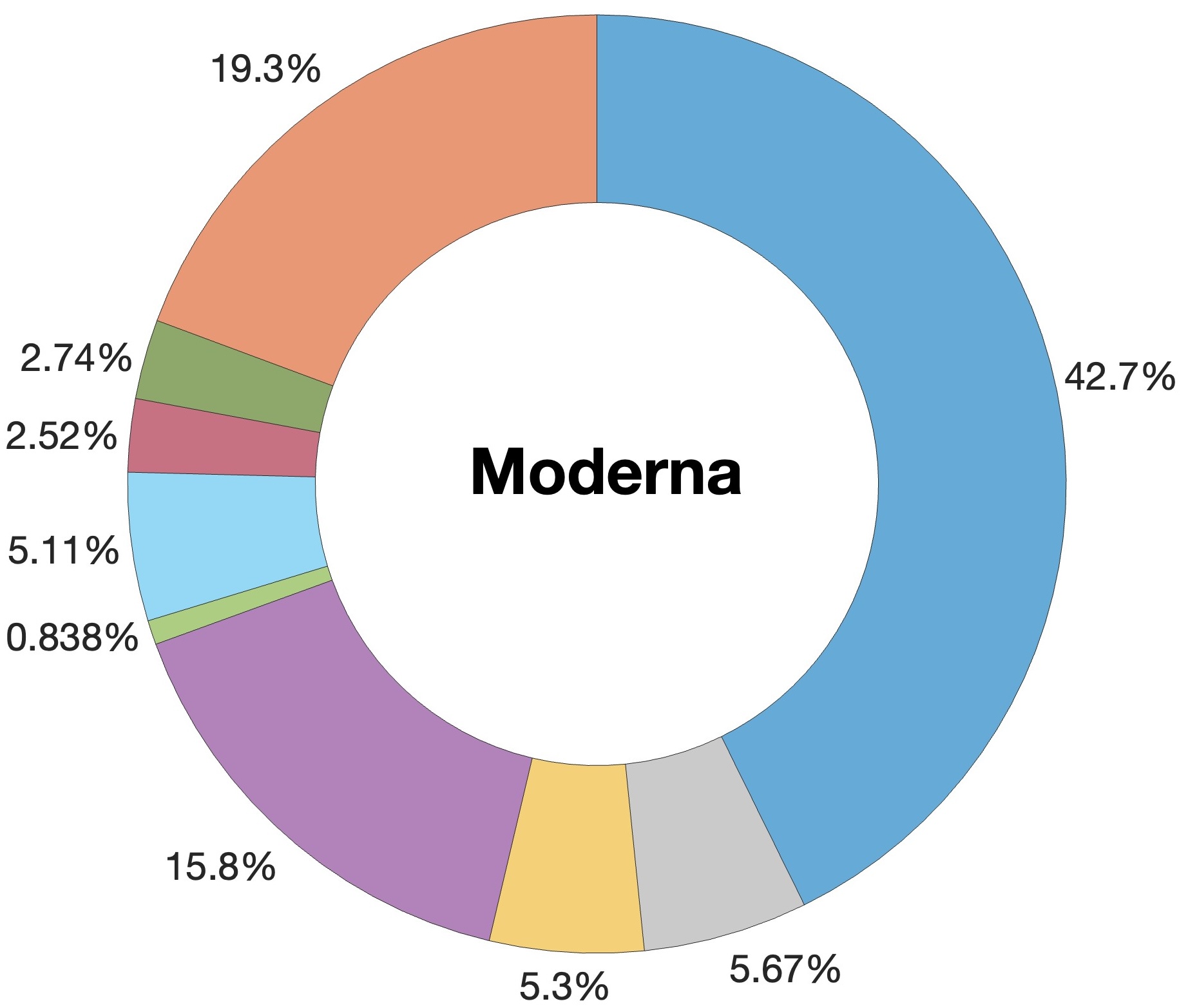}}
\subfigure[Average allocation for sectors]{\includegraphics[width=0.58\textwidth]{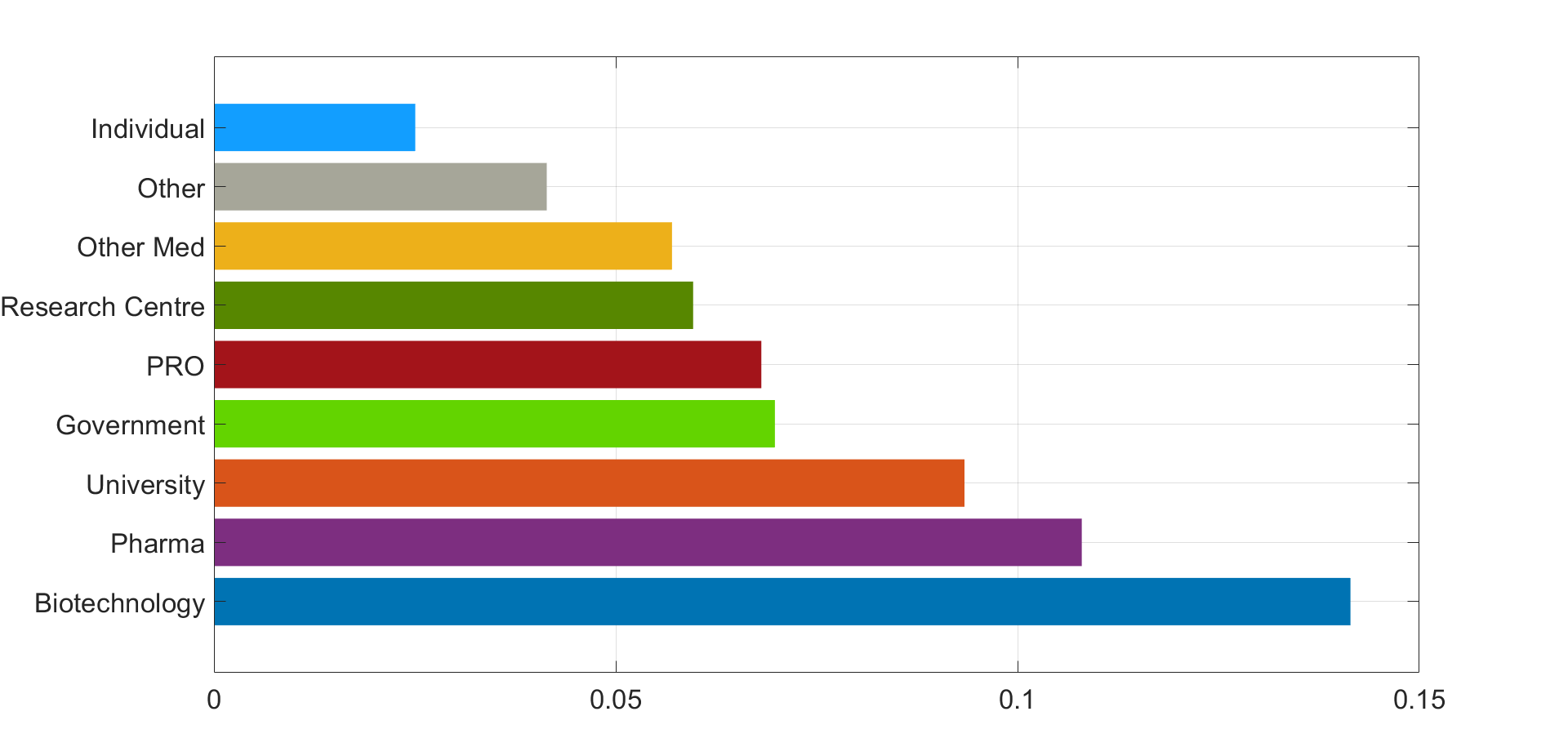}}
\subfigure[Total allocation for sectors]
{\includegraphics[width=0.4\textwidth]{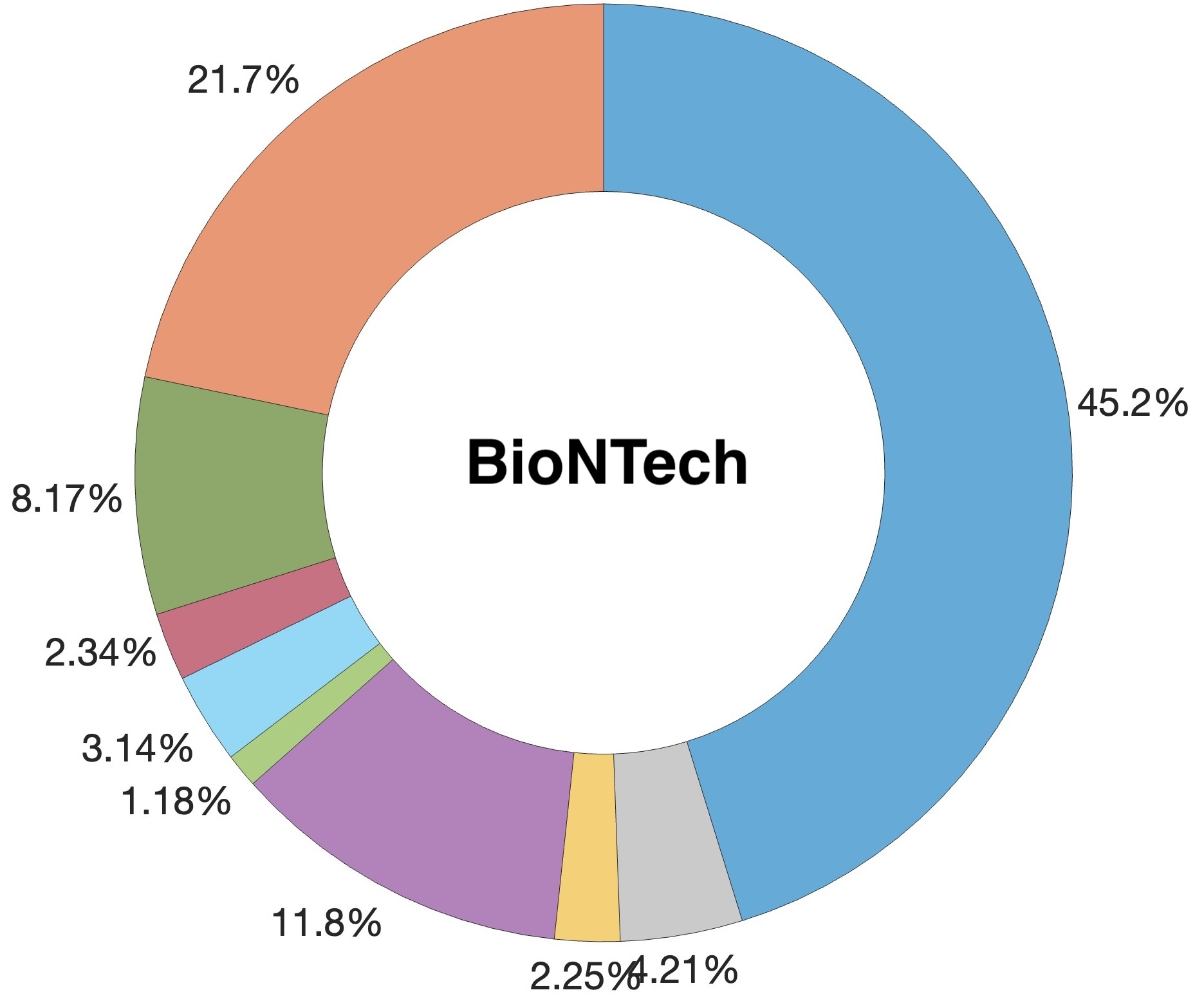}}
\subfigure[Average allocation for sectors]{\includegraphics[width=0.58\textwidth]{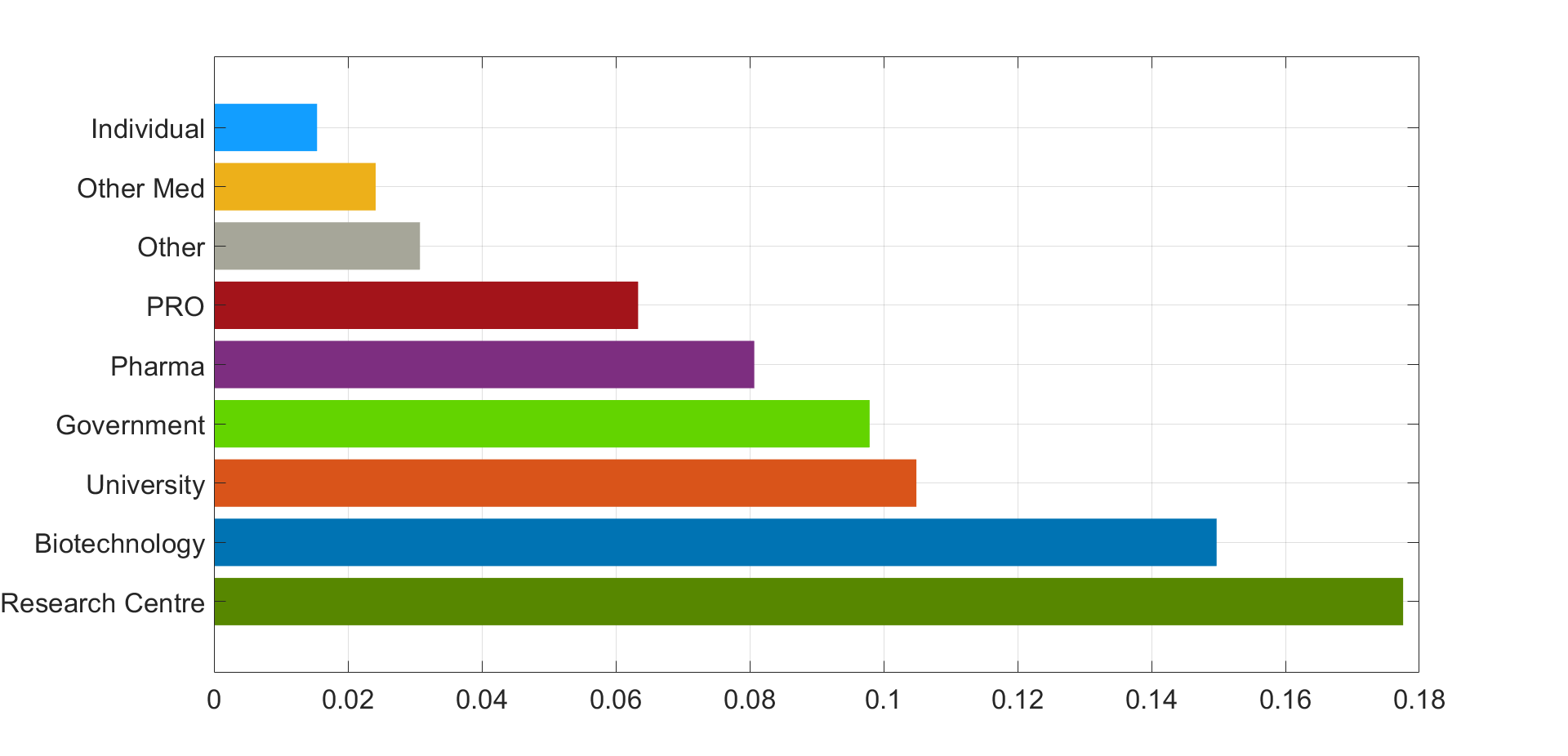}}
\caption{\textbf{Moderna and BioNTech's credit allocation.} Technological credit allocation of Moderna's (top) and BioNTech's (bottom) for sectors according to rule 1). (a),(c) Total credit allocated to each sector; (b),(d) average credit allocated to each sector.}
\label{ModAll1}
\end{figure}

In Figures \ref{ModAll1} we show the total (Figures (a),(c)) and average (Figures (b),(d)) credit allocated to each sector by using Rule 1 with respect to Moderna and BioNTech. This outcome could be interpreted in terms of the technological/scientific contribution given by the different sector to the development of the mRNA Covid-19 vaccines commercialized by Moderna and BioNTech \footnote{Results for the other two rules are qualitatively similar and are available upon request.}. From Figures \ref{ModAll1} (a),(c) it is evident that most of Moderna and BioNTech's credit allocation is directed towards the Biotechnology sector ($42.7\%$ and $45.2\%$, respectively), followed by Universities ($19.3\%$ and $21.7\%$, respectively). Summing up Universities, PROs, Research centers, Individuals and Government, these shares go up to $30\%$ and $37\%$ respectively.
This highlights once again the significant contributions of both the private and public sectors to the mRNA knowledge landscape. Furthermore, it is interesting to observe that Pharmaceutical companies rank third in both rankings.

These results capture two key components: the contribution of each individual patent and the overall number of patents within a given sector. Our approach also allows us to assess a potential size effect by determining whether some sectors appear to have made a limited contribution simply because they have fewer cited patents. To address this, we measure the contribution of each individual patent within its respective sector.

To account for sector size, we report the average credit allocation per patent within each sector (Figures \ref{ModAll1} (b),(d)). Notably, in the case of BioNTech, Research Centers receive the highest average credit allocation, followed by the Biotechnology sector, Universities, and Government institutions. This suggests that, despite having relatively few patents, research centers played a significant role in BioNTech’s COVID-19 knowledge network. In the case of Moderna, the Biotechnology sector holds the highest average credit allocation, followed by the Pharmaceutical sector, Universities, and Government institutions.

Going into detail, in Table \ref{tab:tab_Mod} and Table \ref{tab:tab_Bio} we show the top ten main contributors to Moderna's and BioNTech's inventions, respectively, expressed as a percentage of the final technological value of the two companies' vaccine, according to three rules described before.

\begin{table}[!ht]
  \centering
  \small
    \caption{Moderna's redistribution}
\scalebox{1}{ 
\begin{threeparttable}  
    \begin{tabular}{ll||ll||ll}
\toprule
   \textbf{\textit{Markov}}& \textbf{Share} &\textbf{\textit{Markov+Katz}}&\textbf{Share} &\textbf{\textit{Markov+PageRank}}\\
   \hline
   \hline 
  \text{Curevac} &6.2 \%& \text{Curevac}& 8.4 \% & \text{Curevac} & 9.3\%  \\
  
  \text{Moderna} & 5.2\% & \text{Novartis}& 7.1 \% & \text{Genentech}&8.6\%  \\
  
   \text{Novartis} &4.2\% &\text{Moderna} &6.3\% &\text{Novartis} & 5.8\%  \\
   
   \text{Bind Th.} &2.6\% &\text{Patent Univ Inc.}& 4.5\% & \text{Bioject} & 5.2\%\\
   
   \text{Patent Univ Inc.}&2.1\% & \text{Brickell}&4.5\%&\text{Anti Gene} & 4.1\%  \\
   
   \text{Bioject Inc.} & 2.0\% & \text{Genetech}& 3.9\%& \text{Patent Univ Inc.} & 4.1\%  \\
   
  \text{Brickell}& 1.6\% &\text{MIT}& 3.1\% & \text{Merck\&co.}& 3.3\%  \\
  
   \text{Genentech}& 1.6\% &\text{Inex*}& 2.8\% & \text{Brickell}& 3.3\% \\
   
   \text{Alnylam }&1.5\% & \text{Merck\&co.}&2.8\% & \text{Moderna} &3.0\%  \\
   
   \text{Inex*}& 1.4\% & \text{WARF**}&2.7\% & \text{Calif Univ}& 2.6\%\\
   
\hline \hline
   	\bottomrule
		  \end{tabular}
		\footnotesize
        The table displays the percentage of the final technological value of Moderna’s mRNA vaccine, based on three allocation rules outlined in the paper, assuming that 50\% of the value is directly assigned to Moderna. According to the Markov Rule, CureVac’s contribution is estimated at 3.1\% of the total value. (*Inex/Tekmira/Arbutus; **Wisconsin Alumni Research Foundation)

  \end{threeparttable}}
   \label{tab:tab_Mod}
\end{table} 


\begin{table}[!ht]
  \centering
    \caption{BioNTech's redistribution.}
\scalebox{1}{ 
\begin{threeparttable}  
    \begin{tabular}{ll||ll||ll}
\toprule
   \textbf{\textit{Markov}}& \textbf{Share} &\textbf{\textit{Markov+Katz}}&\textbf{Share} &\textbf{\textit{Markov+PageRank}}\\
   \hline
   \hline 
  \text{Moderna} &11.1 \%& \text{Moderna}& 13.1 \% & \text{Moderna} & 7.4\%  \\
  
  \text{Curevac} & 4.5\% & \text{Brickell}& 7.3 \% & \text{Curevac}&7.0\%  \\
  
   \text{BioNTech} &4.3\% &\text{Curevac} &5.8\% &\text{Brickell} & 6.3\%  \\
   
   \text{Mainz Univ} & 3.9\% &\text{Novartis}& 4.8\% & \text{Genentech} & 5.3\%\\
   
   \text{Tron}&3.7\% & \text{WARF**}&4.5\%&\text{Novartis} & 4.3\%  \\
   
   \text{Novartis} & 3.1\% & \text{BioNTech}& 3.4\%& \text{WARF**} & 4.1\%  \\
   
  \text{Brickell}& 2.5\% &\text{Liposome}& 3.8\% & \text{Merck\&co.}& 3.6\%  \\
  
   \text{DFCI*.}& 1.8\% &\text{Merck\&co.}& 2.6\% & \text{Liposome}& 3.5\% \\
   
   \text{WARF** }&1.7\% & \text{Genentech}&2.5\% & \text{BioNTech} &2.7\%  \\
   
   \text{Liposome}& 1.2\% & \text{DFCI*}&2.4\% & \text{DFCI*}& 2.6\%\\
   
\hline \hline
   	\bottomrule
		  \end{tabular}
		\footnotesize  The table displays the percentage of the final technological value of BioNTech’s mRNA vaccine, based on three allocation rules outlined in the paper, assuming that 50\% of the value is directly assigned to BioNTech. According to the Markov Rule, CureVac’s contribution is estimated at 2.2\% of the total value.
(*Dana-Faber Cancer Institute;**Wisconsin Alumni Research Foundation)
  \end{threeparttable}}
   \label{tab:tab_Bio}
\end{table} 

Table \ref{tab:tab_Mod} shows the top ranking for credit allocation for Moderna. In case of \textit{Markov} rule, Curevac will be assigned the $6.2\%$ of Moderna's technological credit, and Moderna itself will receive the $5.2\%$ from its credit redistribution. Furthermore, independently of the approach used, Curevac and Novartis appear in the top 3 ranking for credit allocation. 
Recalling the different properties emphasized by the three approaches (see section \ref{Allocation}), it becomes evident that the two companies are significant not only for the role they played with respect to Moderna, but also for their broader impact on the entire network. This highlights their global centrality within the mRNA knowledge ecosystem. This is confirmed by the fact that these two companies appear also as main contributors of Biontech's mRNA vaccine technology (See Table \ref{tab:tab_Bio}).
Finally, it is worth mentioning that Moderna is present in all the rankings (although only in position number 9 with rule \#2), suggesting that Moderna appears significant both locally (due to its large number of self-citations) and through its connections to other key nodes. On the other hand, Biontech is absent from the list, indicating that its role in the advancement of Moderna's mRNA vaccine might have been relatively minor.


Notably, four academic institutions emerge as a significant contributors to Moderna's mRNA technology: University Patents Inc., MIT, University of California and Wisconsin Alumni Research Foundation (WARF).  University Patents Inc. was a company in the U.S. dedicated to technology transfer for American universities. Our data includes ten patents assigned to University Patents, Inc., with nine originating from researchers at the University of Colorado. These patents are all part of the Antecedent dataset. Interestingly, seven of these patents have faced litigation at least once, indicating their relevance in the advancement of mRNA vaccine technology. MIT and the University of California are mentioned only in rules \#2 and \#3, indicating that their role extends beyond merely supporting the development of Moderna's technology and encompasses a broader contribution to advancing mRNA vaccine technology as a whole.


Table \ref{tab:tab_Bio} presents the leading rankings for credit distribution concerning Biontech. Under the \textit{Markov} rule, Moderna receives $11.1\%$ of the technological credit from Biontech; Curevac follows as another significant contributor with $4.5\%$, affirming its pivotal role in advancing mRNA vaccine technology. According to the remaining two criteria, Moderna retains its leading position, with Brickell climbing into the top three.

Notably, four scientific institutions emerge as a significant contributors to Biontech's mRNA technology (under rule \#1): Mainz University, Tron, Dana-Faber Cancer Institute, and Wisconsin Alumni Research Foundation. This is in line with the finding in Figure \ref{ModAll1} (d), where individual patents from Research Centers appear to have been particularly important in the Biontech Case.

Finally, one key insight of our analysis displayed in Tables \ref{tab:tab_Mod} and \ref{tab:tab_Bio} is that Moderna's advancements in mRNA technology appear to rely less on Biontech's mRNA developments compared to the reverse situation. This result may have some important implications on the current patent litigation that sees Moderna arguing that BioNTech/Pfizer's vaccine infringes on Moderna’s patents granted prior to the pandemic. 




\section{Conclusion}

A central tenet of science and innovation policy is that public investment in R\&D underpins breakthrough innovations with significant societal benefits. Unlike private R\&D, which is often closely tied to corporate products and patents, publicly funded research is typically disseminated more broadly. This makes it difficult to track its use and determine who benefits from it. Additionally, publicly funded research can have applications that extend far beyond its original scope, potentially taking years or even decades to materialize. This complicates the task of drawing direct links between public sector research activities and commercial outcomes. Given that public investment in science and technology often serves as the basis for important breakthrough innovations — with varying time lags and spillover effects in different areas — a key challenge lies in accurately attribute its contribution to these advances.

In this study, we have addressed this challenge by analyzing the development of mRNA vaccines against COVID-19, which represents a groundbreaking achievement of modern science and technology and embodies the rapid translation of basic research into life-saving innovations. Based on a sample of 151 patent families and exploiting their citations to 2,416 patent families, this paper uses network theory to analyze the innovative ecosystem that fostered this breakthrough innovation and, in particular, the dynamics and key players involved in the development of the mRNA vaccine platform, offering insights into the interplay between different companies and institutions.

We found that universities and public research organizations (PROs) played a  role in the development of the mRNA vaccine platform. Prior to 2010, their patents accounted for 44\% of our Core Dataset and 23\% of the Antecedent Database (which contains the backward patent citations of the Core Dataset). In addition, these institutions consistently ranked at the top of the authority ratings, highlighting their continued importance as a source of fundamental discoveries. Notable contributions came from the University of Pennsylvania, MIT, Max Planck, and the University of British Columbia, underscoring the essential role of academic research in providing the foundational knowledge necessary for mRNA vaccine development.

Based on this evidence, we have proposed a hypothetical redistribution of the innovation breakthrough's credit from the primary beneficiaries of mRNA COVID-19 vaccines — BioNTech and Moderna. Using the structural properties of the citation network, we found that the universities account for about 19.3\% and 21.7\% of an idealized value, respectively. These shares increase to 30\% and 37\% when PROs, research centers, government institutions and individuals are included.

It is important to emphasize that PROs and universities contribute to the innovation process through a variety of direct and indirect means, including the generation and dissemination of foundational discoveries, the development of human capital, and the provision of research infrastructure. Therefore, our results, based on patent citations, represent a significant yet conservative estimate of the overall contribution of the public sector to the successful development of mRNA vaccine platform.

This paper also provides a detailed description of the structure and dynamics of the COVID-19 knowledge network. Our investigation of patent assignees and their citation networks highlights the importance of interactions and knowledge flows between different types of organizations, with Biotechnology firms playing an important and ubiquitous role. Our results reveal the heterogeneous character of the mRNA knowledge network, which is characterized by a dense and increasingly interconnected structure over time. The temporal analysis shows that the number of players and the complexity of interactions increased significantly after 2010, corresponding to the years of intensified research efforts and investments in mRNA technology. During this period, new market players have also emerged and existing market leaders have consolidated their presence, indicating a maturing of the mRNA innovation landscape.

Biotechnology companies, especially in recent years, have emerged as major players, with a significant increase in Core patents and self-citations (e.g. Moderna, Curevac and companies specializing in efficient LNPs). These companies have played a central role in the development and dissemination of mRNA technologies and are involved in extensive citation networks. The centrality measures have further highlighted the importance of Biotechnology companies and underlined their influence and leadership in the field of mRNA.


Our analysis provides a nuanced understanding of the innovative ecosystem of the pharmaceutical and Biotechnology industries (as discussed in Section 2), particularly in relation to the discovery of major breakthroughs. The findings have several policy implications. First, they underscore the critical role of publicly funded research institutions in fundamental scientific discoveries. Policy makers should prioritize funding for these institutions and universities while supporting a broad range of research trajectories. It is noteworthy that for many years mRNA research was considered a niche area with limited immediate applications. This underlines the importance of maintaining a broad spectrum of scientific research without immediate returns.

The innovative ecosystem is characterized by a multitude of actors, and collaboration between the public and private sectors has proven to be essential for translating scientific research into practical applications. To facilitate this, policy should encourage and support such partnerships and ensure that public research is effectively leveraged for innovation (especially in Biotechnology companies). This includes implementing programs to streamline and facilitate licensing and creating mechanisms to mitigate patent disputes, especially in areas with a dense patent landscape and potentially overlapping claims.


Finally, our study suggests that the development of long-term impact assessment methods is necessary to fully appreciate and demonstrate the value of public R\&D investments. Such methods would help to quantify the broader societal benefits of public research beyond its immediate commercial applications and provide a more comprehensive view of its contributions. In this paper, we present a new methodological approach to trace back the underpinnings of breakthrough innovations using credit allocation schemes based on knowledge network analysis.

\subsubsection*{Data Availability}

\noindent Data available upon request.

\subsubsection*{Declaration of competing interest}

\noindent The authors declare that they have no conflicts of interest.

\pagebreak
\setlength\bibsep{0.5pt}
\bibliographystyle{elsarticle-harv}
\bibliography{biblio}

\begin{thebibliography}{89}
\newcommand{\enquote}[1]{``#1''}
\expandafter\ifx\csname natexlab\endcsname\relax\def\natexlab#1{#1}\fi

\bibitem[\protect\citeauthoryear{Agarwal and Gaule}{Agarwal and
  Gaule}{2022}]{agarwal2022}
\textsc{Agarwal, R. and P.~Gaule} (2022): \enquote{What drives innovation?
  Lessons from COVID-19 R\&D,} \emph{Journal of Health Economics}, 102591.

\bibitem[\protect\citeauthoryear{Ahuja and Morris~Lampert}{Ahuja and
  Morris~Lampert}{2001}]{ahuja2001}
\textsc{Ahuja, G. and C.~Morris~Lampert} (2001): \enquote{Entrepreneurship in
  the large corporation: A longitudinal study of how established firms create
  breakthrough inventions,} \emph{Strategic management journal}, 22, 521--543.

\bibitem[\protect\citeauthoryear{Androsavich}{Androsavich}{2024}]{androsavich2024}
\textsc{Androsavich, J.~R.} (2024): \enquote{Frameworks for transformational
  breakthroughs in RNA-based medicines,} \emph{Nature Reviews Drug Discovery},
  1--24.

\bibitem[\protect\citeauthoryear{Barab{\'a}si}{Barab{\'a}si}{2016}]{barabasi2016network}
\textsc{Barab{\'a}si, A.} (2016): \enquote{Network Science. Cambridge
  University Press, Cambridge,} .

\bibitem[\protect\citeauthoryear{Barber{\'a}-Tom{\'a}s, Jim{\'e}nez-S{\'a}ez,
  and Castell{\'o}-Molina}{Barber{\'a}-Tom{\'a}s
  et~al.}{2011}]{barbera2011mapping}
\textsc{Barber{\'a}-Tom{\'a}s, D., F.~Jim{\'e}nez-S{\'a}ez, and
  I.~Castell{\'o}-Molina} (2011): \enquote{Mapping the importance of the real
  world: The validity of connectivity analysis of patent citations networks,}
  \emph{Research policy}, 40, 473--486.

\bibitem[\protect\citeauthoryear{Barbier, Jiang, Zhang, Wooster, and
  Anderson}{Barbier et~al.}{2022}]{barbier2022}
\textsc{Barbier, A.~J., A.~Y. Jiang, P.~Zhang, R.~Wooster, and D.~G. Anderson}
  (2022): \enquote{The clinical progress of mRNA vaccines and immunotherapies,}
  \emph{Nature biotechnology}, 40, 840--854.

\bibitem[\protect\citeauthoryear{Bessen}{Bessen}{2009}]{bessen09}
\textsc{Bessen, J.} (2009): \enquote{NBER PDP Project User Documentation:
  Matching Patent Data to Compustat Firms,} Working {{Paper}}, Unpublished
  working paper, Boston University.

\bibitem[\protect\citeauthoryear{Blondel, Guillaume, Lambiotte, and
  Lefebvre}{Blondel et~al.}{2008}]{blondel2008fast}
\textsc{Blondel, V.~D., J.-L. Guillaume, R.~Lambiotte, and E.~Lefebvre} (2008):
  \enquote{Fast unfolding of communities in large networks,} \emph{Journal of
  statistical mechanics: theory and experiment}, 2008, P10008.

\bibitem[\protect\citeauthoryear{Brin and Page}{Brin and
  Page}{1998}]{brin1998anatomy}
\textsc{Brin, S. and L.~Page} (1998): \enquote{The anatomy of a large-scale
  hypertextual web search engine,} \emph{Computer networks and ISDN systems},
  30, 107--117.

\bibitem[\protect\citeauthoryear{Capponi, Martinelli, and Nuvolari}{Capponi
  et~al.}{2022}]{capponi2022}
\textsc{Capponi, G., A.~Martinelli, and A.~Nuvolari} (2022):
  \enquote{Breakthrough innovations and where to find them,} \emph{Research
  Policy}, 51, 104376.

\bibitem[\protect\citeauthoryear{Chakraborty, Byshkin, and
  Crestani}{Chakraborty et~al.}{2020}]{chakraborty2020patent}
\textsc{Chakraborty, M., M.~Byshkin, and F.~Crestani} (2020): \enquote{Patent
  citation network analysis: A perspective from descriptive statistics and
  ERGMs,} \emph{Plos one}, 15, e0241797.

\bibitem[\protect\citeauthoryear{Cho and Shih}{Cho and
  Shih}{2011}]{cho2011patent}
\textsc{Cho, T.-S. and H.-Y. Shih} (2011): \enquote{Patent citation network
  analysis of core and emerging technologies in Taiwan: 1997--2008,}
  \emph{Scientometrics}, 89, 795--811.

\bibitem[\protect\citeauthoryear{Cockburn and Henderson}{Cockburn and
  Henderson}{1998}]{cockburn1998}
\textsc{Cockburn, I.~M. and R.~M. Henderson} (1998): \enquote{Absorptive
  capacity, coauthoring behavior, and the organization of research in drug
  discovery,} \emph{The journal of industrial economics}, 46, 157--182.

\bibitem[\protect\citeauthoryear{Cockburn, Henderson, and Stern}{Cockburn
  et~al.}{2000}]{cockburn2000}
\textsc{Cockburn, I.~M., R.~M. Henderson, and S.~Stern} (2000):
  \enquote{Untangling the origins of competitive advantage,} \emph{Strategic
  management journal}, 21, 1123--1145.

\bibitem[\protect\citeauthoryear{Dahlin and Behrens}{Dahlin and
  Behrens}{2005}]{dahlin2005}
\textsc{Dahlin, K.~B. and D.~M. Behrens} (2005): \enquote{When is an invention
  really radical?: Defining and measuring technological radicalness,}
  \emph{Research policy}, 34, 717--737.

\bibitem[\protect\citeauthoryear{David, Mowery, and Steinmueller}{David
  et~al.}{2001}]{david2001public}
\textsc{David, P.~A., D.~C. Mowery, and W.~E. Steinmueller} (2001):
  \enquote{Public Policy and Knowledge-Based Economy in the United States,} in
  \emph{Innovation Policy and the Economy, Volume 1}, ed. by A.~Jaffe,
  J.~Lerner, and S.~Stern, Cambridge, MA: MIT Press, 103--142.

\bibitem[\protect\citeauthoryear{DiMasi, Hansen, and Grabowski}{DiMasi
  et~al.}{2003}]{dimasi2003price}
\textsc{DiMasi, J.~A., R.~W. Hansen, and H.~G. Grabowski} (2003): \enquote{The
  price of innovation: new estimates of drug development costs,} \emph{Journal
  of health economics}, 22, 151--185.

\bibitem[\protect\citeauthoryear{Ding, Yan, Frazho, and Caverlee}{Ding
  et~al.}{2009}]{ding2009pagerank}
\textsc{Ding, Y., E.~Yan, A.~Frazho, and J.~Caverlee} (2009): \enquote{PageRank
  for ranking authors in co-citation networks,} \emph{Journal of the American
  Society for Information Science and Technology}, 60, 2229--2243.

\bibitem[\protect\citeauthoryear{Dolgin}{Dolgin}{2021{\natexlab{a}}}]{dolgin2021covid}
\textsc{Dolgin, E.} (2021{\natexlab{a}}): \enquote{How COVID unlocked the power
  of RNA vaccines.} \emph{Nature}, 189--191.

\bibitem[\protect\citeauthoryear{Dolgin}{Dolgin}{2021{\natexlab{b}}}]{dolgin2021tangled}
---\hspace{-.1pt}---\hspace{-.1pt}--- (2021{\natexlab{b}}): \enquote{The
  tangled history of mRNA vaccines,} \emph{Nature}, 597, 318--324.

\bibitem[\protect\citeauthoryear{Dosi}{Dosi}{1982}]{dosi1982technological}
\textsc{Dosi, G.} (1982): \enquote{Technological paradigms and technological
  trajectories: a suggested interpretation of the determinants and directions
  of technical change,} \emph{Research policy}, 11, 147--162.

\bibitem[\protect\citeauthoryear{D’Souza and Snyder}{D’Souza and
  Snyder}{2024}]{d2024can}
\textsc{D’Souza, A. and C.~M. Snyder} (2024): \enquote{Can Operation Warp
  Speed Serve as a Model for Accelerating Innovations Beyond Covid Vaccines?}
  \emph{NBER Chapters}.

\bibitem[\protect\citeauthoryear{{\'E}rdi, Makovi, Somogyv{\'a}ri, Strandburg,
  Tobochnik, Volf, and Zal{\'a}nyi}{{\'E}rdi et~al.}{2013}]{erdi2013prediction}
\textsc{{\'E}rdi, P., K.~Makovi, Z.~Somogyv{\'a}ri, K.~Strandburg,
  J.~Tobochnik, P.~Volf, and L.~Zal{\'a}nyi} (2013): \enquote{Prediction of
  emerging technologies based on analysis of the US patent citation network,}
  \emph{Scientometrics}, 95, 225--242.

\bibitem[\protect\citeauthoryear{Fauci}{Fauci}{2021}]{fauci_story_2021}
\textsc{Fauci, A.~S.} (2021): \enquote{The story behind {COVID}-19 vaccines,}
  \emph{Science}, 372, 109--109.

\bibitem[\protect\citeauthoryear{Fleming}{Fleming}{2001}]{fleming2001}
\textsc{Fleming, L.} (2001): \enquote{Recombinant uncertainty in technological
  search,} \emph{Management science}, 47, 117--132.

\bibitem[\protect\citeauthoryear{Florio}{Florio}{2022}]{florio2022extent}
\textsc{Florio, M.} (2022): \enquote{To what extent patents for Covid-19 mRNA
  vaccines are based on public research and taxpayers’ funding? A case study
  on the privatization of knowledge,} \emph{Industrial and Corporate Change},
  31, 1137--1151.

\bibitem[\protect\citeauthoryear{Florio, Gamba, and Pancotti}{Florio
  et~al.}{2023}]{florio2023mapping}
\textsc{Florio, M., S.~Gamba, and C.~Pancotti} (2023): \enquote{Mapping of
  long-term public and private investments in the development of Covid-19
  vaccines,} \emph{Eur Parliament COVI Committee}, 1--95.

\bibitem[\protect\citeauthoryear{Franzoni, Stephan, and Veugelers}{Franzoni
  et~al.}{2022}]{franzoni2022}
\textsc{Franzoni, C., P.~Stephan, and R.~Veugelers} (2022): \enquote{Funding
  risky research,} \emph{Entrepreneurship and Innovation Policy and the
  Economy}, 1, 103--133.

\bibitem[\protect\citeauthoryear{Galkina~Cleary, Beierlein, Khanuja, McNamee,
  and Ledley}{Galkina~Cleary et~al.}{2018}]{galkina_cleary_contribution_2018}
\textsc{Galkina~Cleary, E., J.~M. Beierlein, N.~S. Khanuja, L.~M. McNamee, and
  F.~D. Ledley} (2018): \enquote{Contribution of {NIH} funding to new drug
  approvals 2010–2016,} \emph{Proceedings of the National Academy of
  Sciences}, 115, 2329--2334.

\bibitem[\protect\citeauthoryear{Gambardella et~al.}{Gambardella
  et~al.}{1995}]{gambardella1995}
\textsc{Gambardella, A. et~al.} (1995): \emph{Science and innovation: The US
  pharmaceutical industry during the 1980s}, Cambridge University Press.

\bibitem[\protect\citeauthoryear{Gaviria and Kilic}{Gaviria and
  Kilic}{2021}]{gaviria2021network}
\textsc{Gaviria, M. and B.~Kilic} (2021): \enquote{A network analysis of
  COVID-19 mRNA vaccine patents,} .

\bibitem[\protect\citeauthoryear{Gittelman}{Gittelman}{2016}]{gittelman2016}
\textsc{Gittelman, M.} (2016): \enquote{The revolution re-visited: Clinical and
  genetics research paradigms and the productivity paradox in drug discovery,}
  \emph{Research Policy}, 45, 1570--1585.

\bibitem[\protect\citeauthoryear{Golosovsky and Solomon}{Golosovsky and
  Solomon}{2017}]{golosovsky2017growing}
\textsc{Golosovsky, M. and S.~Solomon} (2017): \enquote{Growing complex network
  of citations of scientific papers: Modeling and measurements,} \emph{Physical
  Review E}, 95, 012324.

\bibitem[\protect\citeauthoryear{Guan, Yan, and Zhang}{Guan
  et~al.}{2017}]{guan2017impact}
\textsc{Guan, J., Y.~Yan, and J.~J. Zhang} (2017): \enquote{The impact of
  collaboration and knowledge networks on citations,} \emph{Journal of
  Informetrics}, 11, 407--422.

\bibitem[\protect\citeauthoryear{Hall, Jaffe, and Trajtenberg}{Hall
  et~al.}{2005{\natexlab{a}}}]{hall_market_2005}
\textsc{Hall, B., A.~Jaffe, and M.~Trajtenberg} (2005{\natexlab{a}}):
  \enquote{Market value and patent citations,} 36, 16--38, type: Journal
  Article.

\bibitem[\protect\citeauthoryear{Hall, Jaffe, and Trajtenberg}{Hall
  et~al.}{2005{\natexlab{b}}}]{hall2005}
\textsc{Hall, B.~H., A.~Jaffe, and M.~Trajtenberg} (2005{\natexlab{b}}):
  \enquote{Market value and patent citations,} \emph{RAND Journal of
  economics}, 16--38.

\bibitem[\protect\citeauthoryear{Huenteler, Ossenbrink, Schmidt, and
  Hoffmann}{Huenteler et~al.}{2016}]{huenteler2016product}
\textsc{Huenteler, J., J.~Ossenbrink, T.~S. Schmidt, and V.~H. Hoffmann}
  (2016): \enquote{How a product’s design hierarchy shapes the evolution of
  technological knowledge—Evidence from patent-citation networks in wind
  power,} \emph{Research Policy}, 45, 1195--1217.

\bibitem[\protect\citeauthoryear{Hughes}{Hughes}{2001}]{hughes2001}
\textsc{Hughes, S.~S.} (2001): \enquote{Making dollars out of DNA: the first
  major patent in biotechnology and the commercialization of molecular biology,
  1974-1980,} \emph{Isis}, 92, 541--575.

\bibitem[\protect\citeauthoryear{Hummon and Dereian}{Hummon and
  Dereian}{1989}]{hummon1989connectivity}
\textsc{Hummon, N.~P. and P.~Dereian} (1989): \enquote{Connectivity in a
  citation network: The development of DNA theory,} \emph{Social networks}, 11,
  39--63.

\bibitem[\protect\citeauthoryear{Iori, Martinelli, and Mina}{Iori
  et~al.}{2022}]{iori_direction_2022}
\textsc{Iori, M., A.~Martinelli, and A.~Mina} (2022): \enquote{The direction of
  technical change in {AI} and the trajectory effects of government funding,} .

\bibitem[\protect\citeauthoryear{Jaffe and de~Rassenfosse}{Jaffe and
  de~Rassenfosse}{2017}]{jaffe_patent_2017}
\textsc{Jaffe, A.~B. and G.~de~Rassenfosse} (2017): \enquote{Patent citation
  data in social science research: Overview and best practices,} 68,
  1360--1374, place: Hoboken Publisher: Wiley {WOS}:000401545600002.

\bibitem[\protect\citeauthoryear{Kaitin, Bryant, and Lasagna}{Kaitin
  et~al.}{1993}]{kaitin1993role}
\textsc{Kaitin, K.~I., N.~R. Bryant, and L.~Lasagna} (1993): \enquote{The role
  of the research-based pharmaceutical industry in medical progress in the
  United States,} \emph{The Journal of Clinical Pharmacology}, 33, 412--417.

\bibitem[\protect\citeauthoryear{Kajikawa, Ohno, Takeda, Matsushima, and
  Komiyama}{Kajikawa et~al.}{2007}]{kajikawa2007creating}
\textsc{Kajikawa, Y., J.~Ohno, Y.~Takeda, K.~Matsushima, and H.~Komiyama}
  (2007): \enquote{Creating an academic landscape of sustainability science: an
  analysis of the citation network,} \emph{Sustainability Science}, 2,
  221--231.

\bibitem[\protect\citeauthoryear{Kaplan and Vakili}{Kaplan and
  Vakili}{2015}]{kaplan_double-edged_2015}
\textsc{Kaplan, S. and K.~Vakili} (2015): \enquote{The double-edged sword of
  recombination in breakthrough innovation: {The} {Double}-{Edged} {Sword} of
  {Recombination},} \emph{Strategic Management Journal}, 36, 1435--1457.

\bibitem[\protect\citeauthoryear{Karik{\'o}, Buckstein, Ni, and
  Weissman}{Karik{\'o} et~al.}{2005}]{kariko2005suppression}
\textsc{Karik{\'o}, K., M.~Buckstein, H.~Ni, and D.~Weissman} (2005):
  \enquote{Suppression of RNA recognition by Toll-like receptors: the impact of
  nucleoside modification and the evolutionary origin of RNA,} \emph{Immunity},
  23, 165--175.

\bibitem[\protect\citeauthoryear{Kirchdoerfer, Cottrell, Wang, Pallesen,
  Yassine, Turner, Corbett, Graham, McLellan, and Ward}{Kirchdoerfer
  et~al.}{2016}]{kirchdoerfer2016pre}
\textsc{Kirchdoerfer, R.~N., C.~A. Cottrell, N.~Wang, J.~Pallesen, H.~M.
  Yassine, H.~L. Turner, K.~S. Corbett, B.~S. Graham, J.~S. McLellan, and A.~B.
  Ward} (2016): \enquote{Pre-fusion structure of a human coronavirus spike
  protein,} \emph{Nature}, 531, 118--121.

\bibitem[\protect\citeauthoryear{Kleinberg}{Kleinberg}{1999}]{kleinberg1999hubs}
\textsc{Kleinberg, J.~M.} (1999): \enquote{Hubs, authorities, and communities,}
  \emph{ACM computing surveys (CSUR)}, 31, 5--es.

\bibitem[\protect\citeauthoryear{Kuhn}{Kuhn}{1962}]{kuhn_structure_1962}
\textsc{Kuhn, T.~S.} (1962): \enquote{The {Structure} of {Scientific}
  {Revolutions}.} .

\bibitem[\protect\citeauthoryear{Li, Chen, Huang, and Roco}{Li
  et~al.}{2007}]{li2007patent}
\textsc{Li, X., H.~Chen, Z.~Huang, and M.~C. Roco} (2007): \enquote{Patent
  citation network in nanotechnology (1976--2004),} \emph{Journal of
  Nanoparticle Research}, 9, 337--352.

\bibitem[\protect\citeauthoryear{Malaspina}{Malaspina}{2019}]{malaspina2019patent}
\textsc{Malaspina, P.~A.} (2019): \enquote{Patent citation analysis and patent
  damages,} \emph{Chi.-Kent J. Intell. Prop.}, 18, 232.

\bibitem[\protect\citeauthoryear{Malhotra, Zhang, Beuse, and Schmidt}{Malhotra
  et~al.}{2021}]{malhotra2021new}
\textsc{Malhotra, A., H.~Zhang, M.~Beuse, and T.~Schmidt} (2021): \enquote{How
  do new use environments influence a technology's knowledge trajectory? A
  patent citation network analysis of lithium-ion battery technology,}
  \emph{Research Policy}, 50, 104318.

\bibitem[\protect\citeauthoryear{Mansfield}{Mansfield}{1991}]{mansfield_academic_1991}
\textsc{Mansfield, E.} (1991): \enquote{Academic research and industrial
  innovation,} \emph{Research Policy}, 20, 1--12.

\bibitem[\protect\citeauthoryear{Mansfield}{Mansfield}{1995}]{mansfield_academic_1995}
---\hspace{-.1pt}---\hspace{-.1pt}--- (1995): \enquote{Academic {Research}
  {Underlying} {Industrial} {Innovations}: {Sources}, {Characteristics}, and
  {Financing},} \emph{The Review of Economics and Statistics}, 77, 55.

\bibitem[\protect\citeauthoryear{Mansfield}{Mansfield}{1998}]{mansfield1998}
---\hspace{-.1pt}---\hspace{-.1pt}--- (1998): \enquote{Academic research and
  industrial innovation: An update of empirical findings,} \emph{Research
  policy}, 26, 773--776.

\bibitem[\protect\citeauthoryear{Mariani, Medo, and Lafond}{Mariani
  et~al.}{2019}]{mariani2019early}
\textsc{Mariani, M.~S., M.~Medo, and F.~Lafond} (2019): \enquote{Early
  identification of important patents: Design and validation of citation
  network metrics,} \emph{Technological forecasting and social change}, 146,
  644--654.

\bibitem[\protect\citeauthoryear{Martin and Lowery}{Martin and
  Lowery}{2020}]{martin2020mrna}
\textsc{Martin, C. and D.~Lowery} (2020): \enquote{mRNA vaccines: intellectual
  property landscape.} \emph{Nature Reviews Drug Discovery}, 19, 578--579.

\bibitem[\protect\citeauthoryear{McKelvey, Alm, and Riccaboni}{McKelvey
  et~al.}{2003}]{mckelvey2003}
\textsc{McKelvey, M., H.~Alm, and M.~Riccaboni} (2003): \enquote{Does
  co-location matter for formal knowledge collaboration in the Swedish
  biotechnology--pharmaceutical sector?} \emph{Research policy}, 32, 483--501.

\bibitem[\protect\citeauthoryear{Montobbio, Pellegrino, and Sterzi}{Montobbio
  et~al.}{2024}]{Montobbio_et_al_2024}
\textsc{Montobbio, F., G.~Pellegrino, and V.~Sterzi} (2024): \enquote{Patent
  landscape analysis of mRNA COVID-19 Vaccine Technology: Examining the role of
  biotech companies, public research and universities,} in \emph{Handbook of
  Innovation and Intellectual Property Rights. Evolving Scholarship and
  Reflections.}, ed. by W.~Park, Edward Elgar, chap.~24, 415--437.

\bibitem[\protect\citeauthoryear{Mowery, Nelson, Sampat, and Ziedonis}{Mowery
  et~al.}{2004}]{mowery2004ivory}
\textsc{Mowery, D.~C., R.~R. Nelson, B.~N. Sampat, and A.~A. Ziedonis} (2004):
  \emph{Ivory Tower and Industrial Innovation: University-Industry Technology
  Transfer Before and After the Bayh-Dole Act}, Stanford, CA: Stanford
  University Press.

\bibitem[\protect\citeauthoryear{Narin, Hamilton, and Olivastro}{Narin
  et~al.}{1997}]{narin_increasing_1997}
\textsc{Narin, F., K.~S. Hamilton, and D.~Olivastro} (1997): \enquote{The
  increasing linkage between {U}.{S}. technology and public science,}
  \emph{Research Policy}, 26, 317--330.

\bibitem[\protect\citeauthoryear{Narin and Olivastro}{Narin and
  Olivastro}{1992}]{narin1992status}
\textsc{Narin, F. and D.~Olivastro} (1992): \enquote{Status report: linkage
  between technology and science,} \emph{Research policy}, 21, 237--249.

\bibitem[\protect\citeauthoryear{Newman}{Newman}{2018}]{newman2018networks}
\textsc{Newman, M.} (2018): \emph{Networks}, Oxford university press.

\bibitem[\protect\citeauthoryear{Newman}{Newman}{2006}]{newman2006modularity}
\textsc{Newman, M.~E.} (2006): \enquote{Modularity and community structure in
  networks,} \emph{Proceedings of the national academy of sciences}, 103,
  8577--8582.

\bibitem[\protect\citeauthoryear{Ng}{Ng}{2004}]{ng_drugs_2004}
\textsc{Ng, R.} (2004): \emph{Drugs: {From} {Discovery} to {Approval}.}, New
  Jersey: John Wiley.

\bibitem[\protect\citeauthoryear{Orsenigo, Pammolli, and Riccaboni}{Orsenigo
  et~al.}{2001}]{orsenigo2001}
\textsc{Orsenigo, L., F.~Pammolli, and M.~Riccaboni} (2001):
  \enquote{Technological change and network dynamics: lessons from the
  pharmaceutical industry,} \emph{Research policy}, 30, 485--508.

\bibitem[\protect\citeauthoryear{Owen-Smith, Riccaboni, Pammolli, and
  Powell}{Owen-Smith et~al.}{2002}]{owen2002comparison}
\textsc{Owen-Smith, J., M.~Riccaboni, F.~Pammolli, and W.~W. Powell} (2002):
  \enquote{A comparison of US and European university-industry relations in the
  life sciences,} \emph{Management science}, 48, 24--43.

\bibitem[\protect\citeauthoryear{Pallesen, Wang, Corbett, Wrapp, Kirchdoerfer,
  Turner, Cottrell, Becker, Wang, Shi, Kong, Andres, Kettenbach, Denison,
  Chappell, Graham, Ward, and McLellan}{Pallesen
  et~al.}{2017}]{pallesen_immunogenicity_2017}
\textsc{Pallesen, J., N.~Wang, K.~S. Corbett, D.~Wrapp, R.~N. Kirchdoerfer,
  H.~L. Turner, C.~A. Cottrell, M.~M. Becker, L.~Wang, W.~Shi, W.-P. Kong,
  E.~L. Andres, A.~N. Kettenbach, M.~R. Denison, J.~D. Chappell, B.~S. Graham,
  A.~B. Ward, and J.~S. McLellan} (2017): \enquote{Immunogenicity and
  structures of a rationally designed prefusion {MERS}-{CoV} spike antigen,}
  \emph{Proceedings of the National Academy of Sciences}, 114, E7348--E7357.

\bibitem[\protect\citeauthoryear{Pammolli, Riccaboni, and Spelta}{Pammolli
  et~al.}{2021}]{pammolli2021}
\textsc{Pammolli, F., M.~Riccaboni, and A.~Spelta} (2021): \enquote{The network
  origins of Schumpeterian innovation,} \emph{Journal of Evolutionary
  Economics}, 31, 1411--1431.

\bibitem[\protect\citeauthoryear{Pardi, Hogan, Porter, and Weissman}{Pardi
  et~al.}{2018}]{pardi2018}
\textsc{Pardi, N., M.~J. Hogan, F.~W. Porter, and D.~Weissman} (2018):
  \enquote{mRNA vaccines—a new era in vaccinology,} \emph{Nature reviews Drug
  discovery}, 17, 261--279.

\bibitem[\protect\citeauthoryear{Patridge, Gareiss, Kinch, and Hoyer}{Patridge
  et~al.}{2015}]{patridge2015analysis}
\textsc{Patridge, E.~V., P.~C. Gareiss, M.~S. Kinch, and D.~W. Hoyer} (2015):
  \enquote{An analysis of original research contributions toward FDA-approved
  drugs,} \emph{Drug discovery today}, 20, 1182--1187.

\bibitem[\protect\citeauthoryear{Phene, Fladmoe-Lindquist, and Marsh}{Phene
  et~al.}{2006}]{phene2006}
\textsc{Phene, A., K.~Fladmoe-Lindquist, and L.~Marsh} (2006):
  \enquote{Breakthrough innovations in the US biotechnology industry: the
  effects of technological space and geographic origin,} \emph{Strategic
  management journal}, 27, 369--388.

\bibitem[\protect\citeauthoryear{Powell, Koput, and Smith-Doerr}{Powell
  et~al.}{1996}]{powell_interorganizational_1996}
\textsc{Powell, W.~W., K.~W. Koput, and L.~Smith-Doerr} (1996):
  \enquote{Interorganizational {Collaboration} and the {Locus} of {Innovation}:
  {Networks} of {Learning} in {Biotechnology},} \emph{Administrative Science
  Quarterly}, 41, 116.

\bibitem[\protect\citeauthoryear{Radicchi, Fortunato, and Vespignani}{Radicchi
  et~al.}{2011}]{radicchi2011citation}
\textsc{Radicchi, F., S.~Fortunato, and A.~Vespignani} (2011):
  \enquote{Citation networks,} \emph{Models of science dynamics: Encounters
  between complexity theory and information sciences}, 233--257.

\bibitem[\protect\citeauthoryear{Sahin, Karik{\'o}, and T{\"u}reci}{Sahin
  et~al.}{2014}]{sahin2014}
\textsc{Sahin, U., K.~Karik{\'o}, and {\"O}.~T{\"u}reci} (2014):
  \enquote{mRNA-based therapeutics—developing a new class of drugs,}
  \emph{Nature reviews Drug discovery}, 13, 759--780.

\bibitem[\protect\citeauthoryear{Sampat}{Sampat}{2009}]{sampat2009academic}
\textsc{Sampat, B.~N.} (2009): \enquote{Academic patents and access to
  medicines in developing countries,} \emph{American Journal of Public Health},
  99, 9--17.

\bibitem[\protect\citeauthoryear{Sampat and Lichtenberg}{Sampat and
  Lichtenberg}{2011}]{sampat2011respective}
\textsc{Sampat, B.~N. and F.~R. Lichtenberg} (2011): \enquote{What are the
  respective roles of the public and private sectors in pharmaceutical
  innovation?} \emph{Health affairs}, 30, 332--339.

\bibitem[\protect\citeauthoryear{Silvestri, Riccaboni, and
  Della~Malva}{Silvestri et~al.}{2018}]{silvestri2018}
\textsc{Silvestri, D., M.~Riccaboni, and A.~Della~Malva} (2018):
  \enquote{Sailing in all winds: Technological search over the business cycle,}
  \emph{Research Policy}, 47, 1933--1944.

\bibitem[\protect\citeauthoryear{Slaoui and Hepburn}{Slaoui and
  Hepburn}{2020}]{slaoui2020}
\textsc{Slaoui, M. and M.~Hepburn} (2020): \enquote{Developing safe and
  effective Covid vaccines—Operation Warp Speed’s strategy and approach,}
  \emph{New England Journal of Medicine}, 383, 1701--1703.

\bibitem[\protect\citeauthoryear{Stevens, Jensen, Wyller, Kilgore, Chatterjee,
  and Rohrbaugh}{Stevens et~al.}{2011}]{stevens2011role}
\textsc{Stevens, A.~J., J.~J. Jensen, K.~Wyller, P.~C. Kilgore, S.~Chatterjee,
  and M.~L. Rohrbaugh} (2011): \enquote{The role of public-sector research in
  the discovery of drugs and vaccines,} \emph{New England Journal of Medicine},
  364, 535--541.

\bibitem[\protect\citeauthoryear{Toole}{Toole}{2007}]{toole_does_2007}
\textsc{Toole, A.} (2007): \enquote{Does {Public} {Scientific} {Research}
  {Complement} {Private} {Investment} in {Research} and {Development} in the
  {Pharmaceutical} {Industry}?} \emph{The Journal of Law and Economics}, 50,
  81--104.

\bibitem[\protect\citeauthoryear{Toole}{Toole}{2012}]{toole_impact_2012}
---\hspace{-.1pt}---\hspace{-.1pt}--- (2012): \enquote{The impact of public
  basic research on industrial innovation: {Evidence} from the pharmaceutical
  industry,} \emph{Research Policy}, 41, 1--12.

\bibitem[\protect\citeauthoryear{Trajtenberg, Henderson, and Jaffe}{Trajtenberg
  et~al.}{1997}]{trajtenberg1997}
\textsc{Trajtenberg, M., R.~Henderson, and A.~Jaffe} (1997):
  \enquote{University versus corporate patents: A window on the basicness of
  invention,} \emph{Economics of Innovation and new technology}, 5, 19--50.

\bibitem[\protect\citeauthoryear{Van~Raan}{Van~Raan}{2017}]{van2017patent}
\textsc{Van~Raan, A.~F.} (2017): \enquote{Patent citations analysis and its
  value in research evaluation: A review and a new approach to map
  technology-relevant research,} \emph{Journal of Data and Information
  Science}, 2, 13--50.

\bibitem[\protect\citeauthoryear{Verhoeven, Bakker, and Veugelers}{Verhoeven
  et~al.}{2016}]{verhoeven2016}
\textsc{Verhoeven, D., J.~Bakker, and R.~Veugelers} (2016): \enquote{Measuring
  technological novelty with patent-based indicators,} \emph{Research policy},
  45, 707--723.

\bibitem[\protect\citeauthoryear{Veugelers}{Veugelers}{2021}]{veugelers2021mrna}
\textsc{Veugelers, R.} (2021): \enquote{mRNA vaccines: a lucky shot?} Tech.
  rep., Bruegel Working Paper.

\bibitem[\protect\citeauthoryear{Wallace, Larivi{\`e}re, and Gingras}{Wallace
  et~al.}{2012}]{wallace2012small}
\textsc{Wallace, M.~L., V.~Larivi{\`e}re, and Y.~Gingras} (2012): \enquote{A
  small world of citations? The influence of collaboration networks on citation
  practices,} \emph{PloS one}, 7, e33339.

\bibitem[\protect\citeauthoryear{Zucker, Darby, and Brewer}{Zucker
  et~al.}{1998}]{zucker_intellectual_1998}
\textsc{Zucker, L., M.~Darby, and M.~B. Brewer} (1998): \enquote{Intellectual
  {Human} {Capital} and the {Birth} of {U}.{S}. {Biotechnology} {Enterprises},}
  \emph{American Economic Review}, 88, 290--306.

\bibitem[\protect\citeauthoryear{Zucker, Darby, and Armstrong}{Zucker
  et~al.}{2002}]{zucker2002}
\textsc{Zucker, L.~G., M.~R. Darby, and J.~S. Armstrong} (2002):
  \enquote{Commercializing knowledge: University science, knowledge capture,
  and firm performance in biotechnology,} \emph{Management science}, 48,
  138--153.

\bibitem[\protect\citeauthoryear{Zuckerman}{Zuckerman}{2021}]{zuckerman_shot_2021}
\textsc{Zuckerman, G.} (2021): \emph{A {Shot} to {Save} the {World}: {The}
  {Inside} {Story} of the {Life}-or-{Death} {Race} for a {COVID}-19 {Vaccine}},
  United States: Penguin Publishing Group.

\end{thebibliography}

\clearpage
\pagebreak
\begin{appendix}
\begin{center}
\noindent {\Large\textbf{ONLINE APPENDIX}}\\
\vspace{0.4cm}
{\Large\textbf{Leveraging Knowledge Networks: RethinkingTechnological Value Distribution in   Vaccine Innovations}}\\
Rossana Mastrandrea ~~ Fabio Montobbio ~~ Gabriele Pellegrino ~~ Massimo Riccaboni ~~ Valerio Sterzi
\end{center}

\appendix
\let\oldthesection\thesection
\renewcommand\thetable{\oldthesection\arabic{table}}
\renewcommand\thesection{Appendix \Alph{section}}   
\setcounter{table}{0}
\renewcommand\theequation{\oldthesection\arabic{equation}}
\setcounter{equation}{0}

\section{Complex Network Theory}\label{app:AppA}

 A network is mathematically described by $G=(V,E)$, where $V$ is the set of all nodes with size $|V|=N$ and $E$ the set of all edges with size $|E|=L$. The ratio between the observed number of links and the potential ones ($N(N-1)$), namely \textit{density}, quantifies its cohesiveness level. A graph is uniquely defined by the adjacency matrix $A \equiv (a_{ij})_{1 \le i,j \le N}$ with $a_{ij}=1$ if and only if it exists a link between node $i$ and node $j$. It is worth noticing that for the patent citations network generally holds  $a_{ij} \ne a_{ji}$, i.e. the network is directed/asymmetric. 
For the mRNA knowledge network we can consider different level of aggreagation: (i) network of patent citations , whose nodes represent patents and edges indicate the existence of a citation between them; (ii) network of entities, whose nodes are entities (companies, research and government institutions) and edges are weighted by the total number backward/forward citations among them; (iii) network of sectors, whose nodes are sectors and links are weighted according to the total number of backward/forward citations among them.  In the cases (ii) and (iii), the network is defined also by a weighted adjacency matrix, $W \equiv (w_{ij})_{1 \le i,j \le N}$ with generally  $w_{ij} \ne w_{ji}$.

We first investigate some basic properties (summarized in Table \ref{network}) to evaluate the network global organization and the role of nodes. We introduce the node in/out \textit{degree} and \textit{strength}: two local properties computing the total number of incoming/outcoming links of a node and the total number of forward/backward citations of a node, respectively. The distribution of node degrees allows to understand how connections are organized/placed in the network, for example through the presence and number of hubs (i,e., node with many links); while the strength distribution offers some insights about the heterogeneity of citing and cited entities. 
We also computed an higher-order property, the \textit{in-clustering coefficient}: it counts  the number of closed triangles with two links pointing to node $i$ divided by the total number of triplets involving it. This measure sheds light on the tendency of the system to form specific clusters.\footnote{Other clustering coefficients could be computed, according to the directions of the involved links. However, we selected the most relevant one for our scope.}

Node degree and strength can also be thought as local centrality measures (when normalized by the network size and total weight) as they quantify the importance of a node in terms of number of links or weighted links. However, the importance of nodes can been explored also by different quantities: (i) the betweenness centrality; (ii) the in-Katz centrality; (iii) the PageRank centrality; (iv) the Hub\&Authority scores. The betweenness centrality assigns higher value to nodes behaving as bridges helping communication flow in the network. Indeed, it counts how many shortest paths connecting any two nodes in the network pass through node $i$ with respect to all possible shortest paths existing between any pair of nodes in the web. The in-Katz centrality is based on the idea that node importance depends upon the centrality of neighbours pointing to it in a backward fashion. In other terms, it computes the relative influence of a node coming not only from its immediate neighbours, but also through indirect and longer connections in an iterative way. Intuitively, we can imagine that the closest connected nodes have more influence over the node than more distant ones. Thus when combining paths of all lengths, one can introduce an attenuation factor to give more importance to shorter walks with respect to longer ones.

Let $A = (a_{i,j})$ be the adjacency matrix of a directed graph. The Katz centrality of node $i$ is given by:

\begin{equation}\label{KO}
z_i\footnote{In the general formulation the Katz centrality is expressed in an iterative way as $z_i = \alpha \sum_k a_{i,k} \, z_k + \beta$, where $\beta$ is a constant representing some exogenous factors. Some simple computations show that two formulations are equivalent; moreover, as we are not interested in the absolute magnitude of the centrality, but in the ranking of node importance, we can assume without losing of generality $\beta = 1$.} = [(\alpha^0 A^0 + \alpha A + \alpha^2 A^2 + \dots + \alpha^k A^k + \dots )\textbf{1}]_i= \left[\sum_{k=0}^{\infty} (\alpha^k A^k)\textbf{1} \right]_i
\end{equation}

where $\alpha$ is a constant and $\textbf{1}$ is the vector $(1,1,\dots,1)$. 

The series in \eqref{KO} converges when $\alpha < 1/\rho(A)$, where $\rho(A)$ is the maximum eigenvalue of the adjacency matrix $A$, and in that case the Katz centrality in matrix form reads:

\begin{equation}\label{inkz}
z_i = [(I - \alpha A)^{-1}\textbf{1}]_i 
\end{equation}

The Katz status of a node is defined as the number of weighted paths reaching the node in the network: a generalization of the degree measure which counts only paths of length one. Notice that long paths are weighted less than short ones by exploiting the attenuation factor $\alpha$. For small (close to 0) values of $\alpha$ the contribution given by paths longer than one rapidly declines, and thus Katz scores are mainly influenced by short paths (mostly in-degrees). When the damping factor is large, long paths are devalued smoothly, and Katz scores are more influenced the by endogenous topological part of the system (it is recommended to choose $\alpha$ between $0$ and $1 / \rho(A)$).

A potential problem with the Katz centrality is that if a node with high centrality links many others then all those others get high centrality. The centrality gained by an incoming link from an important node should be \textit{diluted} if the important vertex is for example an hub. 

PageRank is an adjustment of Katz centrality taking into account this need of \lq\lq diluting\rq\rq{} the importance received from a node if it is for example an hub \citep{brin1998anatomy}. Indeed, the centrality derived from a node’ neighbors is now proportional to their centrality divided by their out-degree. According to this procedure, vertices pointing to many others transfer only a small amount of centrality to their contacts. 

 The PageRank\footnote{Also in this case, the general formulation is given by $p_i = \alpha \sum_k \frac{a_{k,i}}{d_k}p_k + \beta$.}  of node $i$ is given by:

 \begin{equation}\label{PR}
 p_i = [(I - \alpha D^{-1}A)^{-1}\textbf{1}]_i
 \end{equation}

with $\alpha$ a constant and $D$ a diagonal matrix with $i$-th diagonal element equal to node $i$ out-degree, $d_{i}$. The damping factor $\alpha$ has the same role seen for Katz centrality. In particular, $\alpha$ should be chosen between $0$ and $1 / \rho(D^{-1}A)$, the maximum eigenvalue of the matrix $D^{-1}A$.

The Hub and Authority scores \citep{kleinberg1999hubs} are recursively computed, such that a vertex has high authority centrality if it is pointed to by many hubs, i.e., by many other vertices with high hub centrality, while a vertex with high hub centrality points to many vertices with high authority centrality. 
Specifically, the Hyperlink-Induced Topic Search (HITS) is a link analysis algorithm created by Jon Kleinberg (\cite{kleinberg1999hubs}) to evaluate web pages. In the Internet terminology, a good hub is a page that links to numerous other pages, while a good authority is a page that is linked by many different hubs. In other words, nodes with high authority scores can be considered the ones containing important information about a topic, while hubs are relevant because they point to them. For this reason, in the context of
paper citations, hubs are generally referred as reviews and authorities as \lq\lq relevant papers\rq\rq{} for the topic. In our context, we can interpret as authorities
such entities that released important patents and were therefore cited by
several entities behaving as hubs. Of course, mixed-cases are possible: nodes
showing both high authority and hub scores.
Starting from this idea, the HITS algorithm assigns two scores to each page: the authority score, which measures the value of the page's content, and the hub score, which measures the value of its links to other pages. Generally speaking, the authority score of a node  is the sum of the hub scores of all nodes that point to it; the hub score of a node is the sum of the authority scores of all nodes it points to. 
Mathematically:

\begin{equation}\label{a}
 au_i = \sum_{j} hu_j , \quad  hu_i = \sum_{j} au_j 
\end{equation}

where  $au_i$ is the authority score of node $i$, $hu_i$ is the hub score of node $i$. 
Practically, after having initialized both scores to 1 for all nodes, the iterative update prescribes to compute the hub/authority score of each node according to \eqref{a} and then normalize to prevent overflow and ensure convergence. (typically done by dividing each score by the Euclidean norm of the vector of scores). The procedure continues till to reach a stable set of values.

\begin{longtable}{m{3cm} m{4cm} m{8cm}}\\
 \hline\hline
 \textbf{Network\newline property} & \quad\quad\quad \quad \textbf{Formula}& \textbf{Description}\\
 \hline\hline
 Density & \[\delta = \frac{L}{N(N-1)}\]  & Number of observed edges divided by the total potential connections \\ 
 \hline
 In Degree & \[
k^{in}_i = \sum_{j=1}^N a_{ji} 
\]
& Number of in-coming links\\
\hline
Out Degree & \[
k^{out}_i = \sum_{j=1}^N a_{ij} 
\]
& Number of out-going links\\
\hline
 In Strength & \[
s^{in}_i = \sum_{j=1}^N w_{ji} 
\]
& Number of forward citations\\
\hline
 Out Strength & \[
s^{out}_i = \sum_{j=1}^N w_{ij} 
\]
& Number of backward citations\\
\hline
In-clustering coefficient & 
\[\frac{A^T A^2}{k^{in}(k^{in}-1)}\]
 & Number of closed triangle with two links pointing to node $i$ divided by the total number of triplets involving it \\
 \hline
Betweenness Centrality & \[
b_i = \sum_{s \ne t \ne i} \frac{\sigma^i_{st}}{\sigma_{st}}
\] & Number of shortest paths connecting any pair of nodes and passing through node $i$ divided by the the total number of shortest paths \\

\hline
 Authority centrality & \[x_i = \alpha \sum_j a_{j,i} \,y_j\]  & Defined recursively as the sum of the hub score ($y_j$) of nodes pointing to node $i$. $\alpha$ is a constant\\

\hline

 Hub centrality & \[y_i = \beta \sum_j a_{j,k} \, x_j\]  & Defined recursively as the sum of the authority score ($x_j$) of nodes pointed by node $i$. $\beta$ is a constant \\

\hline
 In-Katz centrality & \[ z^{in}_i = \textbf{1}^T[(I - \alpha A)^{-1}]_i \] & Number of weighted paths reaching the node $i$ discounted by a dumping factor $\alpha$. With $\alpha < 1/\rho(A)$, $\rho(A)$ maximum eigenvalue of the adjacency matrix $A$;\textbf{1} the unit vector \\

 \hline
Page-Rank centrality & \[
 pr_i = [(I - \alpha D^{-1}A)^{-1}\textbf{1}]_i
\] & Variation of the Katz centrality: the centrality
derived from node's $i$ neighbors is proportional to their centrality divided by their out-degree. $\alpha$ is a constant, $D$ is a diagonal matrix with $d_{jj}=k^{out}_j, 1\le j\le N$ \\
 \caption{Network properties}
 \label{network}
\end{longtable}

\section{Community detection}\label{app:AppB}

We perform a community detection analysis to identify groups of nodes more densely connected to each other than to the rest of the network. One of the possible approaches aims to maximize the \textit{modularity} associated to network partitions \citep{newman2006modularity} by comparing the number of edges within communities to the expected number of edges if nodes were randomly connected preserving some local constraints (generally node degree or strength). In this context, we opted for a weighted local constraint, node strength. There is a huge number of algorithms solving the maximization problem, we chose the popular and efficient \lq\lq Louvain method\rq\rq {} \citep{blondel2008fast}. It is worth noting that the modularity score is strongly dependent on the network size, therefore we cannot compare it over the three periods. Moreover, this would be out of the scope, as we are interested more in understanding the organization of the citation networks over time after having identified the best partition according to the Louvain algorithm.

Across the three periods analyzed, we identified ten, nine, and eight communities, respectively. Figure \ref{comm} displays the ratio of communities to the overall network sizes. It highlights a significant difference between the first period and the following two periods, characterized by a large community that encompasses half of the nodes.

\begin{figure}[!ht]
\centering
\subfigure[2006-2010]
{\includegraphics[width=0.35\textwidth]{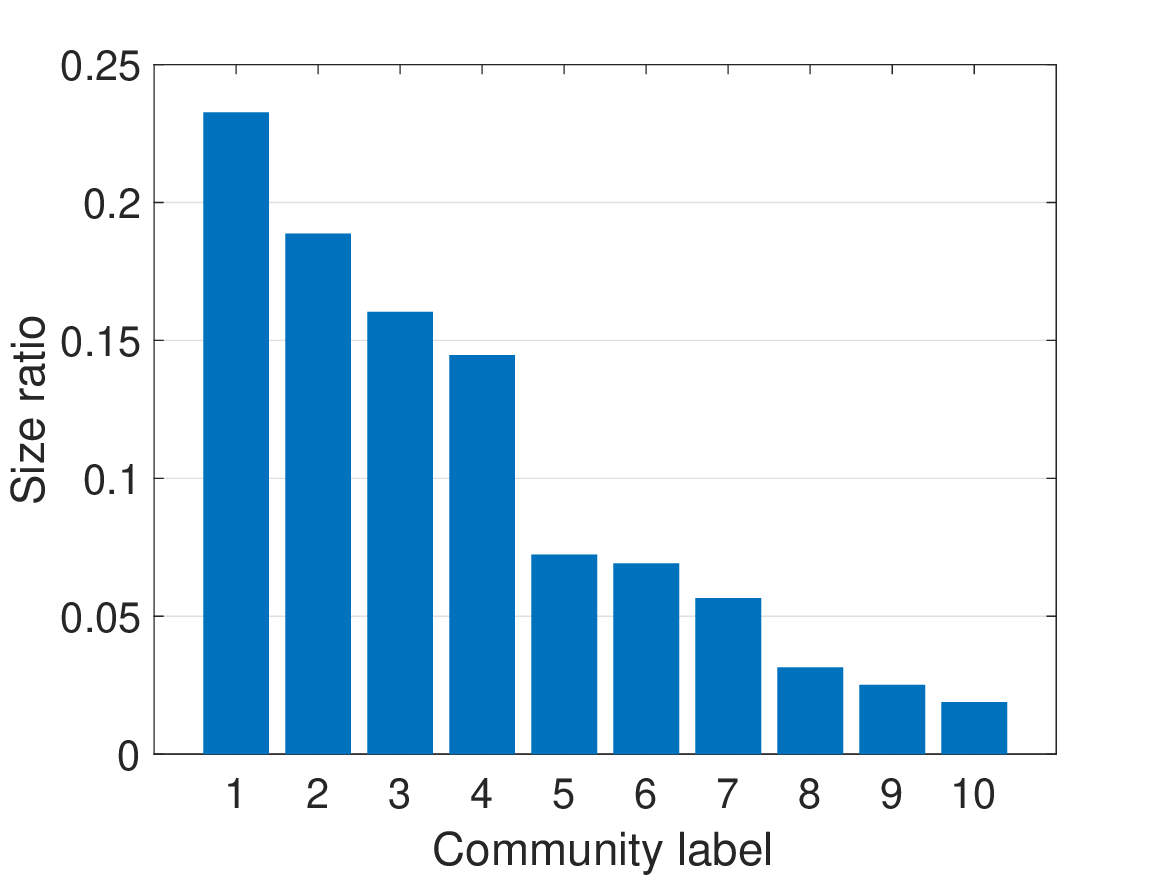}}
\hspace{-7mm}
\subfigure[2011-2015]
{\includegraphics[width=0.35\textwidth]{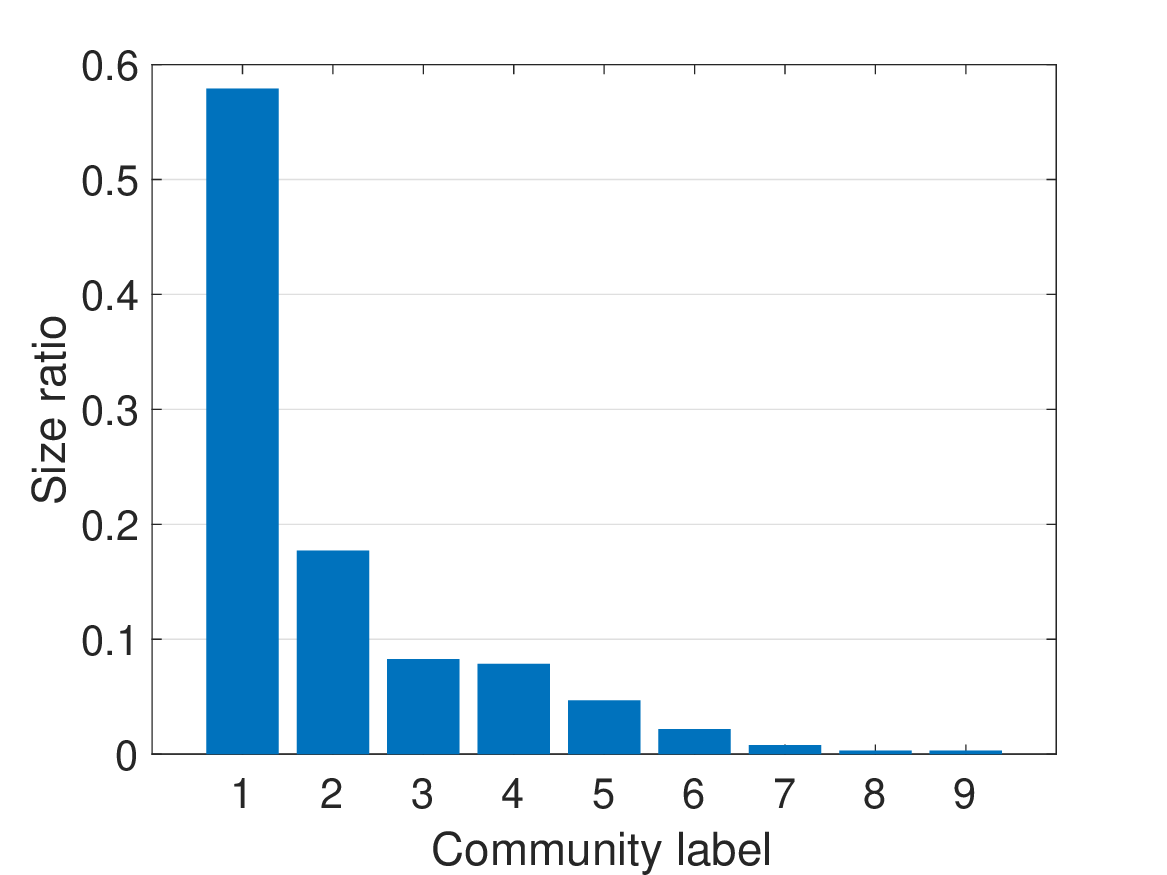}}
\hspace{-7mm}
\subfigure[2016-2020]
{\includegraphics[width=0.35\textwidth]{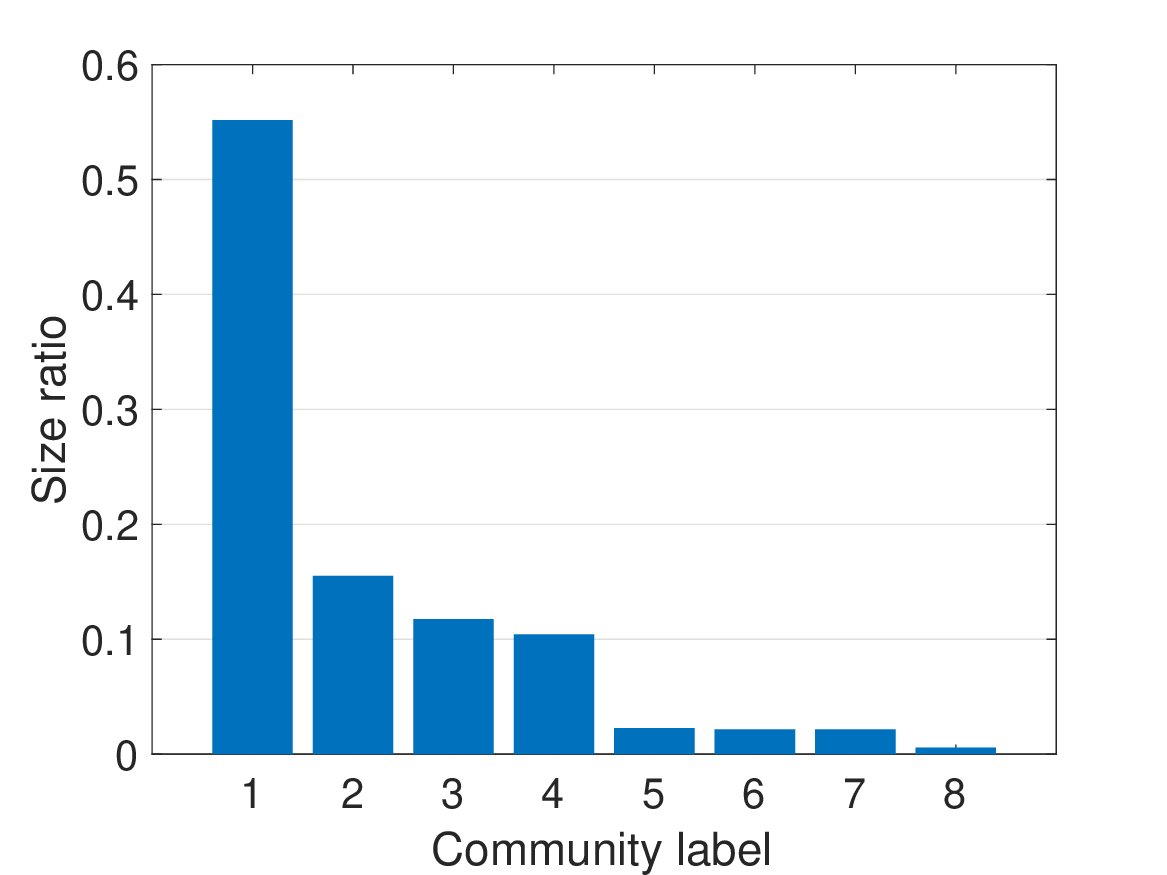}}
\caption{\textbf{Community detection.} Ratio between the communities (identified with the Louvain algorithm) and the network sizes for the three periods under study}
\label{comm}
\end{figure}

\begin{figure}[!ht]
\centering
\subfigure[2006-2010]
{\includegraphics[width=0.3\textwidth]{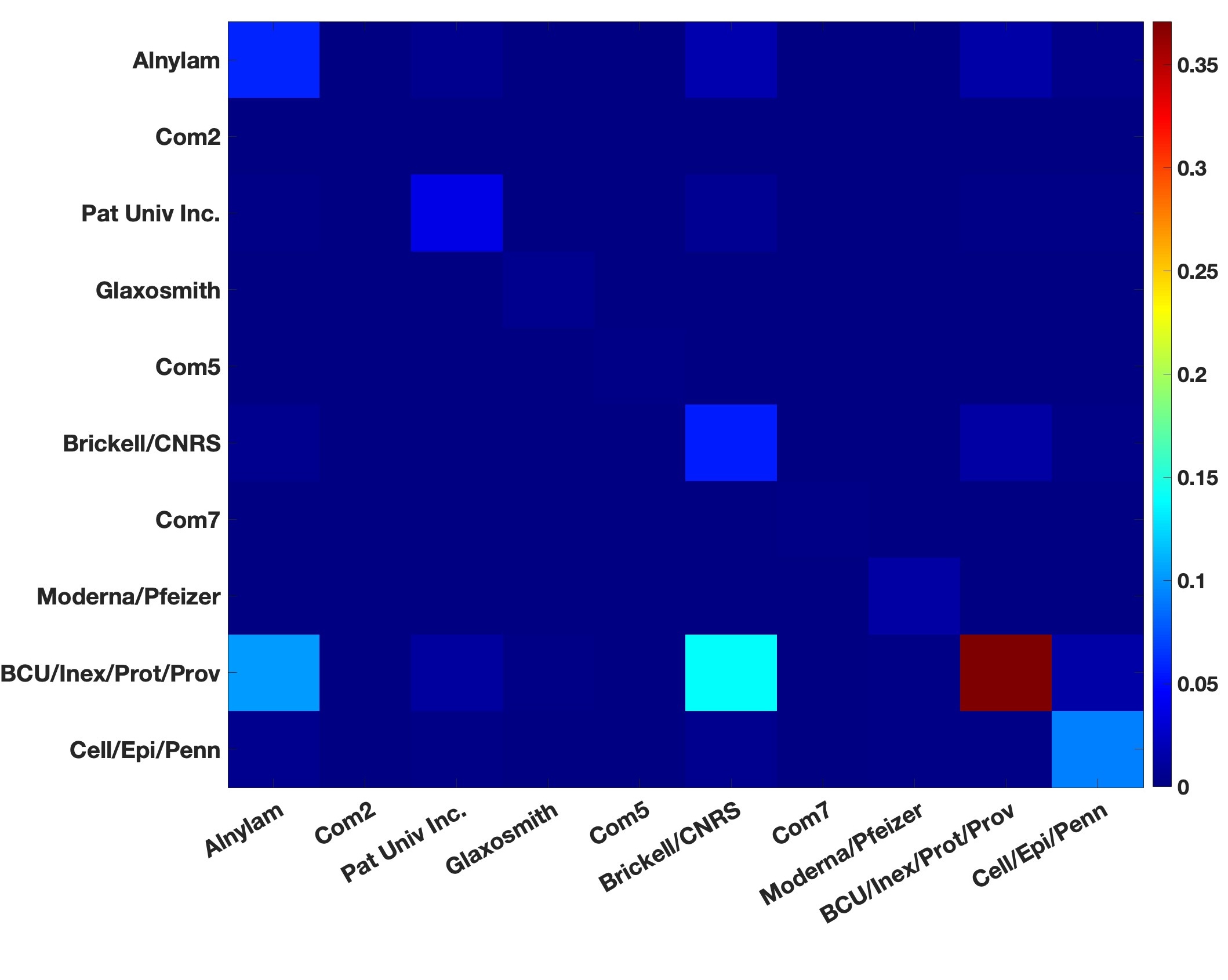}}
\subfigure[2011-2015]
{\includegraphics[width=0.31\textwidth]{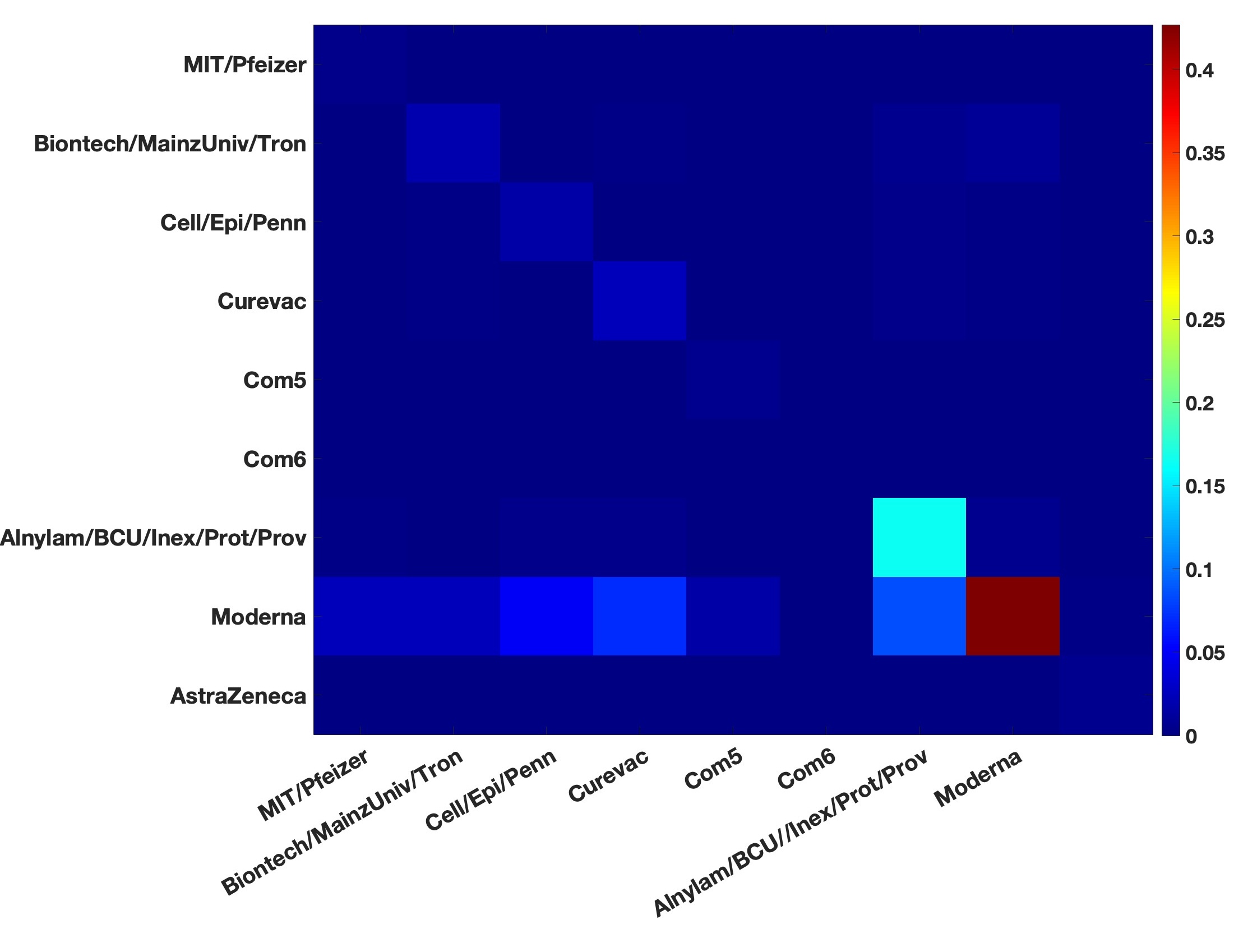}}
\subfigure[2016-2020]
{\includegraphics[width=0.33\textwidth]{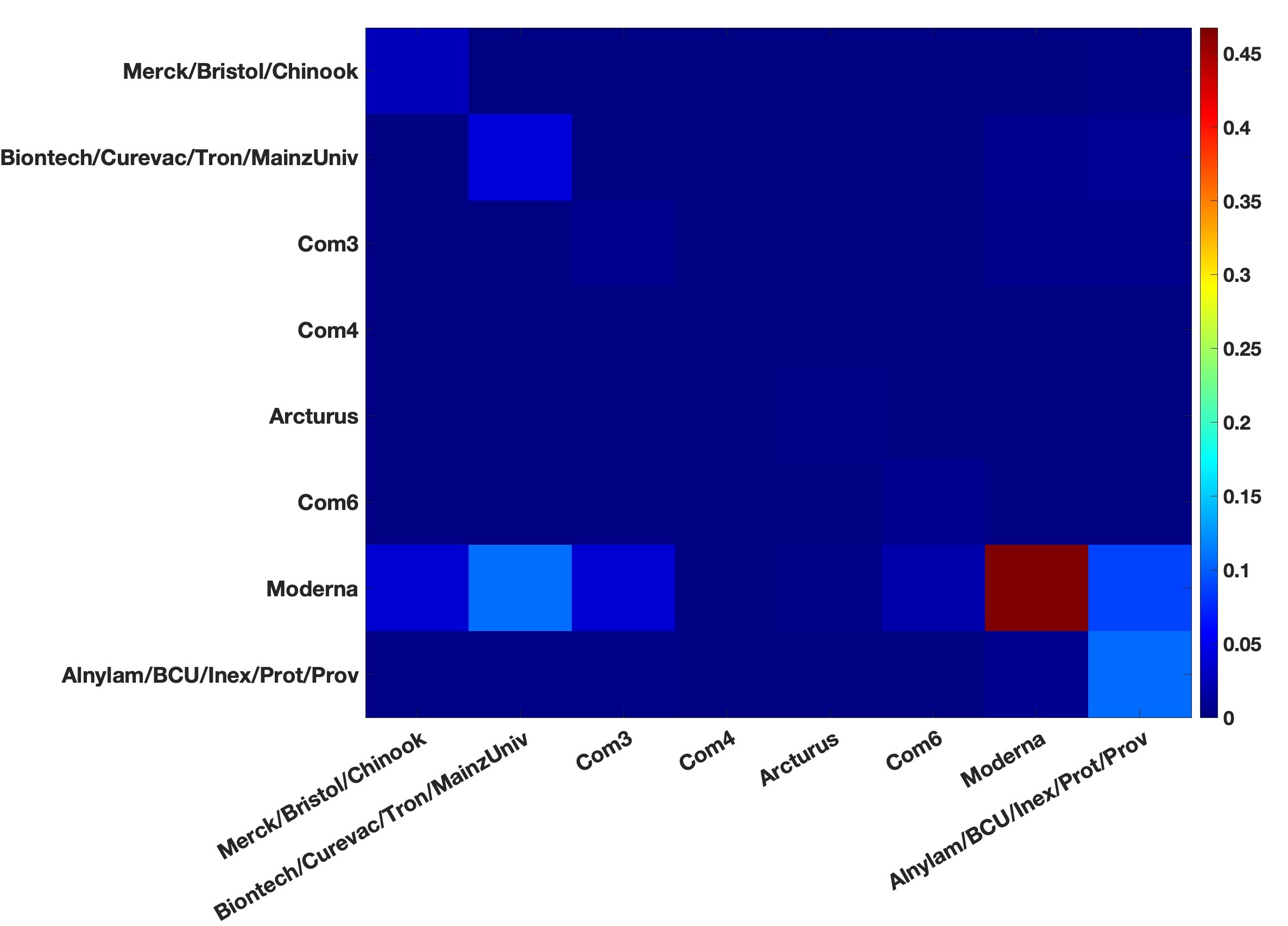}}
\caption{\textbf{Communities densities} Total number of citations within and between communities normalized with respect to the total number of citations for the specific period.}
\label{Dens}
\end{figure}

In Figure \ref{Dens} we show within and between densities of the communities subnetworks for the three periods under study.

In what follows we  investigate the structure and composition of the communities in the three periods, focusing on  the role played by specific actors. 

In the first period, the largest community contains Cellscript, Epicentre and University of Pennsylvania (fig.\ref{comm} (a)). 
This community exhibits an average in-group clustering coefficient of $0.53$, which is higher than the global one of  $0.4$, indicating  strong internal cohesion. It suggests collaborative efforts focused on fundamental research and early technological developments. The presence of the University of Pennsylvania, alongside companies like Cellscript and Epicentre, highlights the importance of academic-industry partnerships in laying the groundwork for mRNA technology. 

The second largest community shows a star-shaped configuration centered in Alnylam (citing $93\%$ of the nodes in the group), which has specialized in RNA interference (RNAi) and therapeutic applications. In fact, the average clustering coefficient appears now to be smaller than the global one ($0.19$). 

The third largest community (Fig.\ref{comm} (b)) shows a pronounced star-like organization centered in Curevac (citing the $98\%$ of nodes in the group, whose $67\%$ has in degree equal to 1, i.e. it is cited only by Curevac) as confirmed by the very small average in-clustering coefficient ($0.02$). Curevac's dominant position in its community, with a nearly absolute star-like structure, underscores its pioneering efforts in mRNA vaccine technology. The small in-clustering coefficient of Curevac's community  further highlights its role as a key innovator.

The fourth largest community comprises British Columbia University, Protiva,  Proviva  and Inex;  (fig.\ref{comm} (b)). The diverse composition of this community highlights collaborative efforts focused on LNP technology. British Columbia University and Biotechnology companies like Protiva, Proviva and Inex were instrumental in developing LNPs, which are crucial for the delivery of mRNA into cells.

\begin{figure}[!ht]
\centering
\subfigure[]
{\includegraphics[width=0.4\textwidth]{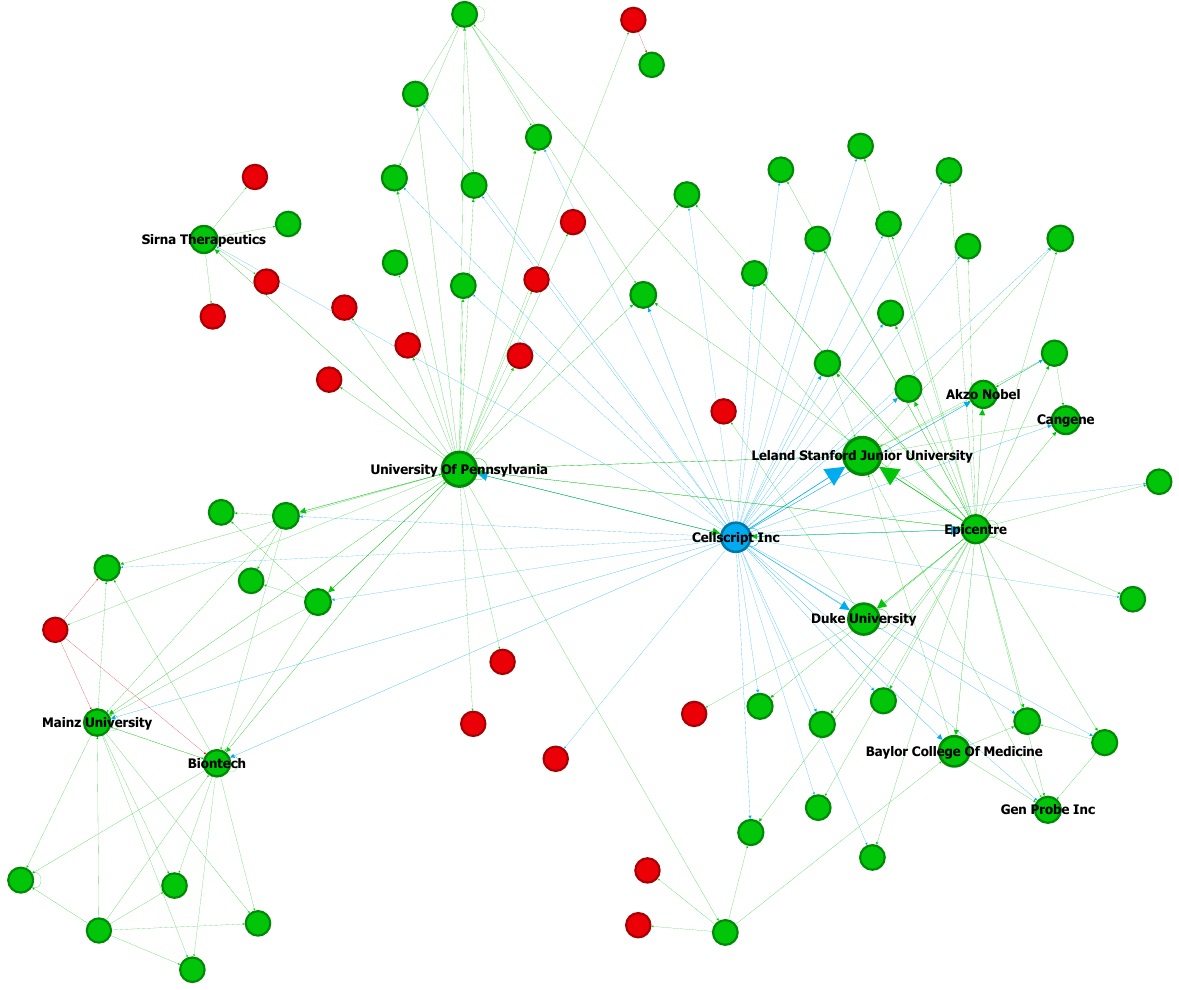}}
\subfigure[]
{\includegraphics[width=0.35\textwidth]{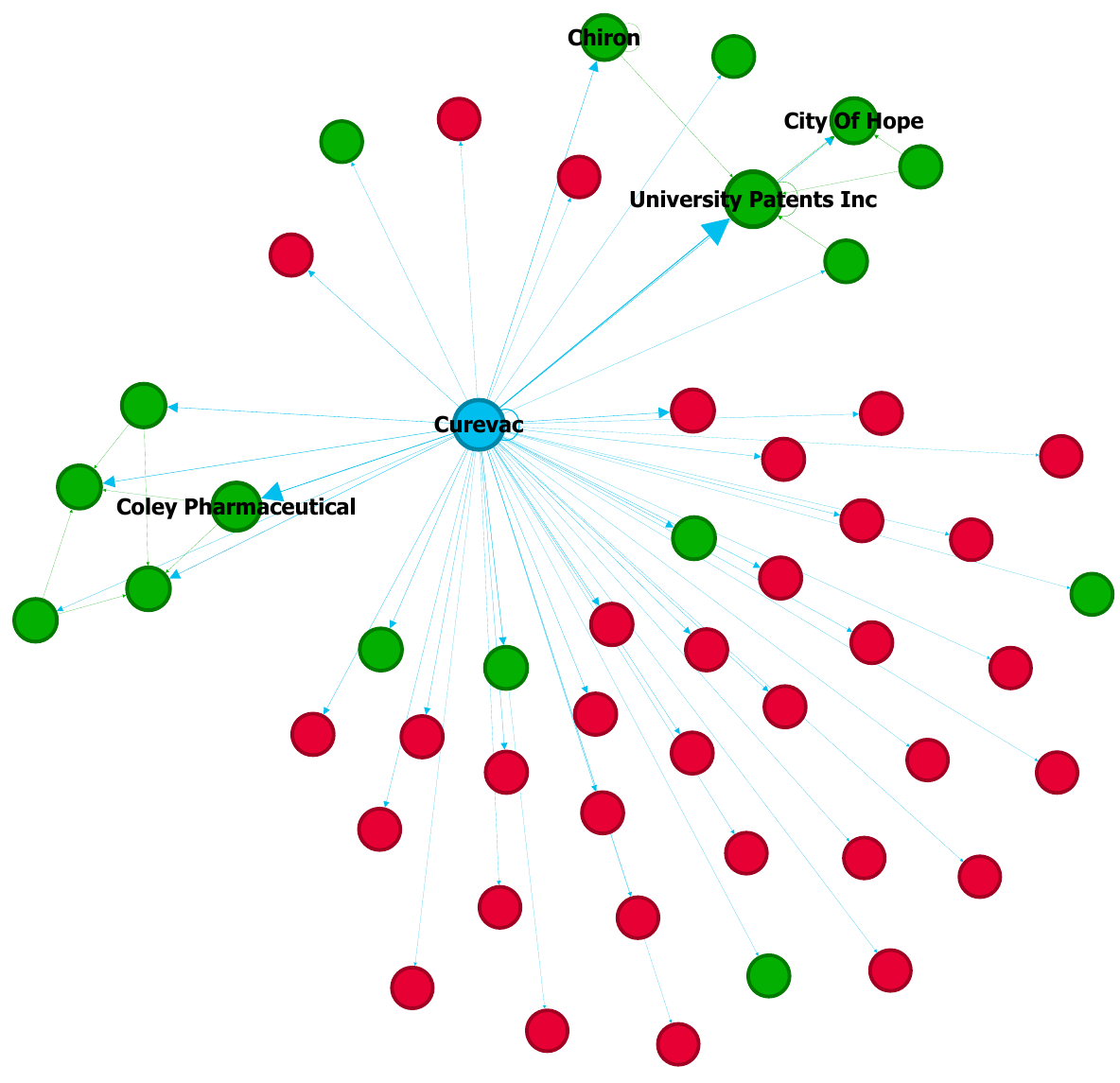}}
\subfigure[]
{\includegraphics[width=0.4\textwidth]{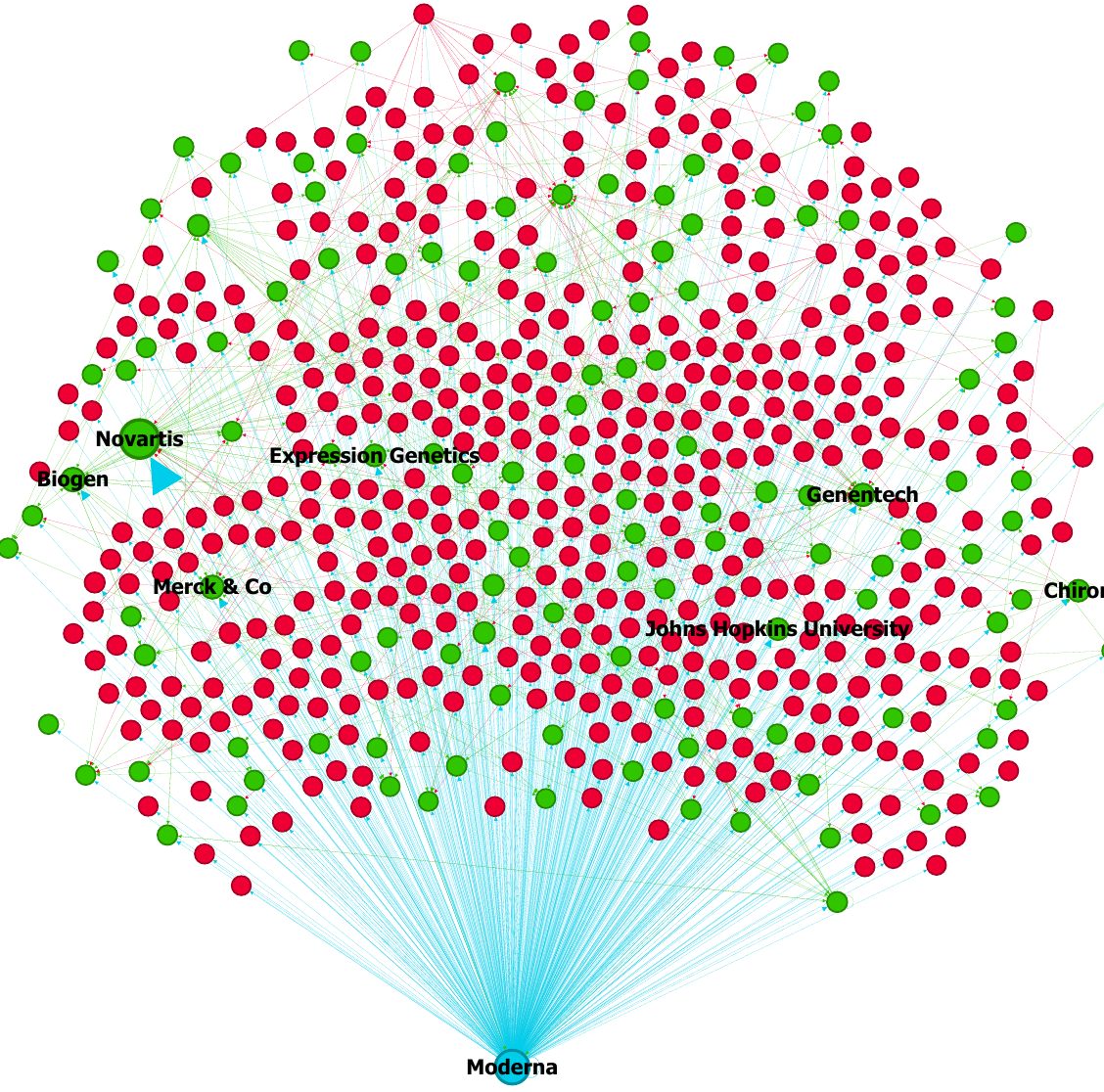}}
\subfigure[]
{\includegraphics[width=0.4\textwidth]{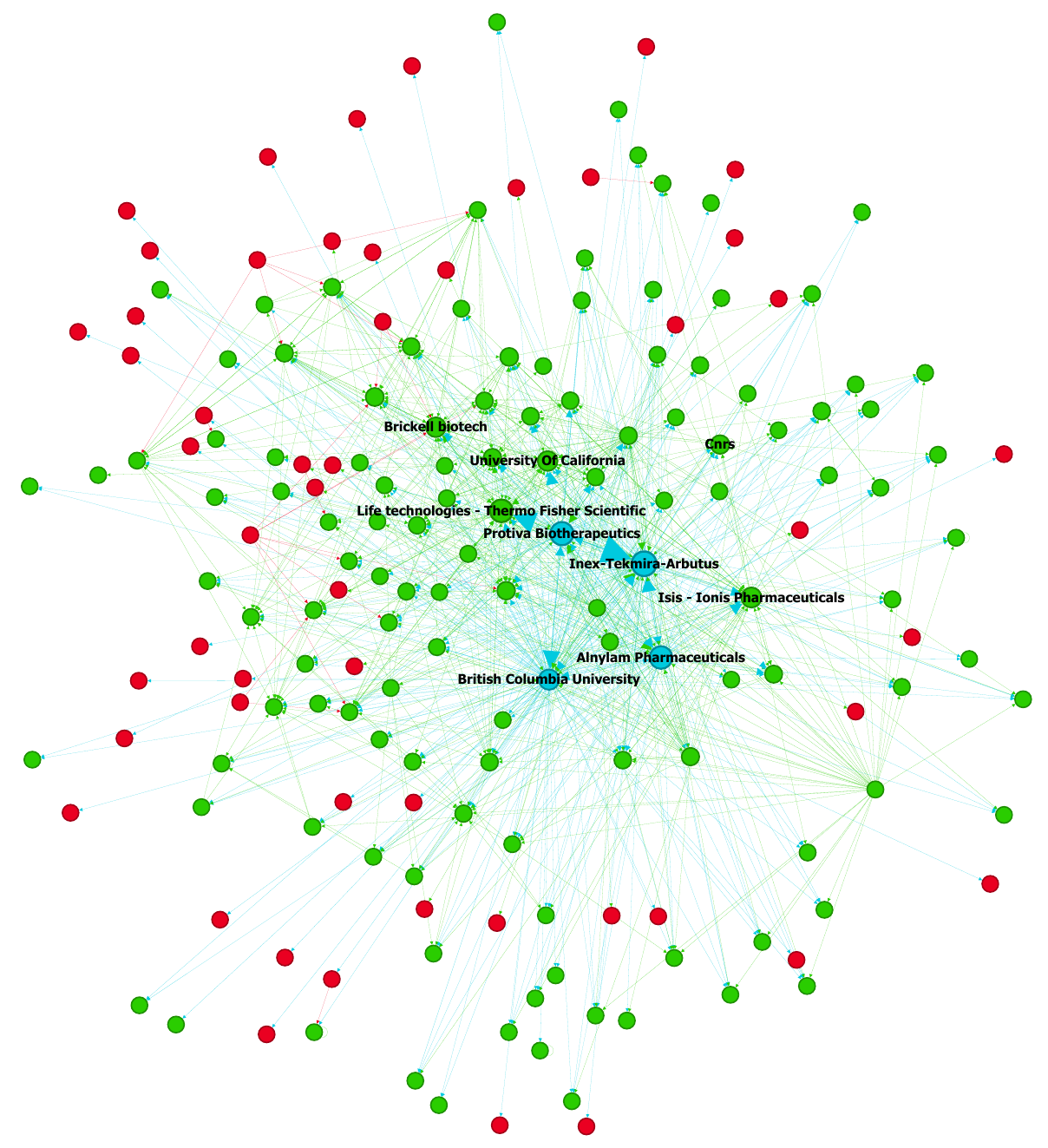}}
\subfigure[]
{\includegraphics[width=0.37\textwidth]{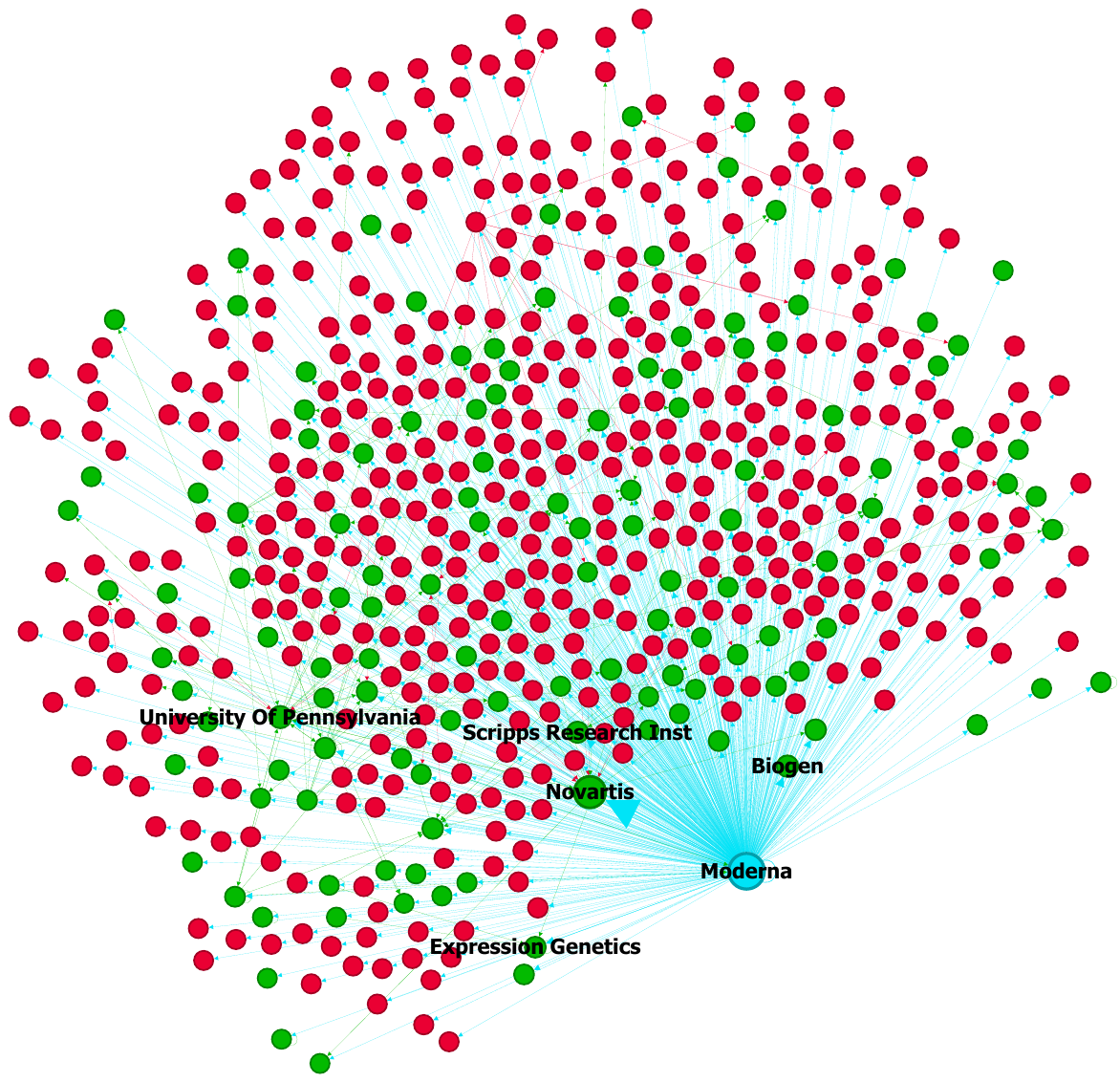}}
\subfigure[]
{\includegraphics[width=0.4\textwidth]{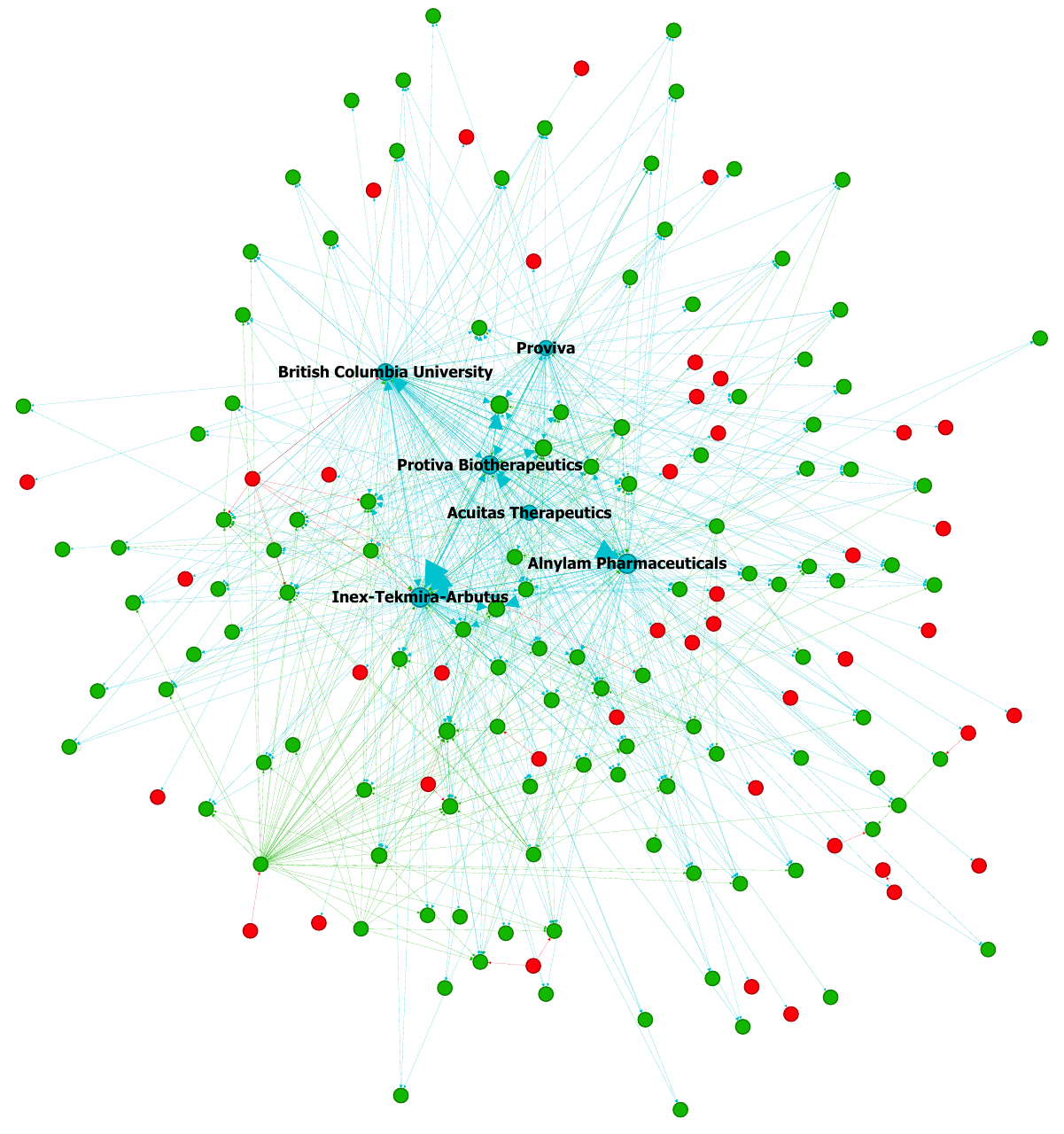}}
\caption{\textbf{Community detection analysis} Two examples of communities for the three periods under study. (a)-(b) First period; (c)-(d) second period; (e)-(f) third period. Node size is proportional to the number of forward citations; nodes color indicates: nodes with the highest out-strength (light blue); nodes receiving only one citation in the whole network (red); nodes receiving more than one citation in the whole network (green).}
\label{comm}
\end{figure}

In the second period, post 2010, the largest community encompasses $58\%$ of entities and presents a star-like organization, with Moderna pointing to the $97\%$ of nodes in the group, followed by Novartis pointing to only $6\%$ of them (Fig.\ref{comm} (c)).  Moderna tends to have a distinctive periphery of cited entities, with $73\%$ of its neighbors having an indegree of one in the entire network, meaning they are cited only by Moderna. The average in-clustering coefficient of the community is $0.09$,  much lower than the global average of $0.3$. During this period, Moderna emerges as the central innovator, reflecting its effort in developing  mRNA-based vaccines and other therapeutic advancments.  The second-largest community comprises  $18\%$ of the nodes and has a  different organizational structure  (fig.\ref{comm} (d)). The top five entities with the highest out-degrees within this community are  British Columbia University, Inex, Alnylam, Protiva and Proviva. The average of the in-clustering coefficient of this community is $0.43$, greater than the global average. The structural organization of this community suggests focused advancements in LNP technology and RNA-based therapeutics. These companies work closely to enhance delivery mechanisms crucial for effective mRNA therapeutics and vaccines. It also highlights a regional concentration of expertise and collaboration in LNP technologies within North America, in Canada in particular, contributing to global mRNA delivery innovations.

The third largest community contains the $8\%$ of actors and demostrates  high clustering, with  an average in-clustering coefficient of $0.79$, much hogher than the global average. The top citing entities are BioNTech, Tron and Mainz University, all based in Germany, idicating  a strong regional hub of innovation in mRNA vaccines within the country. Finally, another community exhibits a quasi star-like configuration centered in Curevac (also located in Germany) and includes highly clustered groups such as Cellscript, Epicenter, and the University of Pennsylvania, with an average in-clustering coefficient equal to 0.76.

In the third period (2016-2020) the largest community has the $55\%$ of entities and includes beside Moderna also Novartis and University of Pennsylvania (fig.\ref{comm} (e)). The organization remains very similar to the second period, with $78\%$ of nodes  cited only by Moderna, and the average in-clustering coefficient at $0.06$, significantly lower than the global average of $0.25$. The second largest community contains $16\%$ of actors and resembles the second largest community of the second period (fig.\ref{comm} (f)), with the top five citing entities being British Columbia University, Inex, Alnylam, Protiva and Proviva, showing high cohesion (the average in-clustering coefficient is , $0.51$, doubling the global average). The third largest component contains the $12\%$ of nodes and appears as a combination of two previously observed groups: a star-like organization centered in Curevac  and a more clustered set containing BioNTech, Tron and Mainz University, with an  average in-clustering coefficient slightly below the global average at $0.15$. Finally, the fourth largest community contains the $10\%$ of actors, with central nodes never observed in the previous periods:  Bristol Meyers Squibb, Chinook Therapeutics and Merck.\&co. having  an average in-clustering coefficient in line with the global average at $0.22$ (fig.\ref{comm} (f)). This  community could suggest an expansion of the field, with  large pharmaceutical companies indicating a more global reach, potentially encompassing entities from North America and beyond. This could mark the entry of traditional large pharma companies into the mRNA space, diversifying the technology's applications.

\begin{figure}[!ht]
\centering
{\includegraphics[width=1\textwidth]{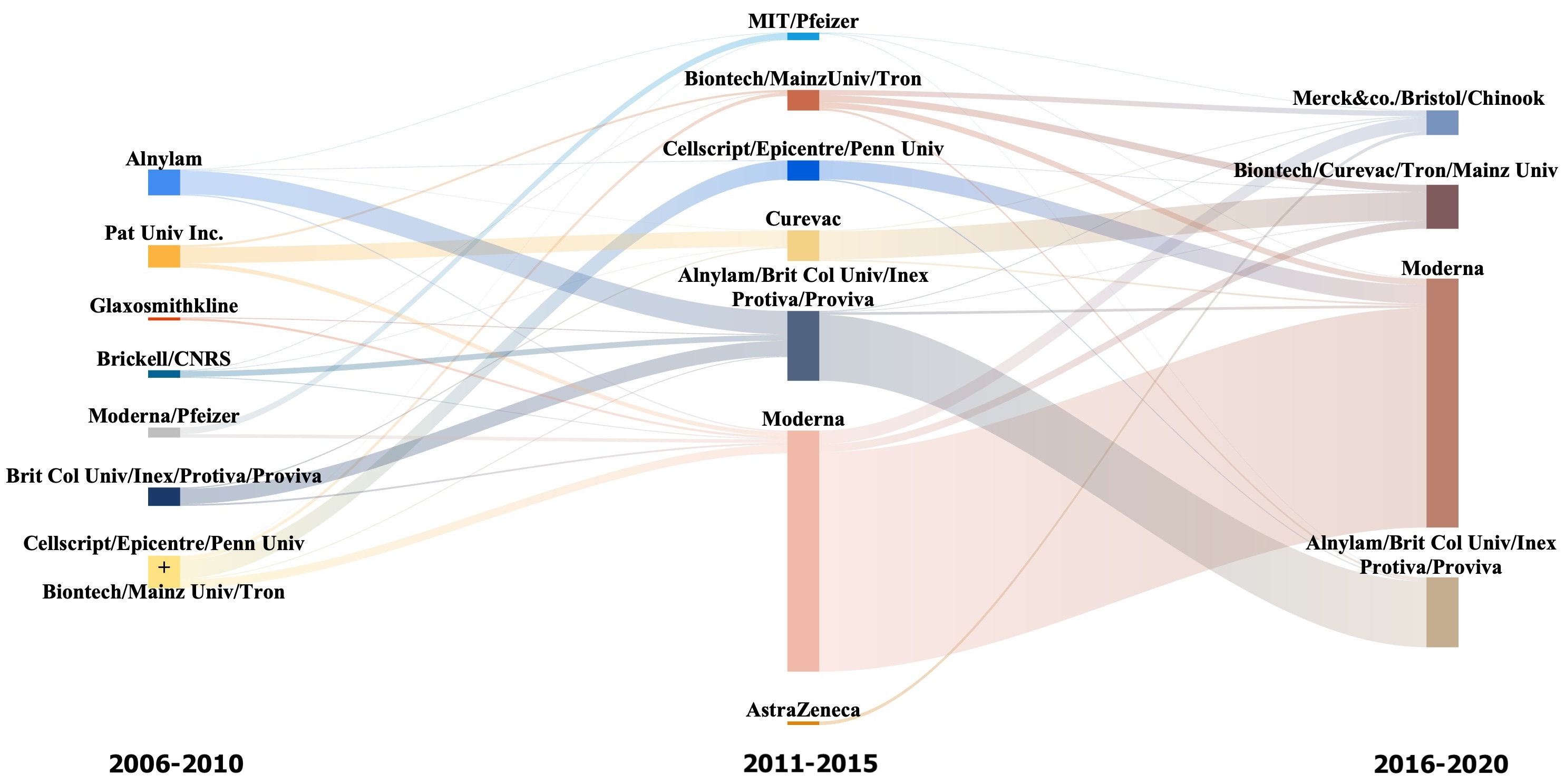}}
\caption{\textbf{Sankey plot.} \lq\lq Movements\rq\rq{} of nodes between communities from one period to the next. Each community is labeled with the names of key-actors  (as highlighted in the community description). The thickness of the flows between communities is proportional to the number of nodes, while the vertical bars represent the size of each community.}
\label{Sankey}
\end{figure}

The Sankey plot in Figure \ref{Sankey} shows the \lq\lq movements\rq\rq{} of nodes between communities from one period to the next. Each community is labelled with the name of the key-actors (as highlighted in the community description). The thickness of flows between communities is proportional to the number of nodes, while the vertical bars represent the size of the community. Notably, nodes that appar exclusively in one community during a certain period are excluded from the visualization. Only the most relevant communities are shown: eight over ten in the first period, representing $92\%$ of nodes; seven over nine in the second period representing $99\%$ of nodes; four over eight in the third period representing $93\%$ of nodes. The figure sheds light on the complex landscape of companies and institutions citations and their changes  over time. 

It is interesting to notice not only the dominant role of Moderna's community from the second period (as already observed before in terms of sizes) but also the composition of this group, which  absorbed  nodes from the Biontech/Mainz University/Tron community (from the first to the second period), while losing some companies in favour of Merck\&co group and the German community around Mainz University. Additionally, the Alnylam and British Columbia University/Inex/Protiva/Proviva communities merged from the first to the second period, with their composition remaining rather stable in the third period. Finally, the group centered in Cellscript, Epicentre and Pennsylvania University appears stable in the first and second periods but it is  absorbed by Moderna's community in the last one.

\clearpage
\section{Credit Allocation}\label{app:AppC}

\setcounter{equation}{0}
\setcounter{table}{0}

In this section we describe the  procedure to allocate credit coming from main actors of mRNA vaccine commercialization to companies and institutions according to their role in the mRNA knowledge network. 

We  consider  the weighted adjacency matrix associated to the mRNA knowledge network aggregated over the whole period under study, $W_{ij}\equiv (w_{ij})_{1\le i,j \le N}$; then, we compute the transition probability matrix associated to it, $M \equiv (m_{ij})_{1 \le i,j \le N}$, and its $d$th powers:

\begin{equation}\label{MK}
m_{ij} = \frac{w_{ij}}{s^{out}_i}, \quad M^d = (m_{ij,d})^{d=1,2,\dots}_{1\le i,j \le N}
\end{equation}
 
 where $m_{ij,d}$ represents the probability to reach node $j$ in $d$ steps for a random walker placed on node $i$ . For $d=1$ the probability associated to each link out going from node $i$ simply corresponds to its weight normalized by the node $i$ out strength (as in \eqref{MK}): the higher the number of backward citations the higher the probability to be the destination of the random walker first step. For $d=2$, $m_{ij,2}=\sum_{l} m_{il}m_{lj}$ gives the sum of all weighted paths of length $2$ going from node $i$ to node $j$ and so on for $d>2$. 
 It is worth noticing that since all directions where a link is present are allowed, the same company can be assigned several shares of node $i$ credit during the redistribution process. The same holds for node $i$ itself, as self loops are not excluded from the system. 
 Once the transition probability matrix and its powers have been computed, its combination with the dumping factor $\beta$ allows to compute the credit assigned to each node according to three different rules:

\begin{enumerate}

 \item Markov approach
 
 \begin{equation}\label{MKV}
 \pi^d_j = \left(\frac{m^d_{ij}}{\sum\limits_{h \in V_i(d)} m^d_{ih}}\right)\beta^d
 \end{equation}
where $V_i(d)$ is the set of all nodes reachable from node $i$ in $d$ steps. It is worth noticing that, except for $d=1$, the rows of $M^d$ do not sum to $1$. However, it is necessary to normalize them as in \eqref{MKV} to properly distribute the share of node $i$ profit assigned at distance level $d$. 

\item Markov and in-Katz centrality

\begin{equation}\label{MKVK}
 \pi^d_j = \left(\frac{m^d_{ij}}{\sum\limits_{h \in V_i(d)}m^d_{ih}}\right)\left( \frac{z_h}{\sum\limits_{h \in V_i(d)} z_h}\right)\beta^d
 \end{equation}

 where $z_i$ is the in-Katz centrality computed on the whole network according to \eqref{inkz}

\item Markov and PageRank centrality

\begin{equation}\label{MKVP}
 \pi^d_j = \left(\frac{m^d_{ij}}{\sum\limits_{h \in V_i(d)} m^d_{ih}}\right)\left(\frac{pr_h}{\sum\limits_{h \in V_i(d)} pr_h}\right)\beta^d
 \end{equation}
where $pr_i$ is the PageRank centrality computed on the whole network according to \eqref{PR}.
\end{enumerate}

\subsubsection*{Example: profit redistribution}

Let us assume that node $i$ has profit $\Pi_i=100$ (coming for example from the commercialization of the mRNA vaccine) and let us assume that half of this profit remains to node $i$ (as the patent's owner and the one who succeeded in the commercialization), while the remaining profit will be redistributed among the other patents' assignees according to their technological and scientific contributions to the development of the vaccine. These quantities can be computed from the mRNA knowledge network using one of the three procedures introduced before (see tables \ref{tab:tab_Mod}, \ref{tab:tab_Bio} for the main contributors of Moderna and BioNTech vaccines development). Hence,  we can introduce a parameter $\alpha$ such that the credit that remains to node $i$ is $\alpha\Pi_i=50$, while $(1-\alpha)\Pi_i=50$ is the credit to be allocated.  Table \ref{profit} reports the credit to be divided among all nodes at distance $d \in\{1,2,3,\dots \}$ from node $i$,   according to our choices of the parameters $\alpha=\beta=0.5$.

\begin{table}[!ht]
\begin{center}
\begin{tabular}{c c} 
 \hline
\textbf{Distance value ($d$)} & \textbf{Credit to allocate}\\ [0.5ex] 
 \hline\hline
1 & 25\\ 
 \hline
 2 & 12.5 \\
 \hline
 3 & 6.25 \\
 \hline
 4 & 3.125 \\
 \hline
 \dots & \dots \\ [1ex] 
 \hline
\end{tabular}
\end{center}
\caption{Credit share assigned to each distance value, $d$, and to be allocated among all nodes at such distance from node $i$.}
\label{profit}
\end{table}

 In other words,  the $25\%$ of $\Pi_i$ will be redistributed among all nodes at distance 1 from node $i$ (i.e., directly cited by it), the $12.5\%$ will be redistributed among all nodes at distance 2 from node $i$ and so on. Generally speaking, the percentage of node $i$ credit to be allocated at distance $d$ from node $i$ is calculated as $\pi_d=\alpha\beta^d\Pi$ (such that $\sum_d^{\infty} \pi_d = \sum_d^{\infty} \alpha\beta^d\Pi_i = \Pi_i$), with $0 < \alpha,\beta < 1$.
Coming back to our example, all nodes at distance $d=1$ from node i, will be allocated a share of the credit assigned to this \lq\lq level\rq\rq{}, i.e. 25, according to their importance computed with one of the three aforementioned quantities. Hence, node $j$ directly pointed by node $i$

\begin{enumerate}

\item will receive $p_{j,1}= 25*(w_{ij}/s^{out}_i)$

\item  will receive  $p_{j,1}= 25*(w_{ij}/s^{out}_i)*z^{norm}_j$

\item will receive $p_{j,1}= 25*(w_{ij}/s^{out}_i)*pr^{norm}_j$

\end{enumerate}

where $z^{norm}_j$ and $pr^{norm}_j$ simply indicates the in-Katz and the PageRank centrality of node $j$, respectively, normalized with respect to the centrality values of all nodes at distance $d$ from node $i$ ($d=1$ in this example).

\clearpage

\end{appendix}

\end{document}